\documentclass{book}
\usepackage[english]{babel}
\usepackage[square,numbers,sort&compress]{natbib}
\usepackage{appendix}
\DeclareOldFontCommand{\bf}{\normalfont\bfseries}{\mathbf}
\DeclareOldFontCommand{\tt}{\normalfont\ttfamily}{}

\usepackage{latexsym}
\usepackage{amsfonts}
\usepackage{amsmath}
\usepackage{xcolor}
\usepackage{graphicx}
\usepackage{mathptmx}      
\usepackage{subfigure}
\usepackage{listings}
\usepackage{grffile}
\usepackage{enumitem}
\usepackage{float}

\definecolor{AliceBlue}{rgb}{0.94,0.97,1.00}
\definecolor{AntiqueWhite1}{rgb}{1.00,0.94,0.86}
\definecolor{AntiqueWhite2}{rgb}{0.93,0.87,0.80}
\definecolor{AntiqueWhite3}{rgb}{0.80,0.75,0.69}
\definecolor{AntiqueWhite4}{rgb}{0.55,0.51,0.47}
\definecolor{AntiqueWhite}{rgb}{0.98,0.92,0.84}
\definecolor{BlanchedAlmond}{rgb}{1.00,0.92,0.80}
\definecolor{BlueViolet}{rgb}{0.54,0.17,0.89}
\definecolor{CadetBlue1}{rgb}{0.60,0.96,1.00}
\definecolor{CadetBlue2}{rgb}{0.56,0.90,0.93}
\definecolor{CadetBlue3}{rgb}{0.48,0.77,0.80}
\definecolor{CadetBlue4}{rgb}{0.33,0.53,0.55}
\definecolor{CadetBlue}{rgb}{0.37,0.62,0.63}
\definecolor{CornflowerBlue}{rgb}{0.39,0.58,0.93}
\definecolor{DarkBlue}{rgb}{0.00,0.00,0.55}
\definecolor{DarkCyan}{rgb}{0.00,0.55,0.55}
\definecolor{DarkGoldenrod1}{rgb}{1.00,0.73,0.06}
\definecolor{DarkGoldenrod2}{rgb}{0.93,0.68,0.05}
\definecolor{DarkGoldenrod3}{rgb}{0.80,0.58,0.05}
\definecolor{DarkGoldenrod4}{rgb}{0.55,0.40,0.03}
\definecolor{DarkGoldenrod}{rgb}{0.72,0.53,0.04}
\definecolor{DarkGray}{rgb}{0.66,0.66,0.66}
\definecolor{DarkGreen}{rgb}{0.00,0.39,0.00}
\definecolor{DarkGrey}{rgb}{0.66,0.66,0.66}
\definecolor{DarkKhaki}{rgb}{0.74,0.72,0.42}
\definecolor{DarkMagenta}{rgb}{0.55,0.00,0.55}
\definecolor{DarkOliveGreen1}{rgb}{0.79,1.00,0.44}
\definecolor{DarkOliveGreen2}{rgb}{0.74,0.93,0.41}
\definecolor{DarkOliveGreen3}{rgb}{0.64,0.80,0.35}
\definecolor{DarkOliveGreen4}{rgb}{0.43,0.55,0.24}
\definecolor{DarkOliveGreen}{rgb}{0.33,0.42,0.18}
\definecolor{DarkOrange1}{rgb}{1.00,0.50,0.00}
\definecolor{DarkOrange2}{rgb}{0.93,0.46,0.00}
\definecolor{DarkOrange3}{rgb}{0.80,0.40,0.00}
\definecolor{DarkOrange4}{rgb}{0.55,0.27,0.00}
\definecolor{DarkOrange}{rgb}{1.00,0.55,0.00}
\definecolor{DarkOrchid1}{rgb}{0.75,0.24,1.00}
\definecolor{DarkOrchid2}{rgb}{0.70,0.23,0.93}
\definecolor{DarkOrchid3}{rgb}{0.60,0.20,0.80}
\definecolor{DarkOrchid4}{rgb}{0.41,0.13,0.55}
\definecolor{DarkOrchid}{rgb}{0.60,0.20,0.80}
\definecolor{DarkRed}{rgb}{0.55,0.00,0.00}
\definecolor{DarkSalmon}{rgb}{0.91,0.59,0.48}
\definecolor{DarkSeaGreen1}{rgb}{0.76,1.00,0.76}
\definecolor{DarkSeaGreen2}{rgb}{0.71,0.93,0.71}
\definecolor{DarkSeaGreen3}{rgb}{0.61,0.80,0.61}
\definecolor{DarkSeaGreen4}{rgb}{0.41,0.55,0.41}
\definecolor{DarkSeaGreen}{rgb}{0.56,0.74,0.56}
\definecolor{DarkSlateBlue}{rgb}{0.28,0.24,0.55}
\definecolor{DarkSlateGray1}{rgb}{0.59,1.00,1.00}
\definecolor{DarkSlateGray2}{rgb}{0.55,0.93,0.93}
\definecolor{DarkSlateGray3}{rgb}{0.47,0.80,0.80}
\definecolor{DarkSlateGray4}{rgb}{0.32,0.55,0.55}
\definecolor{DarkSlateGray}{rgb}{0.18,0.31,0.31}
\definecolor{DarkSlateGrey}{rgb}{0.18,0.31,0.31}
\definecolor{DarkTurquoise}{rgb}{0.00,0.81,0.82}
\definecolor{DarkViolet}{rgb}{0.58,0.00,0.83}
\definecolor{DeepPink1}{rgb}{1.00,0.08,0.58}
\definecolor{DeepPink2}{rgb}{0.93,0.07,0.54}
\definecolor{DeepPink3}{rgb}{0.80,0.06,0.46}
\definecolor{DeepPink4}{rgb}{0.55,0.04,0.31}
\definecolor{DeepPink}{rgb}{1.00,0.08,0.58}
\definecolor{DeepSkyBlue1}{rgb}{0.00,0.75,1.00}
\definecolor{DeepSkyBlue2}{rgb}{0.00,0.70,0.93}
\definecolor{DeepSkyBlue3}{rgb}{0.00,0.60,0.80}
\definecolor{DeepSkyBlue4}{rgb}{0.00,0.41,0.55}
\definecolor{DeepSkyBlue}{rgb}{0.00,0.75,1.00}
\definecolor{DimGray}{rgb}{0.41,0.41,0.41}
\definecolor{DimGrey}{rgb}{0.41,0.41,0.41}
\definecolor{DodgerBlue1}{rgb}{0.12,0.56,1.00}
\definecolor{DodgerBlue2}{rgb}{0.11,0.53,0.93}
\definecolor{DodgerBlue3}{rgb}{0.09,0.45,0.80}
\definecolor{DodgerBlue4}{rgb}{0.06,0.31,0.55}
\definecolor{DodgerBlue}{rgb}{0.12,0.56,1.00}
\definecolor{FloralWhite}{rgb}{1.00,0.98,0.94}
\definecolor{ForestGreen}{rgb}{0.13,0.55,0.13}
\definecolor{GhostWhite}{rgb}{0.97,0.97,1.00}
\definecolor{GreenYellow}{rgb}{0.68,1.00,0.18}
\definecolor{HotPink1}{rgb}{1.00,0.43,0.71}
\definecolor{HotPink2}{rgb}{0.93,0.42,0.65}
\definecolor{HotPink3}{rgb}{0.80,0.38,0.56}
\definecolor{HotPink4}{rgb}{0.55,0.23,0.38}
\definecolor{HotPink}{rgb}{1.00,0.41,0.71}
\definecolor{IndianRed1}{rgb}{1.00,0.42,0.42}
\definecolor{IndianRed2}{rgb}{0.93,0.39,0.39}
\definecolor{IndianRed3}{rgb}{0.80,0.33,0.33}
\definecolor{IndianRed4}{rgb}{0.55,0.23,0.23}
\definecolor{IndianRed}{rgb}{0.80,0.36,0.36}
\definecolor{LavenderBlush1}{rgb}{1.00,0.94,0.96}
\definecolor{LavenderBlush2}{rgb}{0.93,0.88,0.90}
\definecolor{LavenderBlush3}{rgb}{0.80,0.76,0.77}
\definecolor{LavenderBlush4}{rgb}{0.55,0.51,0.53}
\definecolor{LavenderBlush}{rgb}{1.00,0.94,0.96}
\definecolor{LawnGreen}{rgb}{0.49,0.99,0.00}
\definecolor{LemonChiffon1}{rgb}{1.00,0.98,0.80}
\definecolor{LemonChiffon2}{rgb}{0.93,0.91,0.75}
\definecolor{LemonChiffon3}{rgb}{0.80,0.79,0.65}
\definecolor{LemonChiffon4}{rgb}{0.55,0.54,0.44}
\definecolor{LemonChiffon}{rgb}{1.00,0.98,0.80}
\definecolor{LightBlue1}{rgb}{0.75,0.94,1.00}
\definecolor{LightBlue2}{rgb}{0.70,0.87,0.93}
\definecolor{LightBlue3}{rgb}{0.60,0.75,0.80}
\definecolor{LightBlue4}{rgb}{0.41,0.51,0.55}
\definecolor{LightBlue}{rgb}{0.68,0.85,0.90}
\definecolor{LightCoral}{rgb}{0.94,0.50,0.50}
\definecolor{LightCyan1}{rgb}{0.88,1.00,1.00}
\definecolor{LightCyan2}{rgb}{0.82,0.93,0.93}
\definecolor{LightCyan3}{rgb}{0.71,0.80,0.80}
\definecolor{LightCyan4}{rgb}{0.48,0.55,0.55}
\definecolor{LightCyan}{rgb}{0.88,1.00,1.00}
\definecolor{LightGoldenrod1}{rgb}{1.00,0.93,0.55}
\definecolor{LightGoldenrod2}{rgb}{0.93,0.86,0.51}
\definecolor{LightGoldenrod3}{rgb}{0.80,0.75,0.44}
\definecolor{LightGoldenrod4}{rgb}{0.55,0.51,0.30}
\definecolor{LightGoldenrodYellow}{rgb}{0.98,0.98,0.82}
\definecolor{LightGoldenrod}{rgb}{0.93,0.87,0.51}
\definecolor{LightGray}{rgb}{0.83,0.83,0.83}
\definecolor{LightGreen}{rgb}{0.56,0.93,0.56}
\definecolor{LightGrey}{rgb}{0.83,0.83,0.83}
\definecolor{LightPink1}{rgb}{1.00,0.68,0.73}
\definecolor{LightPink2}{rgb}{0.93,0.64,0.68}
\definecolor{LightPink3}{rgb}{0.80,0.55,0.58}
\definecolor{LightPink4}{rgb}{0.55,0.37,0.40}
\definecolor{LightPink}{rgb}{1.00,0.71,0.76}
\definecolor{LightSalmon1}{rgb}{1.00,0.63,0.48}
\definecolor{LightSalmon2}{rgb}{0.93,0.58,0.45}
\definecolor{LightSalmon3}{rgb}{0.80,0.51,0.38}
\definecolor{LightSalmon4}{rgb}{0.55,0.34,0.26}
\definecolor{LightSalmon}{rgb}{1.00,0.63,0.48}
\definecolor{LightSeaGreen}{rgb}{0.13,0.70,0.67}
\definecolor{LightSkyBlue1}{rgb}{0.69,0.89,1.00}
\definecolor{LightSkyBlue2}{rgb}{0.64,0.83,0.93}
\definecolor{LightSkyBlue3}{rgb}{0.55,0.71,0.80}
\definecolor{LightSkyBlue4}{rgb}{0.38,0.48,0.55}
\definecolor{LightSkyBlue}{rgb}{0.53,0.81,0.98}
\definecolor{LightSlateBlue}{rgb}{0.52,0.44,1.00}
\definecolor{LightSlateGray}{rgb}{0.47,0.53,0.60}
\definecolor{LightSlateGrey}{rgb}{0.47,0.53,0.60}
\definecolor{LightSteelBlue1}{rgb}{0.79,0.88,1.00}
\definecolor{LightSteelBlue2}{rgb}{0.74,0.82,0.93}
\definecolor{LightSteelBlue3}{rgb}{0.64,0.71,0.80}
\definecolor{LightSteelBlue4}{rgb}{0.43,0.48,0.55}
\definecolor{LightSteelBlue}{rgb}{0.69,0.77,0.87}
\definecolor{LightYellow1}{rgb}{1.00,1.00,0.88}
\definecolor{LightYellow2}{rgb}{0.93,0.93,0.82}
\definecolor{LightYellow3}{rgb}{0.80,0.80,0.71}
\definecolor{LightYellow4}{rgb}{0.55,0.55,0.48}
\definecolor{LightYellow}{rgb}{1.00,1.00,0.88}
\definecolor{LimeGreen}{rgb}{0.20,0.80,0.20}
\definecolor{MediumAquamarine}{rgb}{0.40,0.80,0.67}
\definecolor{MediumBlue}{rgb}{0.00,0.00,0.80}
\definecolor{MediumOrchid1}{rgb}{0.88,0.40,1.00}
\definecolor{MediumOrchid2}{rgb}{0.82,0.37,0.93}
\definecolor{MediumOrchid3}{rgb}{0.71,0.32,0.80}
\definecolor{MediumOrchid4}{rgb}{0.48,0.22,0.55}
\definecolor{MediumOrchid}{rgb}{0.73,0.33,0.83}
\definecolor{MediumPurple1}{rgb}{0.67,0.51,1.00}
\definecolor{MediumPurple2}{rgb}{0.62,0.47,0.93}
\definecolor{MediumPurple3}{rgb}{0.54,0.41,0.80}
\definecolor{MediumPurple4}{rgb}{0.36,0.28,0.55}
\definecolor{MediumPurple}{rgb}{0.58,0.44,0.86}
\definecolor{MediumSeaGreen}{rgb}{0.24,0.70,0.44}
\definecolor{MediumSlateBlue}{rgb}{0.48,0.41,0.93}
\definecolor{MediumSpringGreen}{rgb}{0.00,0.98,0.60}
\definecolor{MediumTurquoise}{rgb}{0.28,0.82,0.80}
\definecolor{MediumVioletRed}{rgb}{0.78,0.08,0.52}
\definecolor{MidnightBlue}{rgb}{0.10,0.10,0.44}
\definecolor{MintCream}{rgb}{0.96,1.00,0.98}
\definecolor{MistyRose1}{rgb}{1.00,0.89,0.88}
\definecolor{MistyRose2}{rgb}{0.93,0.84,0.82}
\definecolor{MistyRose3}{rgb}{0.80,0.72,0.71}
\definecolor{MistyRose4}{rgb}{0.55,0.49,0.48}
\definecolor{MistyRose}{rgb}{1.00,0.89,0.88}
\definecolor{NavajoWhite1}{rgb}{1.00,0.87,0.68}
\definecolor{NavajoWhite2}{rgb}{0.93,0.81,0.63}
\definecolor{NavajoWhite3}{rgb}{0.80,0.70,0.55}
\definecolor{NavajoWhite4}{rgb}{0.55,0.47,0.37}
\definecolor{NavajoWhite}{rgb}{1.00,0.87,0.68}
\definecolor{NavyBlue}{rgb}{0.00,0.00,0.50}
\definecolor{OldLace}{rgb}{0.99,0.96,0.90}
\definecolor{OliveDrab1}{rgb}{0.75,1.00,0.24}
\definecolor{OliveDrab2}{rgb}{0.70,0.93,0.23}
\definecolor{OliveDrab3}{rgb}{0.60,0.80,0.20}
\definecolor{OliveDrab4}{rgb}{0.41,0.55,0.13}
\definecolor{OliveDrab}{rgb}{0.42,0.56,0.14}
\definecolor{OrangeRed1}{rgb}{1.00,0.27,0.00}
\definecolor{OrangeRed2}{rgb}{0.93,0.25,0.00}
\definecolor{OrangeRed3}{rgb}{0.80,0.22,0.00}
\definecolor{OrangeRed4}{rgb}{0.55,0.15,0.00}
\definecolor{OrangeRed}{rgb}{1.00,0.27,0.00}
\definecolor{PaleGoldenrod}{rgb}{0.93,0.91,0.67}
\definecolor{PaleGreen1}{rgb}{0.60,1.00,0.60}
\definecolor{PaleGreen2}{rgb}{0.56,0.93,0.56}
\definecolor{PaleGreen3}{rgb}{0.49,0.80,0.49}
\definecolor{PaleGreen4}{rgb}{0.33,0.55,0.33}
\definecolor{PaleGreen}{rgb}{0.60,0.98,0.60}
\definecolor{PaleTurquoise1}{rgb}{0.73,1.00,1.00}
\definecolor{PaleTurquoise2}{rgb}{0.68,0.93,0.93}
\definecolor{PaleTurquoise3}{rgb}{0.59,0.80,0.80}
\definecolor{PaleTurquoise4}{rgb}{0.40,0.55,0.55}
\definecolor{PaleTurquoise}{rgb}{0.69,0.93,0.93}
\definecolor{PaleVioletRed1}{rgb}{1.00,0.51,0.67}
\definecolor{PaleVioletRed2}{rgb}{0.93,0.47,0.62}
\definecolor{PaleVioletRed3}{rgb}{0.80,0.41,0.54}
\definecolor{PaleVioletRed4}{rgb}{0.55,0.28,0.36}
\definecolor{PaleVioletRed}{rgb}{0.86,0.44,0.58}
\definecolor{PapayaWhip}{rgb}{1.00,0.94,0.84}
\definecolor{PeachPuff1}{rgb}{1.00,0.85,0.73}
\definecolor{PeachPuff2}{rgb}{0.93,0.80,0.68}
\definecolor{PeachPuff3}{rgb}{0.80,0.69,0.58}
\definecolor{PeachPuff4}{rgb}{0.55,0.47,0.40}
\definecolor{PeachPuff}{rgb}{1.00,0.85,0.73}
\definecolor{PowderBlue}{rgb}{0.69,0.88,0.90}
\definecolor{RosyBrown1}{rgb}{1.00,0.76,0.76}
\definecolor{RosyBrown2}{rgb}{0.93,0.71,0.71}
\definecolor{RosyBrown3}{rgb}{0.80,0.61,0.61}
\definecolor{RosyBrown4}{rgb}{0.55,0.41,0.41}
\definecolor{RosyBrown}{rgb}{0.74,0.56,0.56}
\definecolor{RoyalBlue1}{rgb}{0.28,0.46,1.00}
\definecolor{RoyalBlue2}{rgb}{0.26,0.43,0.93}
\definecolor{RoyalBlue3}{rgb}{0.23,0.37,0.80}
\definecolor{RoyalBlue4}{rgb}{0.15,0.25,0.55}
\definecolor{RoyalBlue}{rgb}{0.25,0.41,0.88}
\definecolor{SaddleBrown}{rgb}{0.55,0.27,0.07}
\definecolor{SandyBrown}{rgb}{0.96,0.64,0.38}
\definecolor{SeaGreen1}{rgb}{0.33,1.00,0.62}
\definecolor{SeaGreen2}{rgb}{0.31,0.93,0.58}
\definecolor{SeaGreen3}{rgb}{0.26,0.80,0.50}
\definecolor{SeaGreen4}{rgb}{0.18,0.55,0.34}
\definecolor{SeaGreen}{rgb}{0.18,0.55,0.34}
\definecolor{SkyBlue1}{rgb}{0.53,0.81,1.00}
\definecolor{SkyBlue2}{rgb}{0.49,0.75,0.93}
\definecolor{SkyBlue3}{rgb}{0.42,0.65,0.80}
\definecolor{SkyBlue4}{rgb}{0.29,0.44,0.55}
\definecolor{SkyBlue}{rgb}{0.53,0.81,0.92}
\definecolor{SlateBlue1}{rgb}{0.51,0.44,1.00}
\definecolor{SlateBlue2}{rgb}{0.48,0.40,0.93}
\definecolor{SlateBlue3}{rgb}{0.41,0.35,0.80}
\definecolor{SlateBlue4}{rgb}{0.28,0.24,0.55}
\definecolor{SlateBlue}{rgb}{0.42,0.35,0.80}
\definecolor{SlateGray1}{rgb}{0.78,0.89,1.00}
\definecolor{SlateGray2}{rgb}{0.73,0.83,0.93}
\definecolor{SlateGray3}{rgb}{0.62,0.71,0.80}
\definecolor{SlateGray4}{rgb}{0.42,0.48,0.55}
\definecolor{SlateGray}{rgb}{0.44,0.50,0.56}
\definecolor{SlateGrey}{rgb}{0.44,0.50,0.56}
\definecolor{SpringGreen1}{rgb}{0.00,1.00,0.50}
\definecolor{SpringGreen2}{rgb}{0.00,0.93,0.46}
\definecolor{SpringGreen3}{rgb}{0.00,0.80,0.40}
\definecolor{SpringGreen4}{rgb}{0.00,0.55,0.27}
\definecolor{SpringGreen}{rgb}{0.00,1.00,0.50}
\definecolor{SteelBlue1}{rgb}{0.39,0.72,1.00}
\definecolor{SteelBlue2}{rgb}{0.36,0.67,0.93}
\definecolor{SteelBlue3}{rgb}{0.31,0.58,0.80}
\definecolor{SteelBlue4}{rgb}{0.21,0.39,0.55}
\definecolor{SteelBlue}{rgb}{0.27,0.51,0.71}
\definecolor{VioletRed1}{rgb}{1.00,0.24,0.59}
\definecolor{VioletRed2}{rgb}{0.93,0.23,0.55}
\definecolor{VioletRed3}{rgb}{0.80,0.20,0.47}
\definecolor{VioletRed4}{rgb}{0.55,0.13,0.32}
\definecolor{VioletRed}{rgb}{0.82,0.13,0.56}
\definecolor{WhiteSmoke}{rgb}{0.96,0.96,0.96}
\definecolor{YellowGreen}{rgb}{0.60,0.80,0.20}
\definecolor{aliceblue}{rgb}{0.94,0.97,1.00}
\definecolor{antiquewhite}{rgb}{0.98,0.92,0.84}
\definecolor{aquamarine1}{rgb}{0.50,1.00,0.83}
\definecolor{aquamarine2}{rgb}{0.46,0.93,0.78}
\definecolor{aquamarine3}{rgb}{0.40,0.80,0.67}
\definecolor{aquamarine4}{rgb}{0.27,0.55,0.45}
\definecolor{aquamarine}{rgb}{0.50,1.00,0.83}
\definecolor{azure1}{rgb}{0.94,1.00,1.00}
\definecolor{azure2}{rgb}{0.88,0.93,0.93}
\definecolor{azure3}{rgb}{0.76,0.80,0.80}
\definecolor{azure4}{rgb}{0.51,0.55,0.55}
\definecolor{azure}{rgb}{0.94,1.00,1.00}
\definecolor{beige}{rgb}{0.96,0.96,0.86}
\definecolor{bisque1}{rgb}{1.00,0.89,0.77}
\definecolor{bisque2}{rgb}{0.93,0.84,0.72}
\definecolor{bisque3}{rgb}{0.80,0.72,0.62}
\definecolor{bisque4}{rgb}{0.55,0.49,0.42}
\definecolor{bisque}{rgb}{1.00,0.89,0.77}
\definecolor{black}{rgb}{0.00,0.00,0.00}
\definecolor{blanchedalmond}{rgb}{1.00,0.92,0.80}
\definecolor{blue1}{rgb}{0.00,0.00,1.00}
\definecolor{blue2}{rgb}{0.00,0.00,0.93}
\definecolor{blue3}{rgb}{0.00,0.00,0.80}
\definecolor{blue4}{rgb}{0.00,0.00,0.55}
\definecolor{blueviolet}{rgb}{0.54,0.17,0.89}
\definecolor{blue}{rgb}{0.00,0.00,1.00}
\definecolor{brown1}{rgb}{1.00,0.25,0.25}
\definecolor{brown2}{rgb}{0.93,0.23,0.23}
\definecolor{brown3}{rgb}{0.80,0.20,0.20}
\definecolor{brown4}{rgb}{0.55,0.14,0.14}
\definecolor{brown}{rgb}{0.65,0.16,0.16}
\definecolor{burlywood1}{rgb}{1.00,0.83,0.61}
\definecolor{burlywood2}{rgb}{0.93,0.77,0.57}
\definecolor{burlywood3}{rgb}{0.80,0.67,0.49}
\definecolor{burlywood4}{rgb}{0.55,0.45,0.33}
\definecolor{burlywood}{rgb}{0.87,0.72,0.53}
\definecolor{cadetblue}{rgb}{0.37,0.62,0.63}
\definecolor{chartreuse1}{rgb}{0.50,1.00,0.00}
\definecolor{chartreuse2}{rgb}{0.46,0.93,0.00}
\definecolor{chartreuse3}{rgb}{0.40,0.80,0.00}
\definecolor{chartreuse4}{rgb}{0.27,0.55,0.00}
\definecolor{chartreuse}{rgb}{0.50,1.00,0.00}
\definecolor{chocolate1}{rgb}{1.00,0.50,0.14}
\definecolor{chocolate2}{rgb}{0.93,0.46,0.13}
\definecolor{chocolate3}{rgb}{0.80,0.40,0.11}
\definecolor{chocolate4}{rgb}{0.55,0.27,0.07}
\definecolor{chocolate}{rgb}{0.82,0.41,0.12}
\definecolor{coral1}{rgb}{1.00,0.45,0.34}
\definecolor{coral2}{rgb}{0.93,0.42,0.31}
\definecolor{coral3}{rgb}{0.80,0.36,0.27}
\definecolor{coral4}{rgb}{0.55,0.24,0.18}
\definecolor{coral}{rgb}{1.00,0.50,0.31}
\definecolor{cornflowerblue}{rgb}{0.39,0.58,0.93}
\definecolor{cornsilk1}{rgb}{1.00,0.97,0.86}
\definecolor{cornsilk2}{rgb}{0.93,0.91,0.80}
\definecolor{cornsilk3}{rgb}{0.80,0.78,0.69}
\definecolor{cornsilk4}{rgb}{0.55,0.53,0.47}
\definecolor{cornsilk}{rgb}{1.00,0.97,0.86}
\definecolor{cyan1}{rgb}{0.00,1.00,1.00}
\definecolor{cyan2}{rgb}{0.00,0.93,0.93}
\definecolor{cyan3}{rgb}{0.00,0.80,0.80}
\definecolor{cyan4}{rgb}{0.00,0.55,0.55}
\definecolor{cyan}{rgb}{0.00,1.00,1.00}
\definecolor{darkblue}{rgb}{0.00,0.00,0.55}
\definecolor{darkcyan}{rgb}{0.00,0.55,0.55}
\definecolor{darkgoldenrod}{rgb}{0.72,0.53,0.04}
\definecolor{darkgray}{rgb}{0.66,0.66,0.66}
\definecolor{darkgreen}{rgb}{0.00,0.39,0.00}
\definecolor{darkgrey}{rgb}{0.66,0.66,0.66}
\definecolor{darkkhaki}{rgb}{0.74,0.72,0.42}
\definecolor{darkmagenta}{rgb}{0.55,0.00,0.55}
\definecolor{darkolive}{rgb}{0.33,0.42,0.18}
\definecolor{darkorange}{rgb}{1.00,0.55,0.00}
\definecolor{darkorchid}{rgb}{0.60,0.20,0.80}
\definecolor{darkred}{rgb}{0.55,0.00,0.00}
\definecolor{darksalmon}{rgb}{0.91,0.59,0.48}
\definecolor{darksea}{rgb}{0.56,0.74,0.56}
\definecolor{darkslate}{rgb}{0.18,0.31,0.31}
\definecolor{darkslate}{rgb}{0.18,0.31,0.31}
\definecolor{darkslate}{rgb}{0.28,0.24,0.55}
\definecolor{darkturquoise}{rgb}{0.00,0.81,0.82}
\definecolor{darkviolet}{rgb}{0.58,0.00,0.83}
\definecolor{deeppink}{rgb}{1.00,0.08,0.58}
\definecolor{deepsky}{rgb}{0.00,0.75,1.00}
\definecolor{dimgray}{rgb}{0.41,0.41,0.41}
\definecolor{dimgrey}{rgb}{0.41,0.41,0.41}
\definecolor{dodgerblue}{rgb}{0.12,0.56,1.00}
\definecolor{firebrick1}{rgb}{1.00,0.19,0.19}
\definecolor{firebrick2}{rgb}{0.93,0.17,0.17}
\definecolor{firebrick3}{rgb}{0.80,0.15,0.15}
\definecolor{firebrick4}{rgb}{0.55,0.10,0.10}
\definecolor{firebrick}{rgb}{0.70,0.13,0.13}
\definecolor{floralwhite}{rgb}{1.00,0.98,0.94}
\definecolor{forestgreen}{rgb}{0.13,0.55,0.13}
\definecolor{gainsboro}{rgb}{0.86,0.86,0.86}
\definecolor{ghostwhite}{rgb}{0.97,0.97,1.00}
\definecolor{gold1}{rgb}{1.00,0.84,0.00}
\definecolor{gold2}{rgb}{0.93,0.79,0.00}
\definecolor{gold3}{rgb}{0.80,0.68,0.00}
\definecolor{gold4}{rgb}{0.55,0.46,0.00}
\definecolor{goldenrod1}{rgb}{1.00,0.76,0.15}
\definecolor{goldenrod2}{rgb}{0.93,0.71,0.13}
\definecolor{goldenrod3}{rgb}{0.80,0.61,0.11}
\definecolor{goldenrod4}{rgb}{0.55,0.41,0.08}
\definecolor{goldenrod}{rgb}{0.85,0.65,0.13}
\definecolor{gold}{rgb}{1.00,0.84,0.00}
\definecolor{gray0}{rgb}{0.00,0.00,0.00}
\definecolor{gray100}{rgb}{1.00,1.00,1.00}
\definecolor{gray10}{rgb}{0.10,0.10,0.10}
\definecolor{gray11}{rgb}{0.11,0.11,0.11}
\definecolor{gray12}{rgb}{0.12,0.12,0.12}
\definecolor{gray13}{rgb}{0.13,0.13,0.13}
\definecolor{gray14}{rgb}{0.14,0.14,0.14}
\definecolor{gray15}{rgb}{0.15,0.15,0.15}
\definecolor{gray16}{rgb}{0.16,0.16,0.16}
\definecolor{gray17}{rgb}{0.17,0.17,0.17}
\definecolor{gray18}{rgb}{0.18,0.18,0.18}
\definecolor{gray19}{rgb}{0.19,0.19,0.19}
\definecolor{gray1}{rgb}{0.01,0.01,0.01}
\definecolor{gray20}{rgb}{0.20,0.20,0.20}
\definecolor{gray21}{rgb}{0.21,0.21,0.21}
\definecolor{gray22}{rgb}{0.22,0.22,0.22}
\definecolor{gray23}{rgb}{0.23,0.23,0.23}
\definecolor{gray24}{rgb}{0.24,0.24,0.24}
\definecolor{gray25}{rgb}{0.25,0.25,0.25}
\definecolor{gray26}{rgb}{0.26,0.26,0.26}
\definecolor{gray27}{rgb}{0.27,0.27,0.27}
\definecolor{gray28}{rgb}{0.28,0.28,0.28}
\definecolor{gray29}{rgb}{0.29,0.29,0.29}
\definecolor{gray2}{rgb}{0.02,0.02,0.02}
\definecolor{gray30}{rgb}{0.30,0.30,0.30}
\definecolor{gray31}{rgb}{0.31,0.31,0.31}
\definecolor{gray32}{rgb}{0.32,0.32,0.32}
\definecolor{gray33}{rgb}{0.33,0.33,0.33}
\definecolor{gray34}{rgb}{0.34,0.34,0.34}
\definecolor{gray35}{rgb}{0.35,0.35,0.35}
\definecolor{gray36}{rgb}{0.36,0.36,0.36}
\definecolor{gray37}{rgb}{0.37,0.37,0.37}
\definecolor{gray38}{rgb}{0.38,0.38,0.38}
\definecolor{gray39}{rgb}{0.39,0.39,0.39}
\definecolor{gray3}{rgb}{0.03,0.03,0.03}
\definecolor{gray40}{rgb}{0.40,0.40,0.40}
\definecolor{gray41}{rgb}{0.41,0.41,0.41}
\definecolor{gray42}{rgb}{0.42,0.42,0.42}
\definecolor{gray43}{rgb}{0.43,0.43,0.43}
\definecolor{gray44}{rgb}{0.44,0.44,0.44}
\definecolor{gray45}{rgb}{0.45,0.45,0.45}
\definecolor{gray46}{rgb}{0.46,0.46,0.46}
\definecolor{gray47}{rgb}{0.47,0.47,0.47}
\definecolor{gray48}{rgb}{0.48,0.48,0.48}
\definecolor{gray49}{rgb}{0.49,0.49,0.49}
\definecolor{gray4}{rgb}{0.04,0.04,0.04}
\definecolor{gray50}{rgb}{0.50,0.50,0.50}
\definecolor{gray51}{rgb}{0.51,0.51,0.51}
\definecolor{gray52}{rgb}{0.52,0.52,0.52}
\definecolor{gray53}{rgb}{0.53,0.53,0.53}
\definecolor{gray54}{rgb}{0.54,0.54,0.54}
\definecolor{gray55}{rgb}{0.55,0.55,0.55}
\definecolor{gray56}{rgb}{0.56,0.56,0.56}
\definecolor{gray57}{rgb}{0.57,0.57,0.57}
\definecolor{gray58}{rgb}{0.58,0.58,0.58}
\definecolor{gray59}{rgb}{0.59,0.59,0.59}
\definecolor{gray5}{rgb}{0.05,0.05,0.05}
\definecolor{gray60}{rgb}{0.60,0.60,0.60}
\definecolor{gray61}{rgb}{0.61,0.61,0.61}
\definecolor{gray62}{rgb}{0.62,0.62,0.62}
\definecolor{gray63}{rgb}{0.63,0.63,0.63}
\definecolor{gray64}{rgb}{0.64,0.64,0.64}
\definecolor{gray65}{rgb}{0.65,0.65,0.65}
\definecolor{gray66}{rgb}{0.66,0.66,0.66}
\definecolor{gray67}{rgb}{0.67,0.67,0.67}
\definecolor{gray68}{rgb}{0.68,0.68,0.68}
\definecolor{gray69}{rgb}{0.69,0.69,0.69}
\definecolor{gray6}{rgb}{0.06,0.06,0.06}
\definecolor{gray70}{rgb}{0.70,0.70,0.70}
\definecolor{gray71}{rgb}{0.71,0.71,0.71}
\definecolor{gray72}{rgb}{0.72,0.72,0.72}
\definecolor{gray73}{rgb}{0.73,0.73,0.73}
\definecolor{gray74}{rgb}{0.74,0.74,0.74}
\definecolor{gray75}{rgb}{0.75,0.75,0.75}
\definecolor{gray76}{rgb}{0.76,0.76,0.76}
\definecolor{gray77}{rgb}{0.77,0.77,0.77}
\definecolor{gray78}{rgb}{0.78,0.78,0.78}
\definecolor{gray79}{rgb}{0.79,0.79,0.79}
\definecolor{gray7}{rgb}{0.07,0.07,0.07}
\definecolor{gray80}{rgb}{0.80,0.80,0.80}
\definecolor{gray81}{rgb}{0.81,0.81,0.81}
\definecolor{gray82}{rgb}{0.82,0.82,0.82}
\definecolor{gray83}{rgb}{0.83,0.83,0.83}
\definecolor{gray84}{rgb}{0.84,0.84,0.84}
\definecolor{gray85}{rgb}{0.85,0.85,0.85}
\definecolor{gray86}{rgb}{0.86,0.86,0.86}
\definecolor{gray87}{rgb}{0.87,0.87,0.87}
\definecolor{gray88}{rgb}{0.88,0.88,0.88}
\definecolor{gray89}{rgb}{0.89,0.89,0.89}
\definecolor{gray8}{rgb}{0.08,0.08,0.08}
\definecolor{gray90}{rgb}{0.90,0.90,0.90}
\definecolor{gray91}{rgb}{0.91,0.91,0.91}
\definecolor{gray92}{rgb}{0.92,0.92,0.92}
\definecolor{gray93}{rgb}{0.93,0.93,0.93}
\definecolor{gray94}{rgb}{0.94,0.94,0.94}
\definecolor{gray95}{rgb}{0.95,0.95,0.95}
\definecolor{gray96}{rgb}{0.96,0.96,0.96}
\definecolor{gray97}{rgb}{0.97,0.97,0.97}
\definecolor{gray98}{rgb}{0.98,0.98,0.98}
\definecolor{gray99}{rgb}{0.99,0.99,0.99}
\definecolor{gray9}{rgb}{0.09,0.09,0.09}
\definecolor{gray}{rgb}{0.75,0.75,0.75}
\definecolor{green1}{rgb}{0.00,1.00,0.00}
\definecolor{green2}{rgb}{0.00,0.93,0.00}
\definecolor{green3}{rgb}{0.00,0.80,0.00}
\definecolor{green4}{rgb}{0.00,0.55,0.00}
\definecolor{greenyellow}{rgb}{0.68,1.00,0.18}
\definecolor{green}{rgb}{0.00,1.00,0.00}
\definecolor{grey0}{rgb}{0.00,0.00,0.00}
\definecolor{grey100}{rgb}{1.00,1.00,1.00}
\definecolor{grey10}{rgb}{0.10,0.10,0.10}
\definecolor{grey11}{rgb}{0.11,0.11,0.11}
\definecolor{grey12}{rgb}{0.12,0.12,0.12}
\definecolor{grey13}{rgb}{0.13,0.13,0.13}
\definecolor{grey14}{rgb}{0.14,0.14,0.14}
\definecolor{grey15}{rgb}{0.15,0.15,0.15}
\definecolor{grey16}{rgb}{0.16,0.16,0.16}
\definecolor{grey17}{rgb}{0.17,0.17,0.17}
\definecolor{grey18}{rgb}{0.18,0.18,0.18}
\definecolor{grey19}{rgb}{0.19,0.19,0.19}
\definecolor{grey1}{rgb}{0.01,0.01,0.01}
\definecolor{grey20}{rgb}{0.20,0.20,0.20}
\definecolor{grey21}{rgb}{0.21,0.21,0.21}
\definecolor{grey22}{rgb}{0.22,0.22,0.22}
\definecolor{grey23}{rgb}{0.23,0.23,0.23}
\definecolor{grey24}{rgb}{0.24,0.24,0.24}
\definecolor{grey25}{rgb}{0.25,0.25,0.25}
\definecolor{grey26}{rgb}{0.26,0.26,0.26}
\definecolor{grey27}{rgb}{0.27,0.27,0.27}
\definecolor{grey28}{rgb}{0.28,0.28,0.28}
\definecolor{grey29}{rgb}{0.29,0.29,0.29}
\definecolor{grey2}{rgb}{0.02,0.02,0.02}
\definecolor{grey30}{rgb}{0.30,0.30,0.30}
\definecolor{grey31}{rgb}{0.31,0.31,0.31}
\definecolor{grey32}{rgb}{0.32,0.32,0.32}
\definecolor{grey33}{rgb}{0.33,0.33,0.33}
\definecolor{grey34}{rgb}{0.34,0.34,0.34}
\definecolor{grey35}{rgb}{0.35,0.35,0.35}
\definecolor{grey36}{rgb}{0.36,0.36,0.36}
\definecolor{grey37}{rgb}{0.37,0.37,0.37}
\definecolor{grey38}{rgb}{0.38,0.38,0.38}
\definecolor{grey39}{rgb}{0.39,0.39,0.39}
\definecolor{grey3}{rgb}{0.03,0.03,0.03}
\definecolor{grey40}{rgb}{0.40,0.40,0.40}
\definecolor{grey41}{rgb}{0.41,0.41,0.41}
\definecolor{grey42}{rgb}{0.42,0.42,0.42}
\definecolor{grey43}{rgb}{0.43,0.43,0.43}
\definecolor{grey44}{rgb}{0.44,0.44,0.44}
\definecolor{grey45}{rgb}{0.45,0.45,0.45}
\definecolor{grey46}{rgb}{0.46,0.46,0.46}
\definecolor{grey47}{rgb}{0.47,0.47,0.47}
\definecolor{grey48}{rgb}{0.48,0.48,0.48}
\definecolor{grey49}{rgb}{0.49,0.49,0.49}
\definecolor{grey4}{rgb}{0.04,0.04,0.04}
\definecolor{grey50}{rgb}{0.50,0.50,0.50}
\definecolor{grey51}{rgb}{0.51,0.51,0.51}
\definecolor{grey52}{rgb}{0.52,0.52,0.52}
\definecolor{grey53}{rgb}{0.53,0.53,0.53}
\definecolor{grey54}{rgb}{0.54,0.54,0.54}
\definecolor{grey55}{rgb}{0.55,0.55,0.55}
\definecolor{grey56}{rgb}{0.56,0.56,0.56}
\definecolor{grey57}{rgb}{0.57,0.57,0.57}
\definecolor{grey58}{rgb}{0.58,0.58,0.58}
\definecolor{grey59}{rgb}{0.59,0.59,0.59}
\definecolor{grey5}{rgb}{0.05,0.05,0.05}
\definecolor{grey60}{rgb}{0.60,0.60,0.60}
\definecolor{grey61}{rgb}{0.61,0.61,0.61}
\definecolor{grey62}{rgb}{0.62,0.62,0.62}
\definecolor{grey63}{rgb}{0.63,0.63,0.63}
\definecolor{grey64}{rgb}{0.64,0.64,0.64}
\definecolor{grey65}{rgb}{0.65,0.65,0.65}
\definecolor{grey66}{rgb}{0.66,0.66,0.66}
\definecolor{grey67}{rgb}{0.67,0.67,0.67}
\definecolor{grey68}{rgb}{0.68,0.68,0.68}
\definecolor{grey69}{rgb}{0.69,0.69,0.69}
\definecolor{grey6}{rgb}{0.06,0.06,0.06}
\definecolor{grey70}{rgb}{0.70,0.70,0.70}
\definecolor{grey71}{rgb}{0.71,0.71,0.71}
\definecolor{grey72}{rgb}{0.72,0.72,0.72}
\definecolor{grey73}{rgb}{0.73,0.73,0.73}
\definecolor{grey74}{rgb}{0.74,0.74,0.74}
\definecolor{grey75}{rgb}{0.75,0.75,0.75}
\definecolor{grey76}{rgb}{0.76,0.76,0.76}
\definecolor{grey77}{rgb}{0.77,0.77,0.77}
\definecolor{grey78}{rgb}{0.78,0.78,0.78}
\definecolor{grey79}{rgb}{0.79,0.79,0.79}
\definecolor{grey7}{rgb}{0.07,0.07,0.07}
\definecolor{grey80}{rgb}{0.80,0.80,0.80}
\definecolor{grey81}{rgb}{0.81,0.81,0.81}
\definecolor{grey82}{rgb}{0.82,0.82,0.82}
\definecolor{grey83}{rgb}{0.83,0.83,0.83}
\definecolor{grey84}{rgb}{0.84,0.84,0.84}
\definecolor{grey85}{rgb}{0.85,0.85,0.85}
\definecolor{grey86}{rgb}{0.86,0.86,0.86}
\definecolor{grey87}{rgb}{0.87,0.87,0.87}
\definecolor{grey88}{rgb}{0.88,0.88,0.88}
\definecolor{grey89}{rgb}{0.89,0.89,0.89}
\definecolor{grey8}{rgb}{0.08,0.08,0.08}
\definecolor{grey90}{rgb}{0.90,0.90,0.90}
\definecolor{grey91}{rgb}{0.91,0.91,0.91}
\definecolor{grey92}{rgb}{0.92,0.92,0.92}
\definecolor{grey93}{rgb}{0.93,0.93,0.93}
\definecolor{grey94}{rgb}{0.94,0.94,0.94}
\definecolor{grey95}{rgb}{0.95,0.95,0.95}
\definecolor{grey96}{rgb}{0.96,0.96,0.96}
\definecolor{grey97}{rgb}{0.97,0.97,0.97}
\definecolor{grey98}{rgb}{0.98,0.98,0.98}
\definecolor{grey99}{rgb}{0.99,0.99,0.99}
\definecolor{grey9}{rgb}{0.09,0.09,0.09}
\definecolor{grey}{rgb}{0.75,0.75,0.75}
\definecolor{honeydew1}{rgb}{0.94,1.00,0.94}
\definecolor{honeydew2}{rgb}{0.88,0.93,0.88}
\definecolor{honeydew3}{rgb}{0.76,0.80,0.76}
\definecolor{honeydew4}{rgb}{0.51,0.55,0.51}
\definecolor{honeydew}{rgb}{0.94,1.00,0.94}
\definecolor{hotpink}{rgb}{1.00,0.41,0.71}
\definecolor{indianred}{rgb}{0.80,0.36,0.36}
\definecolor{ivory1}{rgb}{1.00,1.00,0.94}
\definecolor{ivory2}{rgb}{0.93,0.93,0.88}
\definecolor{ivory3}{rgb}{0.80,0.80,0.76}
\definecolor{ivory4}{rgb}{0.55,0.55,0.51}
\definecolor{ivory}{rgb}{1.00,1.00,0.94}
\definecolor{khaki1}{rgb}{1.00,0.96,0.56}
\definecolor{khaki2}{rgb}{0.93,0.90,0.52}
\definecolor{khaki3}{rgb}{0.80,0.78,0.45}
\definecolor{khaki4}{rgb}{0.55,0.53,0.31}
\definecolor{khaki}{rgb}{0.94,0.90,0.55}
\definecolor{lavenderblush}{rgb}{1.00,0.94,0.96}
\definecolor{lavender}{rgb}{0.90,0.90,0.98}
\definecolor{lawngreen}{rgb}{0.49,0.99,0.00}
\definecolor{lemonchiffon}{rgb}{1.00,0.98,0.80}
\definecolor{lightblue}{rgb}{0.68,0.85,0.90}
\definecolor{lightcoral}{rgb}{0.94,0.50,0.50}
\definecolor{lightcyan}{rgb}{0.88,1.00,1.00}
\definecolor{lightgoldenrod}{rgb}{0.93,0.87,0.51}
\definecolor{lightgoldenrod}{rgb}{0.98,0.98,0.82}
\definecolor{lightgray}{rgb}{0.83,0.83,0.83}
\definecolor{lightgreen}{rgb}{0.56,0.93,0.56}
\definecolor{lightgrey}{rgb}{0.83,0.83,0.83}
\definecolor{lightpink}{rgb}{1.00,0.71,0.76}
\definecolor{lightsalmon}{rgb}{1.00,0.63,0.48}
\definecolor{lightsea}{rgb}{0.13,0.70,0.67}
\definecolor{lightsky}{rgb}{0.53,0.81,0.98}
\definecolor{lightslate}{rgb}{0.47,0.53,0.60}
\definecolor{lightslate}{rgb}{0.47,0.53,0.60}
\definecolor{lightslate}{rgb}{0.52,0.44,1.00}
\definecolor{lightsteel}{rgb}{0.69,0.77,0.87}
\definecolor{lightyellow}{rgb}{1.00,1.00,0.88}
\definecolor{limegreen}{rgb}{0.20,0.80,0.20}
\definecolor{linen}{rgb}{0.98,0.94,0.90}
\definecolor{magenta1}{rgb}{1.00,0.00,1.00}
\definecolor{magenta2}{rgb}{0.93,0.00,0.93}
\definecolor{magenta3}{rgb}{0.80,0.00,0.80}
\definecolor{magenta4}{rgb}{0.55,0.00,0.55}
\definecolor{magenta}{rgb}{1.00,0.00,1.00}
\definecolor{maroon1}{rgb}{1.00,0.20,0.70}
\definecolor{maroon2}{rgb}{0.93,0.19,0.65}
\definecolor{maroon3}{rgb}{0.80,0.16,0.56}
\definecolor{maroon4}{rgb}{0.55,0.11,0.38}
\definecolor{maroon}{rgb}{0.69,0.19,0.38}
\definecolor{mediumaquamarine}{rgb}{0.40,0.80,0.67}
\definecolor{mediumblue}{rgb}{0.00,0.00,0.80}
\definecolor{mediumorchid}{rgb}{0.73,0.33,0.83}
\definecolor{mediumpurple}{rgb}{0.58,0.44,0.86}
\definecolor{mediumsea}{rgb}{0.24,0.70,0.44}
\definecolor{mediumslate}{rgb}{0.48,0.41,0.93}
\definecolor{mediumspring}{rgb}{0.00,0.98,0.60}
\definecolor{mediumturquoise}{rgb}{0.28,0.82,0.80}
\definecolor{mediumviolet}{rgb}{0.78,0.08,0.52}
\definecolor{midnightblue}{rgb}{0.10,0.10,0.44}
\definecolor{mintcream}{rgb}{0.96,1.00,0.98}
\definecolor{mistyrose}{rgb}{1.00,0.89,0.88}
\definecolor{moccasin}{rgb}{1.00,0.89,0.71}
\definecolor{navajowhite}{rgb}{1.00,0.87,0.68}
\definecolor{navyblue}{rgb}{0.00,0.00,0.50}
\definecolor{navy}{rgb}{0.00,0.00,0.50}
\definecolor{oldlace}{rgb}{0.99,0.96,0.90}
\definecolor{olivedrab}{rgb}{0.42,0.56,0.14}
\definecolor{orange1}{rgb}{1.00,0.65,0.00}
\definecolor{orange2}{rgb}{0.93,0.60,0.00}
\definecolor{orange3}{rgb}{0.80,0.52,0.00}
\definecolor{orange4}{rgb}{0.55,0.35,0.00}
\definecolor{orangered}{rgb}{1.00,0.27,0.00}
\definecolor{orange}{rgb}{1.00,0.65,0.00}
\definecolor{orchid1}{rgb}{1.00,0.51,0.98}
\definecolor{orchid2}{rgb}{0.93,0.48,0.91}
\definecolor{orchid3}{rgb}{0.80,0.41,0.79}
\definecolor{orchid4}{rgb}{0.55,0.28,0.54}
\definecolor{orchid}{rgb}{0.85,0.44,0.84}
\definecolor{palegoldenrod}{rgb}{0.93,0.91,0.67}
\definecolor{palegreen}{rgb}{0.60,0.98,0.60}
\definecolor{paleturquoise}{rgb}{0.69,0.93,0.93}
\definecolor{paleviolet}{rgb}{0.86,0.44,0.58}
\definecolor{papayawhip}{rgb}{1.00,0.94,0.84}
\definecolor{peachpuff}{rgb}{1.00,0.85,0.73}
\definecolor{peru}{rgb}{0.80,0.52,0.25}
\definecolor{pink1}{rgb}{1.00,0.71,0.77}
\definecolor{pink2}{rgb}{0.93,0.66,0.72}
\definecolor{pink3}{rgb}{0.80,0.57,0.62}
\definecolor{pink4}{rgb}{0.55,0.39,0.42}
\definecolor{pink}{rgb}{1.00,0.75,0.80}
\definecolor{plum1}{rgb}{1.00,0.73,1.00}
\definecolor{plum2}{rgb}{0.93,0.68,0.93}
\definecolor{plum3}{rgb}{0.80,0.59,0.80}
\definecolor{plum4}{rgb}{0.55,0.40,0.55}
\definecolor{plum}{rgb}{0.87,0.63,0.87}
\definecolor{powderblue}{rgb}{0.69,0.88,0.90}
\definecolor{purple1}{rgb}{0.61,0.19,1.00}
\definecolor{purple2}{rgb}{0.57,0.17,0.93}
\definecolor{purple3}{rgb}{0.49,0.15,0.80}
\definecolor{purple4}{rgb}{0.33,0.10,0.55}
\definecolor{purple}{rgb}{0.63,0.13,0.94}
\definecolor{red1}{rgb}{1.00,0.00,0.00}
\definecolor{red2}{rgb}{0.93,0.00,0.00}
\definecolor{red3}{rgb}{0.80,0.00,0.00}
\definecolor{red4}{rgb}{0.55,0.00,0.00}
\definecolor{red}{rgb}{1.00,0.00,0.00}
\definecolor{rosybrown}{rgb}{0.74,0.56,0.56}
\definecolor{royalblue}{rgb}{0.25,0.41,0.88}
\definecolor{saddlebrown}{rgb}{0.55,0.27,0.07}
\definecolor{salmon1}{rgb}{1.00,0.55,0.41}
\definecolor{salmon2}{rgb}{0.93,0.51,0.38}
\definecolor{salmon3}{rgb}{0.80,0.44,0.33}
\definecolor{salmon4}{rgb}{0.55,0.30,0.22}
\definecolor{salmon}{rgb}{0.98,0.50,0.45}
\definecolor{sandybrown}{rgb}{0.96,0.64,0.38}
\definecolor{seagreen}{rgb}{0.18,0.55,0.34}
\definecolor{seashell1}{rgb}{1.00,0.96,0.93}
\definecolor{seashell2}{rgb}{0.93,0.90,0.87}
\definecolor{seashell3}{rgb}{0.80,0.77,0.75}
\definecolor{seashell4}{rgb}{0.55,0.53,0.51}
\definecolor{seashell}{rgb}{1.00,0.96,0.93}
\definecolor{sienna1}{rgb}{1.00,0.51,0.28}
\definecolor{sienna2}{rgb}{0.93,0.47,0.26}
\definecolor{sienna3}{rgb}{0.80,0.41,0.22}
\definecolor{sienna4}{rgb}{0.55,0.28,0.15}
\definecolor{sienna}{rgb}{0.63,0.32,0.18}
\definecolor{skyblue}{rgb}{0.53,0.81,0.92}
\definecolor{slateblue}{rgb}{0.42,0.35,0.80}
\definecolor{slategray}{rgb}{0.44,0.50,0.56}
\definecolor{slategrey}{rgb}{0.44,0.50,0.56}
\definecolor{snow1}{rgb}{1.00,0.98,0.98}
\definecolor{snow2}{rgb}{0.93,0.91,0.91}
\definecolor{snow3}{rgb}{0.80,0.79,0.79}
\definecolor{snow4}{rgb}{0.55,0.54,0.54}
\definecolor{snow}{rgb}{1.00,0.98,0.98}
\definecolor{springgreen}{rgb}{0.00,1.00,0.50}
\definecolor{steelblue}{rgb}{0.27,0.51,0.71}
\definecolor{tan1}{rgb}{1.00,0.65,0.31}
\definecolor{tan2}{rgb}{0.93,0.60,0.29}
\definecolor{tan3}{rgb}{0.80,0.52,0.25}
\definecolor{tan4}{rgb}{0.55,0.35,0.17}
\definecolor{tan}{rgb}{0.82,0.71,0.55}
\definecolor{thistle1}{rgb}{1.00,0.88,1.00}
\definecolor{thistle2}{rgb}{0.93,0.82,0.93}
\definecolor{thistle3}{rgb}{0.80,0.71,0.80}
\definecolor{thistle4}{rgb}{0.55,0.48,0.55}
\definecolor{thistle}{rgb}{0.85,0.75,0.85}
\definecolor{tomato1}{rgb}{1.00,0.39,0.28}
\definecolor{tomato2}{rgb}{0.93,0.36,0.26}
\definecolor{tomato3}{rgb}{0.80,0.31,0.22}
\definecolor{tomato4}{rgb}{0.55,0.21,0.15}
\definecolor{tomato}{rgb}{1.00,0.39,0.28}
\definecolor{turquoise1}{rgb}{0.00,0.96,1.00}
\definecolor{turquoise2}{rgb}{0.00,0.90,0.93}
\definecolor{turquoise3}{rgb}{0.00,0.77,0.80}
\definecolor{turquoise4}{rgb}{0.00,0.53,0.55}
\definecolor{turquoise}{rgb}{0.25,0.88,0.82}
\definecolor{violetred}{rgb}{0.82,0.13,0.56}
\definecolor{violet}{rgb}{0.93,0.51,0.93}
\definecolor{wheat1}{rgb}{1.00,0.91,0.73}
\definecolor{wheat2}{rgb}{0.93,0.85,0.68}
\definecolor{wheat3}{rgb}{0.80,0.73,0.59}
\definecolor{wheat4}{rgb}{0.55,0.49,0.40}
\definecolor{wheat}{rgb}{0.96,0.87,0.70}
\definecolor{whitesmoke}{rgb}{0.96,0.96,0.96}
\definecolor{white}{rgb}{1.00,1.00,1.00}
\definecolor{yellow1}{rgb}{1.00,1.00,0.00}
\definecolor{yellow2}{rgb}{0.93,0.93,0.00}
\definecolor{yellow3}{rgb}{0.80,0.80,0.00}
\definecolor{yellow4}{rgb}{0.55,0.55,0.00}
\definecolor{yellowgreen}{rgb}{0.60,0.80,0.20}
\definecolor{yellow}{rgb}{1.00,1.00,0.00}

\definecolor{urlorange}{HTML}{ff8000}
\usepackage{listings}
\lstdefinestyle{mystyle}{
    language=Python,
    backgroundcolor=\color{gray90},
    commentstyle=\color{red},
    keywordstyle=\color{blue},
    numberstyle=\tiny\color{gray50},
    stringstyle=\color{purple2},
    basicstyle=\ttfamily,
    breakatwhitespace=false,
    breaklines=true,
    captionpos=t,
    keepspaces=true,
    numbers=left,
    numbersep=5pt,
    showspaces=false,
    showstringspaces=false,
    showtabs=false,
    tabsize=2,
}

\newcommand{\onlinecite}{\cite}
\newcommand\br{\mathbf{r}}
\DeclareMathOperator\sign{sign}
\newcommand\PHASE{part}

\newif\ifCHAPPREFACE\CHAPPREFACEfalse
\newif\ifCHAPONE\CHAPONEfalse
\newif\ifCHAPTWO\CHAPTWOfalse
\newif\ifCHAPYOUNG\CHAPYOUNGfalse
\newif\ifCHAPNEUTRON\CHAPNEUTRONfalse
\newif\ifCHAPEPRB\CHAPEPRBfalse
\newif\ifFULLBOOK\FULLBOOKfalse
\newif\ifSKIPMARTIN\FULLBOOKtrue

\newcommand\ba{\mathbf{a}}
\newcommand\bb{\mathbf{b}}

\newcommand\Cac{1}
\newcommand\Cad{2}
\newcommand\Cbc{3}
\newcommand\Cbd{4}

\newcommand{\doi}[1]{}

\usepackage{hyperref}
\hypersetup{
    colorlinks=true,
    citecolor=red,
    linkcolor=blue,
    filecolor=darkgreen,
    urlcolor=urlorange,
    pdftoolbar=false,
    pdfmenubar=false,
    pdftitle={Event-by-event simulation of fundamental quantum physics experiments},
    pageanchor=true,
    hypertexnames=true,
    }

\newcommand\HOMEURL{https://hansderaedt.synology.me}
\newcommand\BOOK{{book}}

\begin{document}
\frontmatter

\title{\huge
Simulating Quantum Physics Experiments One Event at a Time
}

\author{Hans De Raedt\\ \\
J\"ulich Supercomputing Centre,
Forschungszentrum J\"ulich, \\D-52425 J\"ulich, Germany\\
deraedthans@gmail.com\\
\HOMEURL/DES}
\date{\today}
\maketitle

\chapter*{Abstract}
\begin{quote}
This text explores the fundamental divide between empirical laboratory data and continuous theoretical physics models,
framing the "quantum measurement paradox" as an ontological error,
specifically E.T. Jaynes' ``mind projection fallacy''
where abstract mathematical tools like wavefunctions are mistaken for physical reality.
While quantum theory successfully predicts ensemble averages,
it remains inherently silent on individual, event-by-event detection statistics.
To resolve these conceptual mysteries without resorting to ad-hoc explanations like wavefunction collapse,
the text introduces Event-By-Event Simulation (EBES).
As a modular, discrete-event simulation framework utilizing causally-driven Deterministic Learning Machines (DLMs),
EBES bypasses continuous time and pre-specified probability distributions.
Instead, it generates discrete data from the bottom up in a strictly localized, cause-and-effect manner.
Ultimately, EBES provides a computationally efficient methodology that mirrors actual laboratory observations,
modeling phenomena that lie entirely beyond the reach of standard quantum theory.
\end{quote}

\tableofcontents



\chapter*{Preface}
\addcontentsline{toc}{chapter}{Preface}
It has been well established since the early development of quantum theory
that it does not describe individual, isolated events,
such as the arrival of a single electron at a specific coordinate on a detection screen.
Instead, quantum theory provides a mathematical recipe for computing the probabilities of observing those events.

This book demonstrates that the results of fundamental experiments such as the double-slit and Einstein-Podolsky-Rosen-Bohm
experiments can be explained without quantum theory's abstract concepts.
Instead, the author relies on elementary geometry, high-school algebra, and common sense.
While this approach requires a personal computer to simulate the event-by-event processes seen in the lab,
the payoff is significant: the ``weirdness'' of quantum symbols evaporates and is replaced by a clear, logical framework.

If detectors report individual clicks instead of an intensity distribution,
the interference pattern of a the double-slit experiment builds up event by event.
According to
\href{https://www.feynmanlectures.caltech.edu/III_01.html#Ch1-S1}{Feynman (Vol. 3, p.1-1)}
this phenomenon (the interference pattern) is impossible, absolutely impossible, to explain in any classical way,
and has in it the heart of quantum mechanics. In reality, it contains the only mystery.
A web-based (Javascript) event-by-event simulation demo of \href{\HOMEURL/DES/demo/young.html}{Young's two-slit interference experiment}
can be found here.
This demo shows that Feynman's statement was premature.

Through the first two and final chapters, the paradigm shift comes to life:
fundamental quantum physics experiments are reimagined as games played by people.
Central to this approach is the \href{\HOMEURL/DES/default.html}{event-by-event simulation (EBES) software},
which implements the specific rules for each game.
By running the simulations, the reader will see firsthand that event-by-event simulations faithfully reproduce
experimental results and theoretical predictions.
To demonstrate that the software does not rely on wave equations or ``built-in'' weirdness,
the Python code, \href{\HOMEURL/DES/software.html}{downloadble here},
is discussed in detail.

To avoid any misunderstanding, the EBES approach discussed in this book
is not a proposal for a ``sub-quantum'' theory.
For instance, it does not address the defining hallmark of quantum physics: quantization.
Instead, the purpose of EBES is to show that by shifting the conceptual framework,
we can model laboratory experiments at a fully granular, event-level description that quantum theory,
by its very structure, does not provide.

Most of the material presented in this book has appeared in standard scientific journals;
see the list in Appendix~\ref{EBESBIB}.
What is new here are the formulations of experiments in terms of games
and the accompanying EBES software, implemented in Python.

This book focuses on experiments that are traditionally ``explained'' through wave-particle duality
or quantum entanglement,
notions that becomes unnecessary within the EBES framework.
Applied to optics, the EBES framework demonstrates that a consistent particle-level description
of light can account for the observed phenomena, including interference, diffraction,
and quantum entanglement.
Whether this insight can be leveraged to solve computational electrodynamics problems
more efficiently is an open question.


\mainmatter


\chapter{Introduction}
\section{Generalities}

This \BOOK\ adopts the perspective that a formidable barrier exists between empirical laboratory data,
which are discrete and represented by finite-digit numbers,
and the mathematical models of theoretical physics used to interpret them~\cite{RAED23}.
For these models to yield results comparable to experimental observations, they must be supplemented by a bridge:
a procedure that transforms abstract mathematical constructs, such as real numbers (such the square root of two)
or partial derivatives, into discrete data points.

In the realms of Newtonian mechanics, Maxwell's electrodynamics, and relativistic mechanics,
this process typically involves the discretization of variables such as position, velocity, and light intensity.
Alternatively, these models can be formulated directly in a discrete framework,
as is necessary when simulating them on a digital computer.
While this discretization is a straightforward component of the mathematical apparatus,
its utility does not imply that natural phenomena are inherently ``governed'' by continuum models.
The discretization process is ``part of the apparatus'' and not necessarily a ``law of nature'',
highlighting that theoretical models are maps of the territory, not the territory itself.
As Albert Einstein remarked in his 1921 address to the Prussian Academy of Sciences:
\noindent
\begin{center}
\framebox{
\parbox[t]{0.9\hsize}{%
As far as the laws of mathematics refer to reality, they are not certain; and as
far as they are certain, they do not refer to reality..
}}%
\end{center}
Historically, quantum theory emerged from the need to explain macroscopic phenomena like black-body radiation and atomic spectra,
where data is not recorded in an event-by-event fashion.
While the accuracy with which quantum theory predicts these spectra is tantalizing,
that same quantitative precision is often diminished when describing the raw statistics of individual detection events,
unless the model is supplemented by factors external to the core theory.

When empirical data only permit statistical interpretations,
quantum theory may become the indispensable tool of the trade.
However, establishing a connection to event-by-event empirical data requires a procedure
to sample from the probability distributions that the theory provides.
At this juncture, simple discretization is no longer sufficient.
Crucially, such a sampling procedure lies outside the realm of quantum theory proper (see Appendix~\ref{NOEVENTS}).
In this light, quantum theory may be viewed as incomplete, but only if the description
is required to account for individual observable events.
Conversely, the theory is complete in its ability to predict ensemble averages
measured in the laboratory.%
\par
The actual occurrence of an observed event cannot be explained by quantum theory.
The challenge of reconciling a mathematical formalism,
which is inherently silent on individual events,
with the experimental reality that every observation yields a single,
definite outcome is known as the quantum measurement paradox~\cite{HOME97}.
While widely regarded as the most fundamental and unresolved problem in the foundations of quantum theory,
this ``paradox'' effectively disappears if one avoids the ``ontological trap'',
the error of mistaking a mathematical model for the physical process that generates experimental data.
When we conflate the abstract formalism of a wave function with the underlying reality,
we invite an endless stream of paradoxes and puzzles.
These mysteries do not reside in nature itself;
they exist only because we have confused our conceptual tools with the world they are intended to describe.%
\par
The standard ``fix'' for this discrepancy is to invoke the elusive wavefunction collapse,
a concept that remains as mysterious as Schr\"odinger's cat more than a century after its inception.
Attempting to incorporate a concept into a theory that is fundamentally incompatible
with that theory's own mathematical structure is reminiscent of Greek mythology:
when the cause of an event could not be discovered,
it was conveniently attributed to the goddess Tyche.
In this sense, ``collapse'' serves not as a physical explanation, but as a placeholder for our own ignorance.

The tendency to assume that our own mental constructs, thoughts,
and sensations are objective realities existing in the physical world was termed the ``mind projection fallacy''
by E.T. Jaynes~\cite{JAYN03}[p. 22].
If we strictly refrain from attaching ontological meaning to the mathematical symbols
of quantum theory, limiting their interpretation solely to averages and correlations,
the foundation for paradoxes and puzzles evaporates.
Such ``mysteries'' only emerge when we ignore the omnipresence of the mind projection fallacy,
mistakenly attributing the properties of our abstract mathematical tools to the physical reality they are intended to describe.

\section{Change of paradigm: event-by-event simulation}\label{sectionEBES}
The Event-By-Event Simulation (EBES) approach treats individual events as the primary entities of the model.
Just as in laboratory experiments, the direct output of an EBES is a collection of discrete events.
This framework is a specific realization of \href{https://en.wikipedia.org/wiki/Discrete-event_simulation}{discrete-event simulation},
where the final distribution of outcomes is built up sequentially.
While the resulting patterns may agree with the predictions of quantum theory,
the EBES achieves this while rigorously adhering to Einstein's criterion of local causality.

In an EBES, there is no ``spooky action at a distance''.
Instead, events are generated by (non-Newtonian) dynamical systems in a strictly cause-and-effect manner.
These systems are modular, constructed by ``stitching together''
units similar to Lego\textsuperscript{\textregistered} blocks.
Unlike static blocks, however, certain units exhibit adaptive behavior,
possessing a primitive form of learning capability.
These units are called Deterministic Learning Machines (DLMs)~\cite{RAED05b}.

The EBES detailed in this \BOOK\ is a systematic, computer-based methodology for constructing and analyzing models
that generate data in the exact same manner as quantum physics experiments: event by event.
A defining characteristic of the EBES approach is that the state of the entire system remains
static between consecutive events.
Consequently, ``EBES time'' does not flow continuously; it ``jumps'' directly from the timestamp of
one event to the next.
This discrete temporal structure significantly enhances the computational efficiency of the methodology,
bypassing the need to calculate the system's evolution during periods of inactivity.

\noindent
\begin{center}
\framebox{
\parbox[t]{0.9\hsize}{%
It is crucial to recognize that an EBES does not rely on any pre-specified probability distributions,
such as those derived from Maxwell's electrodynamics, quantum theory, or other probabilistic models,
as its input.
Unlike standard approaches that ``sample'' from an existing wave intensity or probability,
the EBES generates its events autonomously from the bottom up.
It is precisely this feature that empowers the EBES to model phenomena that lie beyond the reach of standard quantum theory.}}%
\end{center}

\begin{figure}[!htp]
\centering
\includegraphics[width=0.90\hsize]{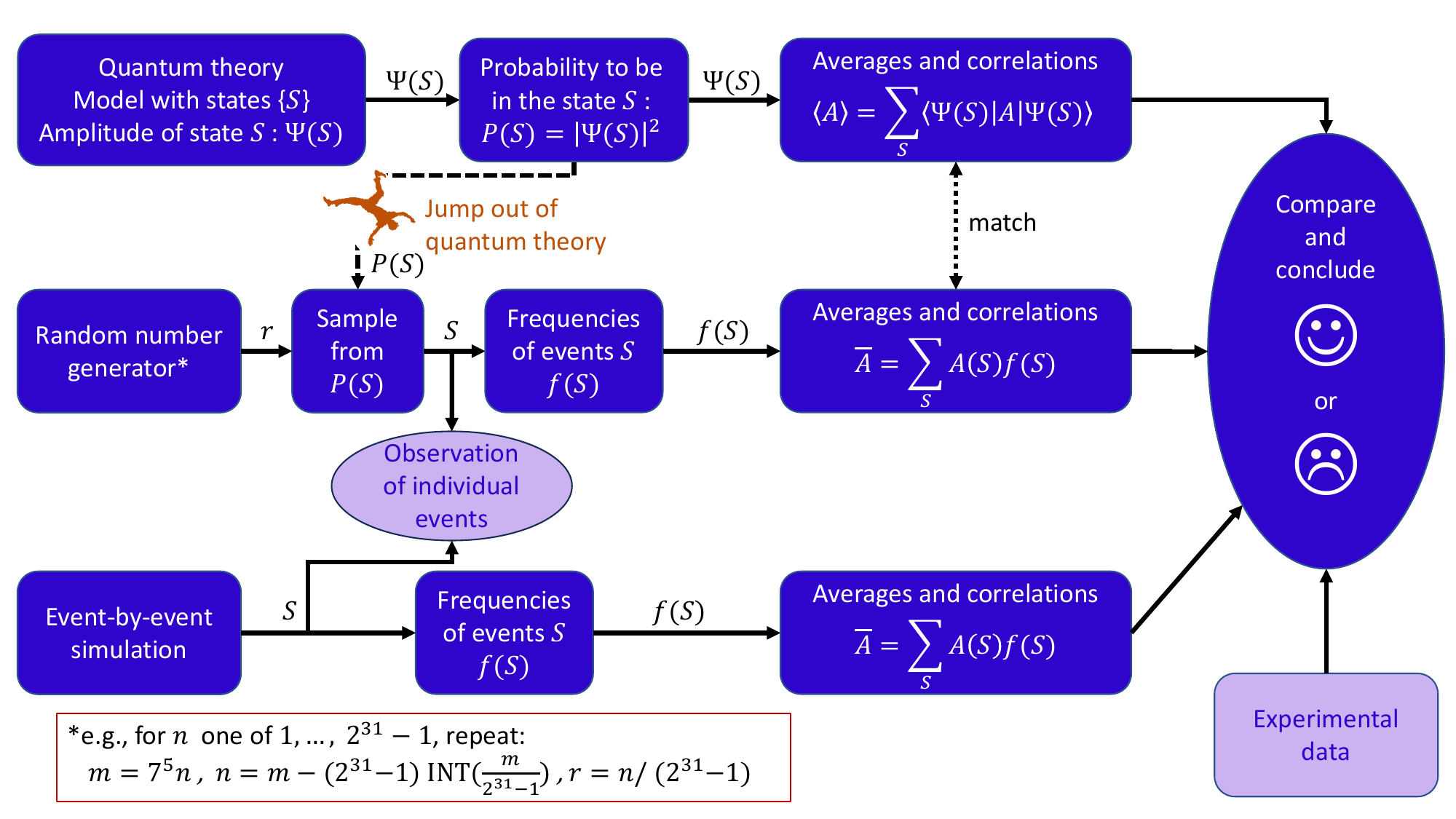}
\caption{Graphical representation of the relations between quantum theory,
event-by-event simulation (as applied to quantum physics experiments), actual
observation of individual events, and experimental data.
The function $\mathrm{INT}(x)$ keeps only the integer part of the number $x$, e.g., $\mathrm{INT}(3/2)=1$.
The random number $r$ takes values between zero and one.
}
\label{qtdiagram}
\end{figure}

Figure~\ref{qtdiagram} illustrates the structural relationship between quantum theory,
event-by-event simulations, and empirical observations.
Although the ``experimental data'' box is positioned in the bottom corner,
this placement should not be mistaken for a lack of importance;
rather, it identifies these data as the final arbiter of scientific validity.

The unidirectional arrows connecting the theoretical and computational models to the ``experimental data''
box emphasize a critical epistemological point:
the applicability of a model can be established only by comparing its generated averages and correlations
with analyzed empirical results.
If this comparison is deemed unsatisfactory, based on necessarily subjective criteria,
the model is revealed as inapplicable to that specific phenomenon.
Crucially, an inapplicable model is not inherently ``wrong'' or ``useless'' as a mathematical construct;
it simply fails to map onto the physical reality in question.

This brings us to the final row of Fig.~\ref{qtdiagram},
which signifies a complete paradigm shift compared to the first row.
In the EBES approach, models are constructed by treating individual events
as the fundamental entities of the theory.
The simulation produces a raw collection of events,
directly mirroring the data streams observed in laboratory experiments.
Notably, no pre-specified probability distribution serves as input.
The resulting statistics are not ``sampled'' from a theory; they emerge entirely from the underlying dynamics of the model.

EBES software produces data in excellent agreement with a wide range of fundamental quantum experiments (see appendix~\ref{FUNDA}),
including photon-based studies~\cite{MICH11a},
neutron interferometry~\cite{RAED12b},
tests of uncertainty relations~\cite{RAED14a},
and Stern-Gerlach experiments~\cite{RAED22}.
It effectively replicates Einstein-Podolsky-Rosen-Bohm experiments~\cite{RAED20a}
and quantum walks~\cite{WILL20b}, etc.,
all without invoking quantum theory.
The success of the EBES approach should not be misconstrued as an argument ``against'' quantum theory;
as a model for describing the distribution of empirical data,
quantum theory possesses unparalleled descriptive power~\cite{RAED14b,RAED23}.
Rather, it is only when we attempt to attach ``reality'' to the abstract symbols
of the quantum formalism that interpretations defying common sense begin to emerge~\cite{RAED23}.

\section{Structure of this \BOOK}

While three chapters frame fundamental quantum physics experiments as interactive games played by human participants,
the remaining chapters focus more concisely on the experimental setups,
standard theoretical frameworks, and the corresponding EBES.
A central feature of this \BOOK\ is the \href{\HOMEURL/DES/default.html}{EBES software},
which implements the specific rules for each of these games.
By running the software, the reader will find that EBES faithfully reproduces experimental results,
aligning perfectly with both theoretical predictions and, where available, raw laboratory data.
To demonstrate that the software does not rely on wave equations or hidden ``weirdness'',
we provide a detailed discussion of the Python source code, \href{\HOMEURL/DES/software.html}{downloadble here}.
Technical nuances and conceptual deep-dives are reserved for the appendices,
while standard historical context and textbook descriptions are available through cited literature and embedded links.
The other chapters demonstrate the versatility of the EBES framework by applying
it to a broader range of fundamental quantum experiments.

\begin{subappendices}

\section{Fundamental quantum physics experiments}\label{FUNDA}

Quantum optics is a field of physics that explores the behavior of photons,
the fundamental, discrete ``packets'' of light,
and their interactions with matter at the atomic and molecular level.
While light is described as a wave, quantum optics allows us to probe its dual nature by
highlighting its particle-like characteristics.
Through experiments involving interference, diffraction, entanglement, and delayed-choice scenarios,
researchers have exposed the often counter-intuitive realities of interpretation of quantum theory,
allegedly challenging our traditional understanding of causality and observation.

The most defining feature of quantum optics is the discrete nature of observation:
detection events occur one by one.
This sequential arrival of ``clicks'',
each signifying the registration of a single photon,
stands in stark contrast to the continuous patterns of light intensity captured by classical sensors.
In fundamental quantum physics experiments, phenomena such as interference, diffraction,
and entanglement are not observed instantaneously.
Instead, they are reconstructed over time, built up through the cumulative collection of these individual,
detection events.

Interferometer experiments using neutrons yield patterns remarkably similar to those produced by photons.
Since neutrons are traditionally categorized as material particles, neutron interferometry
provides a robust platform for exploring their wave-like behavior~\cite{RAUC15}.
For the scope of this \BOOK, both neutron interference and single-electron double-slit
experiments~\cite{MERL76, TONO89, TONO98, BACH13}
are classified as fundamental quantum physics experiments.
These demonstrations underscore the reality of wave-particle duality,
a profound ``weirdness'' that emerges
when we attempt to physically interpret the abstract mathematical formalisms of quantum theory.

\section{Quantum theory cannot produce events}\label{NOEVENTS}

As previously noted, quantum theory provides only the probability distribution $P(S)$ of an event $S$.
While Monte Carlo techniques can sample from this distribution to produce a series of events
that comply with the specified probability~\cite{PRES03},
this process, sketched in the second row of Fig.~\ref{qtdiagram},
relies on an essential external element: the (pseudo-)random number generator.

The red box contains a well-known random number generator algorithm.
This illustrates that the arithmetic rules governing such an algorithm do not fit within the frameworks of quantum theory,
probability theory, or Maxwell's electrodynamics.
These theories describe the distribution of events (or intensities), but they do not, and cannot,
describe what causes an individual event to be observed.
Indeed, orthodox quantum theory employs the concept of ``irreducible randomness''
to effectively declare that such a causal mechanism does not exist.

Repeatedly generating random numbers $r$ to sample from $P(S)$ yields a frequency distribution $f(S)$.
In practice, because the number of random numbers generated within a finite timeframe is limited,
$f(S)$ will inevitably differ from $P(S)$ due to statistical fluctuations.
These fluctuations diminish as the sample size increases.
Consequently, if we set aside these statistical variances,
the averages and correlations derived from the frequency distribution $f(S)$ will match their counterparts in quantum theory.

\noindent
\begin{center}
\framebox{
\parbox[t]{0.9\hsize}{%
To head off any misunderstandings: the event-by-event algorithms discussed in this \BOOK\ do not require prior
knowledge of a quantum-theoretical probability distribution $P(S)$.
On the contrary, these simulations generate a frequency distribution $f(S)$,
which,under appropriate conditions, is found to agree with the predictions of $P(S)$.
}}%
\end{center}

In summary, the central message is that while quantum theory successfully yields probability distributions,
it remains unable to provide a mechanism for generating individual events.
This is only a problem if one insists on the belief that observed events occur because Nature is ``following'' a mathematical model.
There is no logical reason, only a deep-seated conviction, why theorems derived within a specific model
must have a bearing on the reality represented by those events.

Attributing such power to mathematical abstractions is a classic instance of the mind projection fallacy~\cite{JAYN03}.
A similar critique is developed in the book ``The Blind Spot''
by A. Frank, M. Gleiser, and E. Thompson~\cite{BLINDSPOT},
where the authors argue that scientific theories inevitably reflect the limits
and biases of our representational stance rather than revealing the
intrinsic structure of Nature itself.

Instead of imposing our imagination on the world and thereby implicitly appealing
to a kind of magic, we adopt a more modest stance.
Our imagination yields theories, quantum theory among them, that successfully describe the statistics of well-designed experiments.
That success is remarkable, but it does not imply that the theory uncovers the mechanism by which events come to be.
Taking this view means acknowledging that any attempt to construct models capable of describing the occurrence of events,
rather than merely their statistics, requires stepping outside the established framework of theoretical physics.

\end{subappendices}
\label{INTRO}


\chapter{Young's two-beam interference experiment}\label{YOUNGINTRO}
Young's interference experiment, sketched in Fig.~\ref{YoungFig1},
played a major role in the general acceptance of the wave theory of light.
It probably is the first of many double-slit experiments.
Without going into much detail here, the basic elements of this experiment are~\cite{BORN64}:

\begin{figure}[H]
\centering
\includegraphics[width=0.9\hsize]{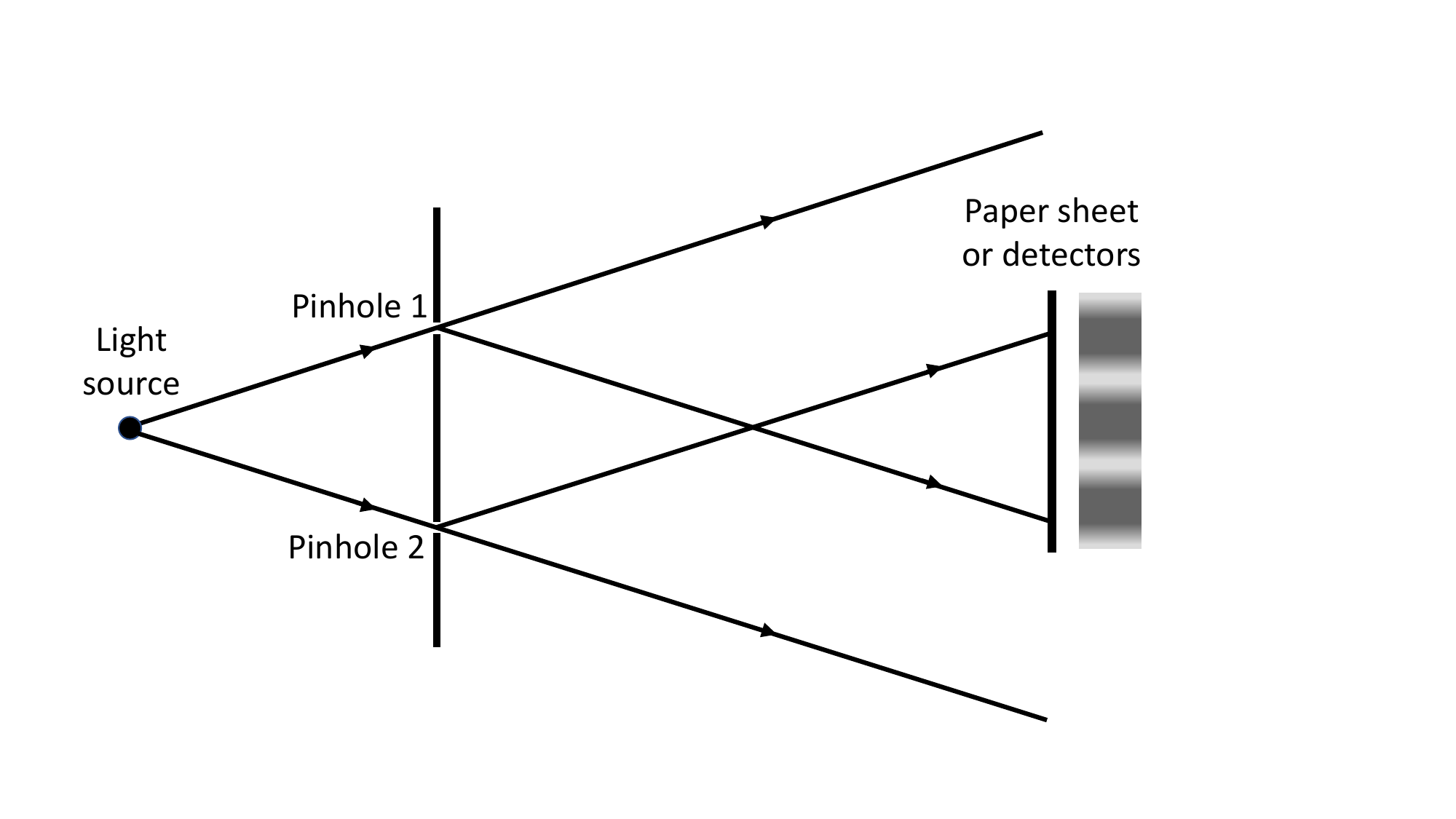}
\caption{
Conceptual, two-dimensional view of Young's two-beam interference experiment~\cite{BORN64}.
The light emerging from a source is transmitted and scattered by the two pinholes, resulting
in light propagating in different directions, as illustrated by the arrows.
The light illuminating a piece of paper (represented by the tick vertical line)
produces an alternating pattern of light and dark stripes, that is an interference pattern.
}
\label{YoungFig1}
\end{figure}

\begin{enumerate}
\item
Light impinges on two pinholes with their centers separated by some distance.
\item
The pinholes act as virtual sources producing waves.
\item
The waves emerging from the pinholes hit a photo plate.
\item
The observed intensity shows dark and light fringes, the characteristic signature
of interference.
\end{enumerate}

If detectors report individual clicks instead of an intensity distribution, the interference pattern builds up event by event.
According to
\href{https://www.feynmanlectures.caltech.edu/III_01.html#Ch1-S1}{Feynman (Vol. 3, p.1-1)}
this phenomenon is impossible, absolutely impossible, to explain in any classical way,
and has in it the heart of quantum mechanics. In reality, it contains the only mystery.

A laboratory experiment similar to Young's employs a single-photon source
and a bi-prism instead of a sheet of paper to split the beam of photons~\cite{JACQ05}.
Videos of the event-by-event build up of the interference pattern can found
\href{https://link.springer.com/article/10.1140/epjd/e2005-00201-y#App1}{here}.
The reader might by surprised by the visual similarity with the EBES dynamic graphics (see below).
The EBES of this experiment (which produces patterns that are different than
those of Young's experiment) can be found in Ref.~\cite{JIN10b}.
Readers who want to have a quick glance at the EBES of the Young experiment
can find a web-based demo \href{\HOMEURL/DES/demo/young.html}{here}.

The sections that follow, demonstrate that experiments with humans (or elephants) can show
the same interference patterns as Young's experiments with light or individual photons.

\section{A game with people}\label{YOUNGGAME}

Imagine playing the YOUNG game, illustrated in Fig.~\ref{YoungFig4}.
The rules of this game are the following:

\begin{enumerate}
\item
Walkers leave the room one-by-one at random intervals,
right at the center of one of the randomly chosen, open doors.
\item
At the door opening, the walker chooses at random, one of the so-called observers,
represented by the dashes of the vertical dashed line.
\item
Observers are distributed regularly over the vertical line, as indicated by the vertical dashed line.
The observers have an equal chance of being chosen.
Observers are not allowed to communicate with each other.
\item
At the door opening the walker picks up a mechanical, analog stopwatch, starts it and begins walking towards
the chosen observer.

\begin{figure}[H]
\centering
\includegraphics[width=0.85\hsize]{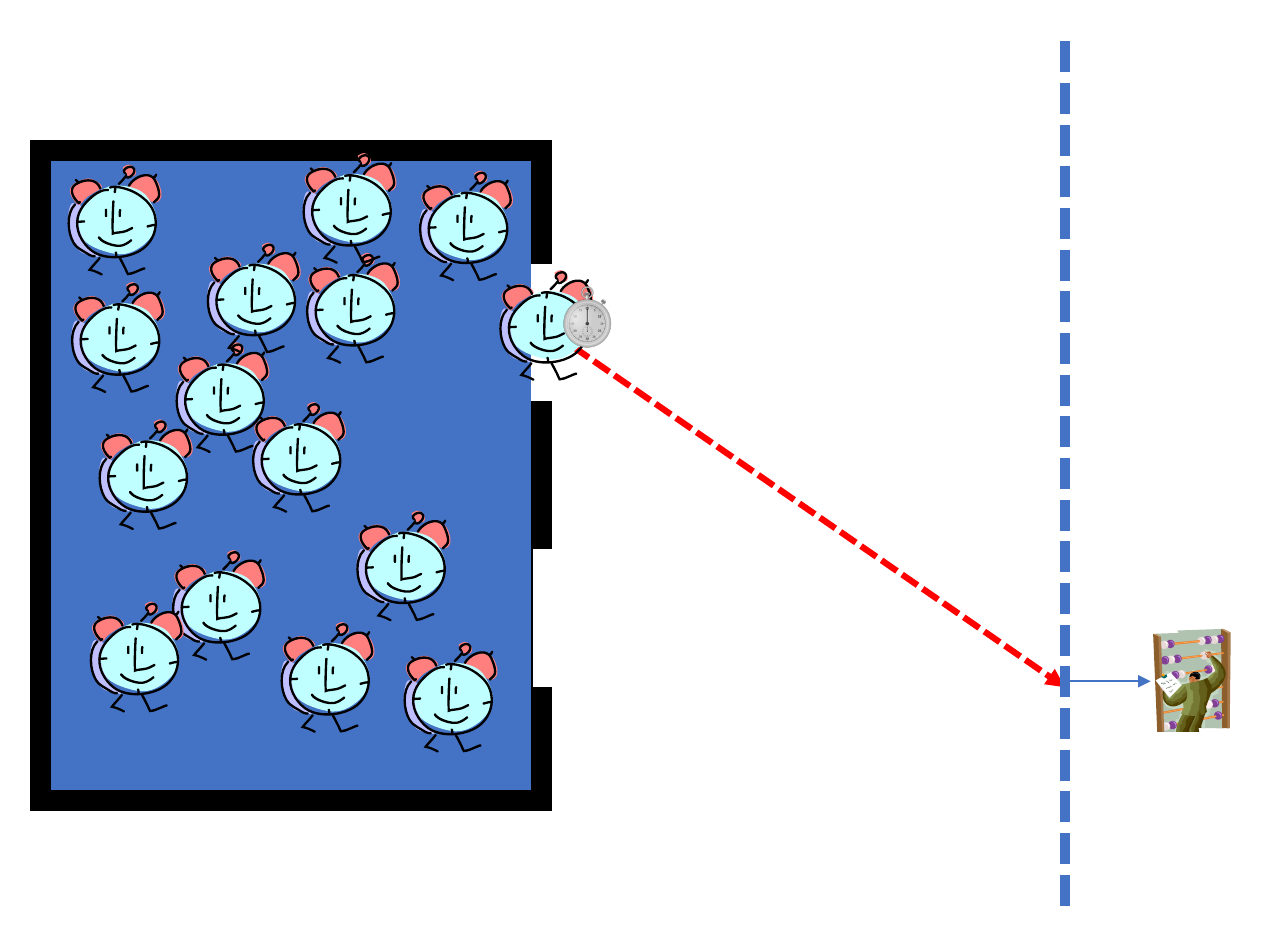} 
\caption{
Picture of a game that shows that people
can build up an interference pattern one-by-one without ever communicating with each other.
The dashes in the leftmost vertical lines represent human observers who may or may
not send a signal to their respective companion, represented by the person with an abacus.
}
\label{YoungFig4}
\end{figure}

\noindent
{\color{red}
\hypertarget{WARNING}{WARNING:}
There is a risk of over interpreting this analogy (see Section~\ref{PICTURES}).
The ``mechanical analog stopwatch'' is merely a convenient metaphor,
a mental model for the information the walker transmits to the observer.
A digital stopwatch would serve the purpose just as effectively.
Alternatively, the walker could use a mobile device to estimate distance rather than duration.
The essential requirement is simply that the walker can convey two things to the observer:
\begin{enumerate}
\item
    The magnitude (distance or duration).
\item
    The resolution (the fundamental scale of length or time).
\end{enumerate}
While the image of walkers or particles carrying ``clocks'' with rotating hands may be intuitively appealing,
it remains a pedagogical device rather than a physical necessity.
}

\item
All walkers move with exactly the same velocity in a straight line, as indicated by the dashed line with the arrow.
\item
When a walker meets the chosen observer, the latter reads off the position of the hand of the stopwatch
and uses this position to update (by rules to be specified) the internal state of the data processing device belonging to
this observer.
The walker then disappears from the scene.
\item
All observers must use the same rules to update the internal state of their own data processing device.
\item
Each observer has a dedicated companion, indicated by the person with the abacus.
Companions are not allowed to communicate with each other.
\item
The data processing device tells the observer to send a signal to his/her companion or not.
\item
The companion counts the number of signals received from the observer.
The communication between an observer and his/her companion is only through the signal
that an observer sends to his/her (and only his/her) companion.
\item
Only after a walker has disappeared from the scene, the next walker is allowed to proceed through a door.
This requirement is only needed to eliminate the possibility walkers communicate among each other.
\item
After many walkers have met an observer, the number of signals registered by
the companions is plotted versus the companion position.
\end{enumerate}

Clearly, the rules of this game are such that the distribution of counts is
build up one walker at a time and that there is no direct communication between
different walkers, different observers, and different companions.
Also note the elementary requirement
that all data processing devices operate in exactly the same manner.

\begin{center}
\framebox{
\parbox[t]{0.9\hsize}{%
KEY QUESTION: which rules should each data processing device employ in order
that the counts registered by the companions shows a distribution
that would normally be associated with interference of waves?
}}%
\end{center}

\section{Event-by-event simulation of the YOUNG game}\label{YOUNGRESULTS}

Instead of playing the YOUNG game with real people, it is much more convenient
to use a computer to simulate this game.
Section~\ref{COMPUTERCODE} discusses in some detail how the YOUNG game with people
is implemented as an event-by-event simulation.
Here, the focus is on the EBES results for a few representative cases
obtained by using the Python source code has been downloaded.
Using the standalone executables only differs in starting the code,
see \href{\HOMEURL/DES/quickstart.html}{quick start} for more details,

Assuming that Python3 and a few additional packages are installed and operational,
see \href{\HOMEURL/DES/installation.html}{software installation} for more details,
the procedure to run a simulation is as follows.
\begin{enumerate}
\item
Open a command window on your computer and change the directory to the directory that
contains the quantum optics demos.
\begin{figure}[!htp]
\centering
\includegraphics[width=0.95\hsize]{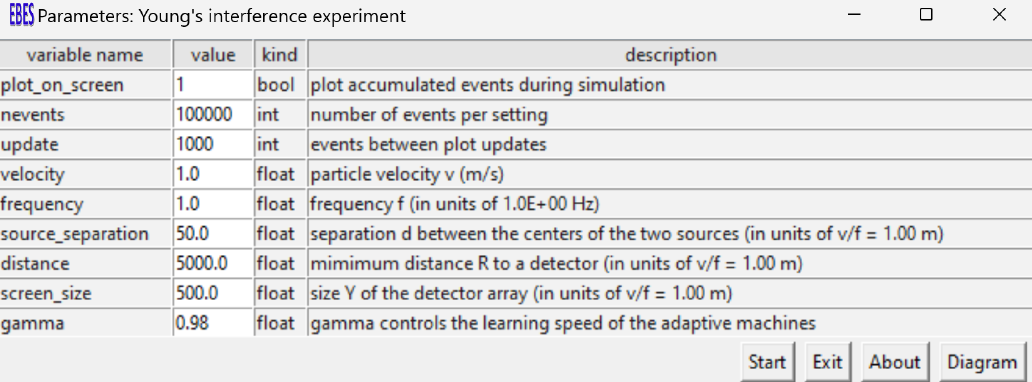} 
\caption{
Graphical interface to control the event-by-event simulation of the YOUNG game.
By default, the program starts with parameters that apply to humans, see below.
}
\label{YoungFig5}
\end{figure}
\item
Windows users, type 
``{\small \tt py run\_youngs\_interference\_experiment.py}''
(adding the {\tt -w} option activates noninteractive mode).
Linux or Mac users type {\tt python3} instead of {\tt py}.
The window with default parameter settings, shown in Fig.~\ref{YoungFig5}, appears,
initializing the parameters with values (see Fig.~\ref{YoungFig5}),
appropriate for humans (see below for other options).
The walking velocity is $3.6\,\mathrm{km}/\mathrm{h}$ and the hand of the clock
makes a full turn in $1\,\mathrm{s}$, corresponding to a frequency $f$ of $1\,\mathrm{Hz}$.
Referring to the schematic shown in Fig.~\ref{YoungFig2},
the separation $d$ between doors is $50\,\mathrm{m}$,
the shortest distance $R$ between the sources and the line of observers is
$5\,\mathrm{km}$ and two hundredth observers are distributed over a line running from  $-Y$ to $Y$
where $Y=500\,\mathrm{m}$.

\begin{figure}[H]
\centering
\includegraphics[width=0.80\hsize]{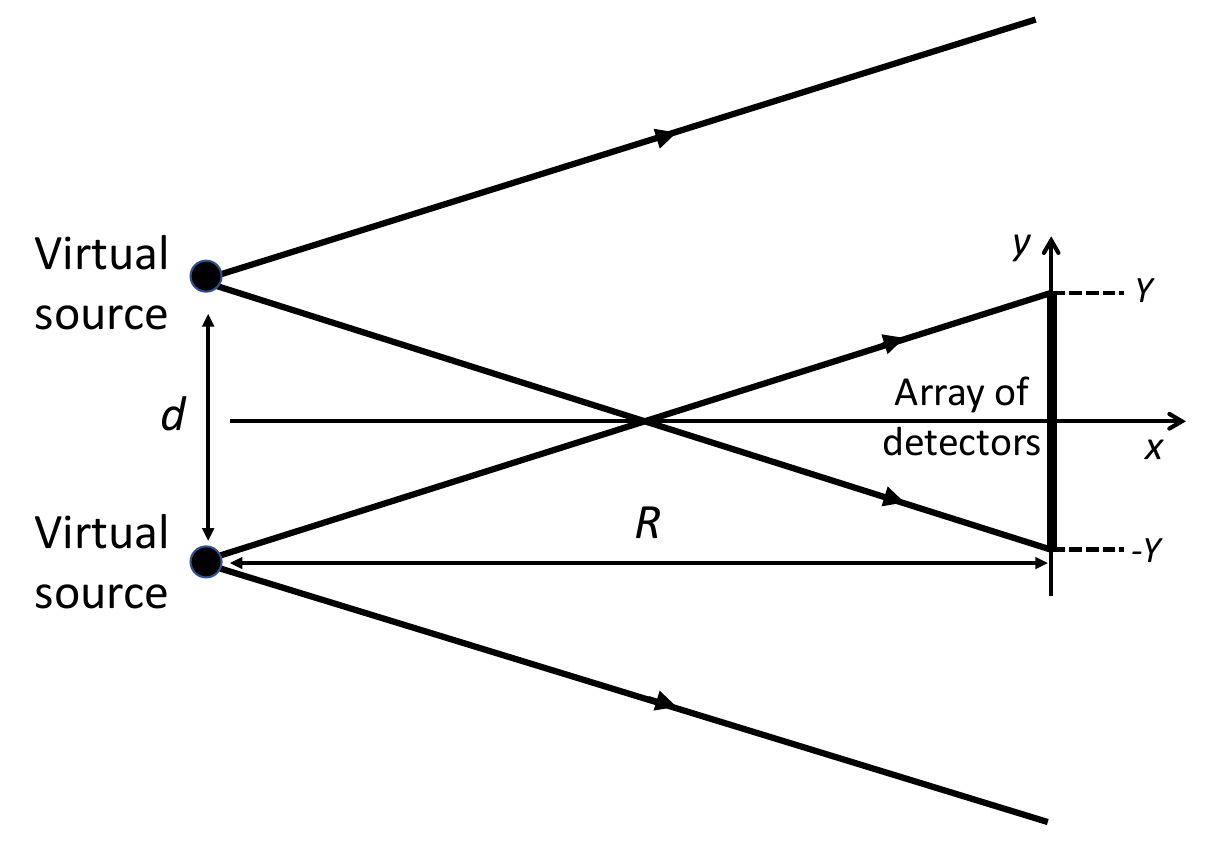} 
\caption{
Geometrical representation of the YOUNG game, also used for the wave mechanical
description of Young's two-beam interference experiment~\cite{BORN64}, see also appendix~\ref{secYOUNG1}.
Virtual point sources replace the doors in Fig.~\ref{YoungFig4}.
They are separated by a distance $d$ and send out, one-by-one, messengers/particles
that carry the time-of-flight information.
The $y$ coordinate determines the point of observation at the screen (the observers in Fig.~\ref{YoungFig4})
which is located a distance $R$ away from the plane of the sources.
}
\label{YoungFig2}
\end{figure}

\begin{figure}[H]
\centering
\includegraphics[width=0.80\hsize]{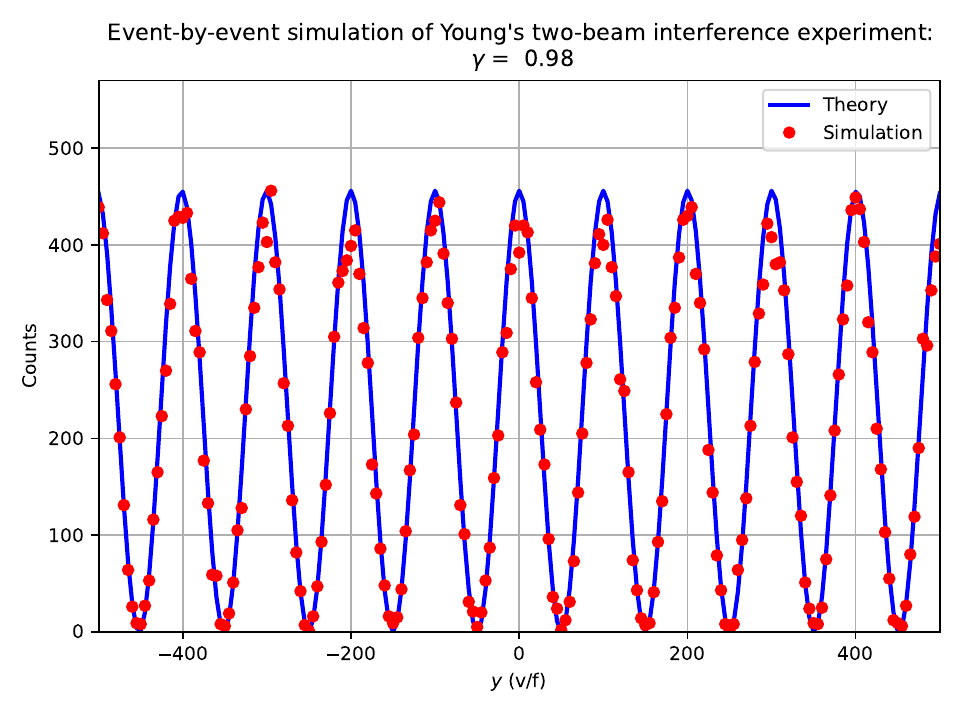}
\caption{
The theoretically expected result \ifFULLBOOK(solid line, see Eq.~(\ref{YOUNGTHEORY}) in Appendix~\ref{secYOUNG1})\ \fi
and the final counts (red bullets) obtained by the event-by-event simulation of the YOUNG game.
}
\label{youngresults2}
\end{figure}

\item
Press RETURN or click on the START button.
\item
Another window with a plot appears, showing the counts generated by the companions as more people leave the room.
The window is updated each time 1000 (variable name = update) walkers have met an observer.
After all 100000 (variable name = nevents) walkers have met an observer, the simulation
stops and draws a solid line on top of the histogram.
This solid line is the theoretically expected result\ifFULLBOOK\ (see Eq.~(\ref{YOUNG6a}) in Appendix~\ref{secYOUNG1})\ \fi.
For simplicity, the maximum of the solid line is set equal to the maximum count.
\item
Click on the small window asking you to continue.
\item
Figure~\ref{youngresults2} appears, showing the wave mechanical, theoretically expected result (solid line)
and the final counts (red bullets).
The differences between the solid line and the counts disappear if the total number
of walkers is increased.
\item
To verify this, click on the small window asking you to continue,
change the number of events to 1000000 and click on start (or press RETURN).
It may take a while for this simulation is finished (Python is not designed to compute fast).
The final result is shown in Fig.~\ref{youngresults3}.
\item
To verify that the same code also correctly simulates the game with one instead
of two doors, set the variable {\tt source\_separation} to zero.
This will not close the door but move the center of the two doors to the $(x=0,y=0)$
position 
so that they exactly overlap and effectively form one door.
\item
Restart the simulation.
The final result is shown in Fig.~\ref{youngresults4}.
\end{enumerate}

\begin{figure}[!htp]
\centering
\includegraphics[width=0.90\hsize]{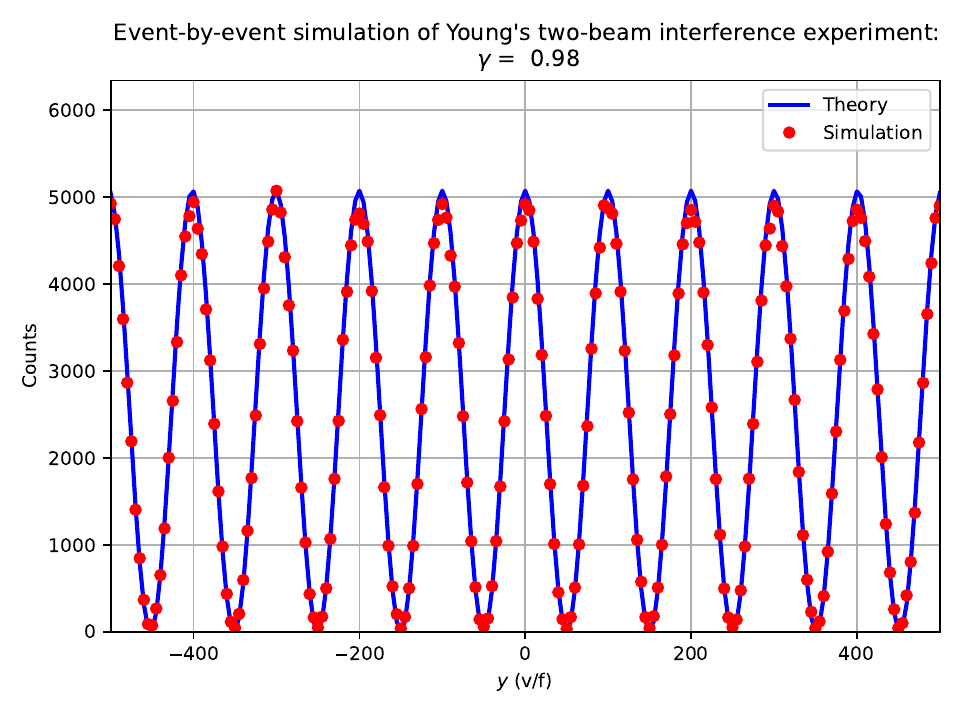}
\caption{
The same as Fig.~\ref{youngresults2} except that 1000000 instead of 100000 walkers were used.
}
\label{youngresults3}
\end{figure}

\begin{figure}[!htp]
\centering
\includegraphics[width=0.90\hsize]{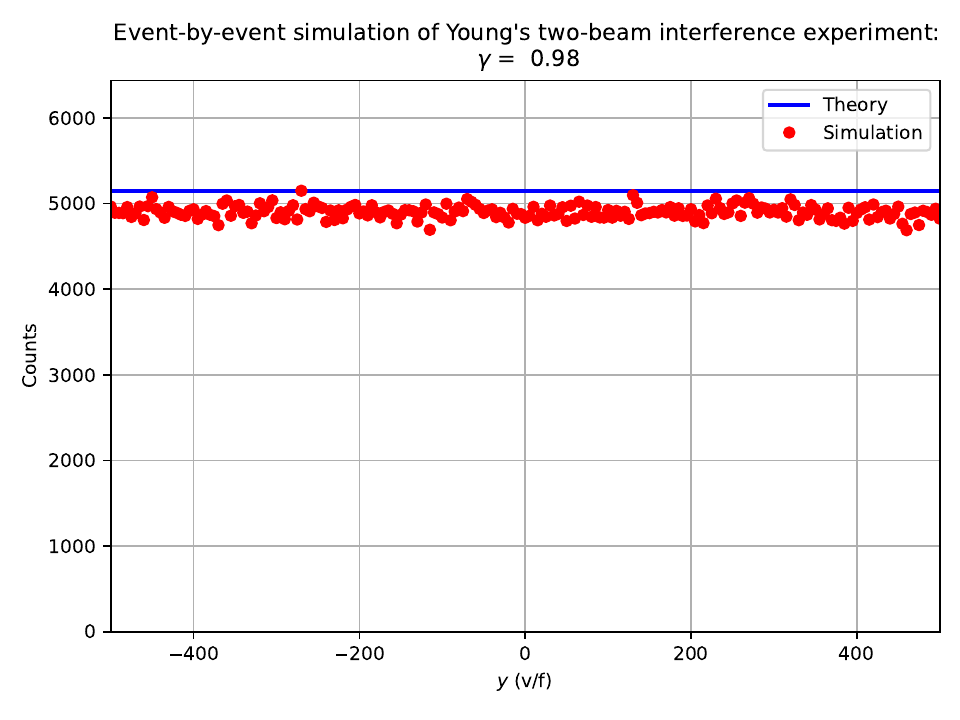}
\caption{
The same as Fig.~\ref{youngresults3} except that there is only one instead of
two doors.
}
\label{youngresults4}
\end{figure}

Figure~\ref{youngresults3} shows beyond any doubt that simulating
the YOUNG game with people according to the (up)date rules laid out in
section~\ref{YOUNGGAME} produces, walker-by-walker (or event-by-event),
the interference pattern \ifFULLBOOK(see Eq.~(\ref{YOUNG6a}) in Appendix~\ref{secYOUNG1})\ \fi that results from the wave theoretical
treatment of Young's two-beam interference experiment with light~\cite{BORN64}.

Figure~\ref{youngresults4} convincingly demonstrates that the same simulation code also reproduces
the expected results (i.e., no interference pattern) for the game with people
with only door or, equivalently, in the wave theoretical treatment of the experiment with one beam of light.
The minor difference between the theoretical results (horizontal line at 5000 counts) and the simulation data
is due to the fact that because of the adaptive nature of the event-base processor,
not all walkers that arrive at the detector increase the detection  count.

For convenience, the software automatically adjust the distances $d$, $R$, and $Y$ if the values
of the velocity and frequency are changed.
For instance, by changing the frequency $f$ from $1\,\mathrm{Hz}$ to  $1/60\,\mathrm{Hz}$ (appropriate for a standard
watch), the distances $d$, $R$, and $Y$ increase by a factor 60.
For this choice of parameters, in order that humans reach an observer at all,
they have to be extremely well-trained and may suffer a lot
to cover the distance of at least $300\,\mathrm{km}$, another good reason to resort to computer simulation.

Instead of playing the game with parameters that are appropriate to humans,
starting the code by adding the option -LIGHT,
that is by typing\hfil\break {\tt py run\_youngs\_interference\_experiment.py -LIGHT},
selects a set of parameters that are appropriate for
experiments with green light.
Of course, the user can always change parameters if desired.

\section{Python code for simulating the YOUNG game}\label{COMPUTERCODE}  

The software accompanying this \BOOK\ have been written in Python3, using
{\tt tkinter} for the graphical user interface (GUI).
The GUI and plotting of the simulation results take by far the largest part of Python code,
a part which is not discussed in this \BOOK.

\newcommand\code{{\bf experiment/youngs\_interference\_experiment.py}} 
In this section and similar sections in the following chapters, references to files
appear in bold (e.g., \code) and have to be read relative to the (sub)directory (by default
{\bf quantumoptics}) in which all the EBES software has been installed,
see the \href{\HOMEURL/DES/usersguide.html}{usersguide}.

\begin{table}[!ht]
\centering
\caption{%
Correspondence between the actors in the YOUNG game and the objects used in the Python code.
}
\begin{tabular}{cc}
\noalign{\medskip}
\hline\hline\noalign{\smallskip}
game with people       & event-by-event simulation \\ 
\hline\noalign{\smallskip}
walker                 & messenger/particle        \\ 
door                   & virtual source            \\ 
distance travelled     & distance travelled        \\ 
stopwatch              & internal clock            \\ 
stopwatch time         & message                   \\ 
observer + companion   & detector                  \\ 
distribution of counts & graphical output          \\ 
\hline\noalign{\smallskip}
\end{tabular}
\label{YOUNGtab1}
\end{table}

As there is a one-to-one mapping between the objects and their functions in the simulation code
and the actors and their action in the game with people (see section~\ref{YOUNGGAME}),
this mapping may help to understand how the simulation code is build and functions.
For convenience, this correspondence is summarized in Table~\ref{YOUNGtab1}.

To prevent potential misunderstanding, it should be noted that within the context of the code,
terms such as ``particle'' are purely functional labels for an object or class.
This nomenclature does not reference the concept of a particle as understood in everyday parlance or physics.
Consequently, the human participants in the YOUNG game are represented as ``particles'' within the codebase.

With this correspondence in mind, the level of abstraction that is required to translate the game
into a computer code is rather low.
The keys elements of the simulation code are the objects and their interactions.
Listing~\ref{Lmain} presents the code snippet (slightly modified from \code\ for typesetting purposes) that performs the simulation.
For the precise meaning of some of the variables used, see Fig.~\ref{YoungFig2}.

The first step (line number 2) is to create an {\tt 2*Ndetector+1} array of detector objects of the type
MessageSensitiveDetector (to be discussed later).
The variable {\tt Ndetector} (default set to 100) is hard-coded in \hfill\break\code\ but its value is easy to change.

The statement in line number 4 initiates a loop over {\tt nevents} events.
The value of {\tt nevents} is taken from the GUI parameter window (see Fig.~\ref{YoungFig5}).

Line 6 creates the object named {\tt p} which is an instance of the class {\tt Particle},
to be discussed in detail later.
For now, it suffices to know that (i) at the end of the loop, the object {\tt p} is destroyed (a Python feature),
(ii) with each iteration, a new object {\tt p} is being created, and (iii) upon creation all objects {\tt p}
have their clock time $t$, frequency $f$ and velocity $v$ set to zero, {\tt f}, and {\tt c}, respectively.
The values of {\tt f} and {\tt c} are taken from the GUI parameter window (see Fig.~\ref{YoungFig5}).

Lines 8 and 10 use the system-supplied uniform random number generator
to select the virtual source {\tt port} from which the particle starts its journey
and a detector {\tt k} to which the particle will travel.
Lines 12 and 14 employ elementary Euclidean geometry to compute the distance between the port and detector
along the $y$-direction, the path length and the time of flight, multiplied by $2\pi$.

Line 16 recalls the value of $\tau$ to update the message (part of the particle's internal state),
carried by the particle. The object-oriented-programming feature of Python is of great use here.
Finally, line 18 implements a similar construct to pass all the information carried by the particle
to the detector. An explanation of what {\tt .process(p,port)} does is given later.

{\scriptsize
\lstset{style=mystyle,linewidth=0.95\textwidth,caption={Code that simulates the YOUNG game, see section~\ref{YOUNGGAME}.},label={Lmain}}
\hbox{\hbox to 0.3cm{}
\begin{lstlisting}
# create an array of detector objects, all instances of the class MessageSensitiveDetector
detector = [MessageSensitiveDetector(gamma=self.settings.options.gamma,rng=self.rng) for it in range(2*Ndetector+1)]
# loop over all events
for i in range(nevents):
#  create the object p, an instance of the class Particle
    p = Particle(t=0,f=f,v=c)
#  randomly pick the virtual source from which the particle starts its journey
    port = self.rng.integers(2)  # a random integer 0 or 1
#  randomly pick a point where the particle will arrive
    k = self.rng.integers(0,2*Ndetector+1)  # a random integer 0, ... , 2 * Ndetector
#  and compute displacement in the y direction
    Y = ((k-Ndetector) * Ymax) / Ndetector -  d2 * (2 * port - 1)  # the y coordinate (relative to the source)
#  compute the path length, time of flight and then multiply by 2 * pi
    tau = twopi * sqrt( X * X + Y * Y) * p.frequency / p.velocity
#  change the particle's message according to time of flight
    p.process(cos(tau), sin(tau))
#  select detector and process the data carried by the particle
    detector[k].process(p,port)  # the variable port is not used in this application
\end{lstlisting}
}
}
\medskip

\renewcommand\code{{\bf dlm/detectors.py}}

How detectors are created and how they process the message
carried by the particle is discussed next.
Listing~\ref{Ldet} gives a code snippet taken from \code.
As before, the focus is on the essentials, not on technical details,
e.g. the {\tt self.} etc., which are Python-language specific.

{\scriptsize
\lstset{style=mystyle,linewidth=0.95\textwidth,caption={Python class definition for creating detector objects.},label={Ldet}}
\hbox{\hbox to 0.3cm{}
\begin{lstlisting}
class MessageSensitiveDetector(Detectors):

    def reset(self, gamma):
        self.count = 0
        self.x = np.zeros(2)
        self.gamma = gamma

    def process(self, particle, *args):
        self.x = self.gamma * self.x + (1 - self.gamma) * particle.message
        if self.rng.random() < self.x[0]**2 + self.x[1]**2:
           self.count += 1  # detector click

\end{lstlisting}\label{LISTING\thelstlisting}
}
}
\medskip

The ``class'' {\tt MessageSensitiveDetector} refers to another, more general class
{\tt Detectors} that contains code that is common to three kinds of detectors available in the simulation software.
As show by line 2 of Listing~\ref{Lmain}, one of them, namely {\tt MessageSensitiveDetector}, is used
to create instances of detector objects.
Upon creation, the function {\tt reset} is called (by {\tt Detectors}, see \code).
It serves to set the detection count to zero, initialize the internal
state (represented by a two-dimensional array {\tt x}), and store the value of the parameter $\gamma$
for later use.

The statement {\tt detector.process(p,port)} on line 18 of Listing~\ref{Lmain},
refers to the function {\tt process} defined in {\tt MessageSensitiveDetector}.
This particular function {\tt process} is called
with the object {\tt p} (an instance of a particle) as argument.
The second argument {\tt *args} is not used by this particular function {\tt process}.
Line 9 of Listing~\ref{Ldet} uses part of the data stored in the particle object
to implement the update rule Eq.~(\ref{UPDATErule0}).
Here, {\tt x[0]} and {\tt x[1]} correspond to $x_0$ and $x_1$ in Eq.~(\ref{UPDATErule0}), respectively.
Lines 10 and 11 implement the last part of the update rule Eq.~(\ref{UPDATErule2}).

\renewcommand\code{{\bf dlm/particle.py}}
Finally, the code creating particle objects and processing the messages carried by these objects will
be discussed.
Listing~\ref{Ldet} gives a code snippet taken from \code.
The class {\tt Particle} contains all the code to create and manipulate instances of a particle object.

{\scriptsize
\lstset{style=mystyle,linewidth=0.95\textwidth,caption={Python class definition for creating particle objects.},label={Lpar}}
\hbox{\hbox to 0.3cm{}
\begin{lstlisting}
class Particle:

    def __init__(self, t=0, f=1, v=1):
        self.time = t       # time of flight
        self.frequency = f  # frequency of stopwatch
        self.velocity = v   # velocity of the particle
        x = twopi * t * f
        self.message = array([cos(x), sin(x)])  # message is a 2D unit vector

    def process(self, c=1.0, s=0.0):
        self.message[0] = c
        self.message[1] = s
\end{lstlisting}
}
}
\medskip

Line 6 of Listing~\ref{Ldet}, creates the object named {\tt p}
which is an object of the class {\tt Particle}.
This ``creation'' entails calling the function {\tt \_\_init\_\_} in {\tt class Particle}.
The result of this operation is an object of type `{\tt Particle}
that has the properties {\tt time}, {\tt frequency}, {\tt velocity}, and {\tt message}.
The latter is a two-dimensional array of floating point numbers.
The values of these properties are taken from the arguments of the {\tt Particle} function in the calling code
(line 6 of Listing~\ref{Ldet}) or from the default values {\tt t=0}, {\tt f=1}, and {\tt  v=1}
if a corresponding argument was not specified.

The statement {\tt p.process(cos(tau),sin(tau))} on line 16 of Listing~\ref{Lmain},
refers to the function {\tt process} defined in {\tt Particle}.
When this particular {\tt process} is called, the default values
of {\tt c=1.0} and {\tt s=0.0} are replaced by the supplied values of {\tt cos(tau)} and {\tt sin(tau)},
respectively and subsequently used to update the content of the two-dimensional array {\tt message}.

\section{Where are the waves?}\label{NOWAVES}

Interference is a phenomenon traditionally associated with wave behavior.
However, the YOUNG game involves no physical waves, nor does the event-by-event algorithm possess any ``knowledge''
of the interference patterns derived from wave equations.
The pattern is not a predefined blueprint; rather, it is an emergent property of independent, localized events.

The reader is challenged to identify even a single line in the Python code suggesting that the program solves a wave equation.
In this simulation, all ``particles'' and ``detectors'' operate independently, and their interactions are strictly local.

Unlike water waves, there is no underlying collective mechanism or medium.
It is only the final distribution of counts,
accumulated as walkers arrive one at a time,
that resembles the patterns physicists typically associate with wave phenomena.

\section{Light versus water waves}

The theoretical model used to describe Young's two-beam interference experiment~\cite{BORN64} (see Appendix~\ref{secYOUNG1})
is fundamentally agnostic regarding the nature of the waves themselves.
Under appropriate conditions, this abstraction applies equally to water waves,
provided the parameters, such as frequency $f$ and velocity $v$, are adjusted accordingly.

However, in the case of water waves, interference is not a property of the individual molecules.
Instead, it characterizes the collective behavior of the medium:
the local density of molecules is higher at the crests than at the troughs.
Here, the ``waves'' are macroscopic excitations of a material surface,
a manifestation of the collective movement of water molecules rather than the independent action of the particles themselves.

\begin{figure}[!htp]
\centering
\includegraphics[width=0.4\hsize]{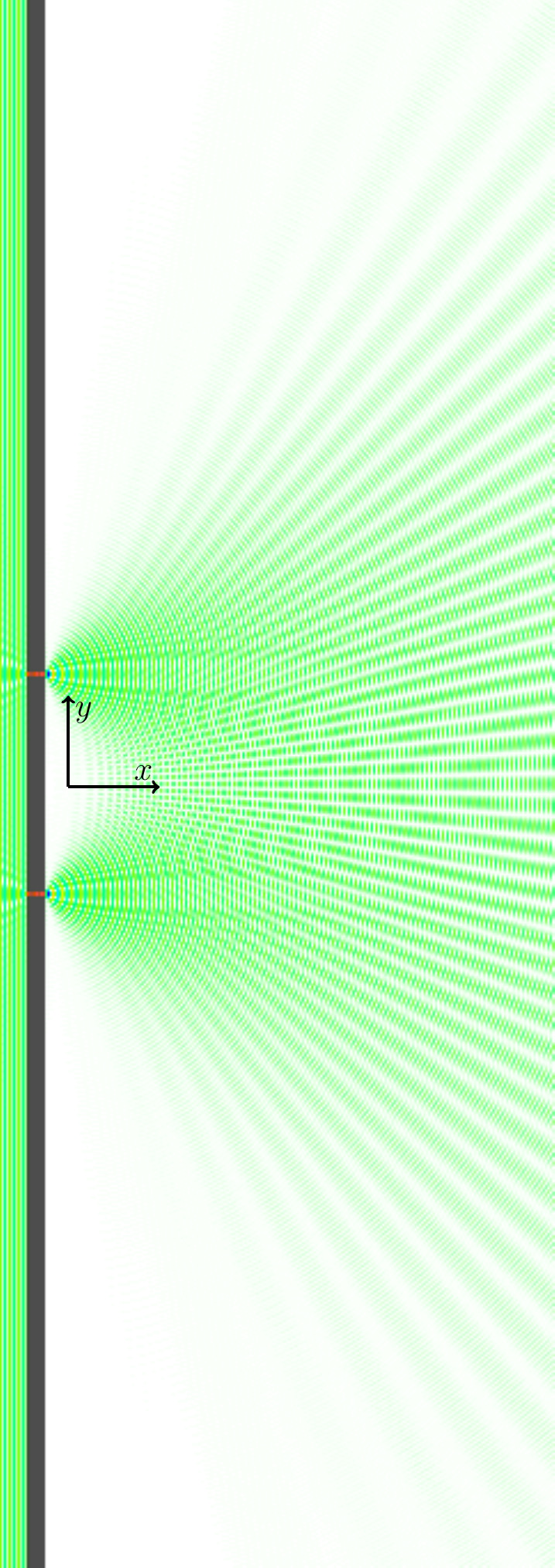} 
\caption{
Computer generated picture of a plane wave passing though a model of a metal foil
with two slits~\cite{RAED12c}.
The amplitude of the wave is color coded, from red representing the maximum amplitude
to blue representing the minimum amplitude.
The incident light is monochromatic and has wavelength $\lambda=v/f$, 
the width of both slits
$a=\lambda/2$
and the distance between the slit centers
$d=24\lambda$.
The slits are cut out of a $2\lambda$ thick metal foil,
the optical properties being modeled by a Drude model~\cite{BORN64,TAFL05} with a complex refractive index
${\widehat n}=n+i\kappa$ where $n=1.7$ and the attenuation index (extinction coefficient) $\kappa=1.8$.
The simulation was performed by TDME3D (in house software), a finite-difference time-domain
Maxwell equation solver~\cite{TAFL05}, using a spatial resolution of $\lambda/100$.
}
\label{YoungFig3}
\end{figure}

In the case of light, the prevailing consensus in physics is that no medium,
such as the hypothetical luminiferous ether, is required for transmission.
Light propagates through a vacuum without the need for an underlying substance.
Paradoxically, while light requires no medium to travel, it must interact with a material substance to be observed.

It is straightforward to demonstrate that two beams of light do not interact with one another directly:
if two beams are made to intersect, they pass through each other without any measurable effect or mutual interference.
This observation underscores the fact that light-on-light scattering is negligible in the classical (i.e., excluding
quantum electrodynamics effects) regime;
interference only becomes manifest when the combined field interacts with a third party, the material substance of a detector.

In Young's original experiment, for instance, the interference pattern became visible only upon striking a piece of paper,
the material medium.
Modern, more sophisticated experiments substitute the paper with photosensitive plates, photomultipliers, or photodiodes.
In every case, the conclusion remains the same: an interference pattern is only manifest when the light interacts with matter.

The presence of a material substance is fundamental to the observation of interference in both water and light.
In water, the wave itself is a structured pattern of local excess and deficit,
a displacement of the medium.
However, for light, the modern view holds that no such substance (i.e., no ether) exists.
Instead, a beam of light is understood as a collection of photons,
massless particles traveling at the speed of light.

The collective motion of these photons is governed by Maxwell's equations,
a system of partial differential equations~\cite{BORN64}
whose solutions in free space represent waves propagating at with the speed of light $c$.
Mathematically, waves are represented by fields that take assign definite values  $\psi = \psi(\br,t)$
at every point in space $\br=(x,y,z)$ and times $t$.
Consequently, while light requires no medium to propagate,
the mathematical framework used to describe its collective behavior remains inherently wave-like.

As an example, Fig.~\ref{YoungFig3} shows the $z$-component of the electric field $E_y(x,y,t=120\lambda/c)$,
obtained by solving Maxwell's equation for a two-slit geometry which extends from $-\infty$ to $+\infty$
in the $z$ direction.
The incident plane wave is generated by a current source parallel to the $y$-axis
in Fig.~\ref{YoungFig3}) and, before it hits the material containing the two slits,
propagates along the $x$-direction.

With this perspective in mind, computer-generated visualizations such as Fig.~\ref{YoungFig3},
which depict the ``interference'' of light waves across extended regions of space, are somewhat misleading.
In reality, these images represent the local light intensity that would be observed at a specific coordinate
if, and only if, a suitable material object, such as a detector,
were placed at that point.
Without the interaction between light and matter,
the ``pattern'' remains a mathematical abstraction rather than a physical observation.

\begin{figure}[!htp]
\centering
\includegraphics[width=0.95\hsize]{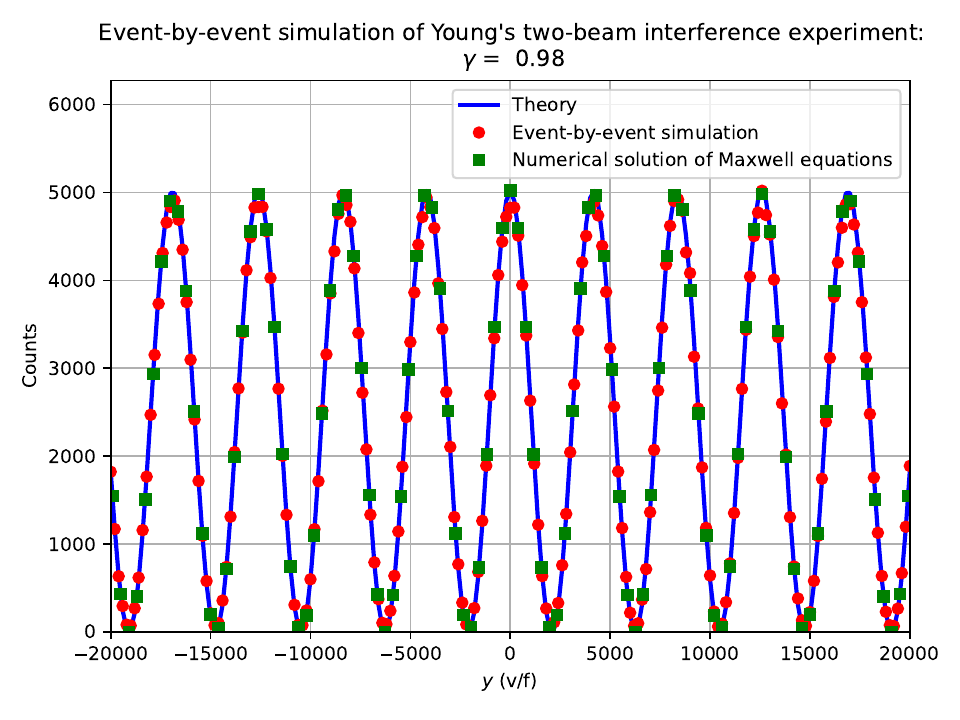}
\caption{
Comparison of the far-field intensity distributions obtained from the approximate analytical model (based on Fig.~\ref{YoungFig2}),
the event-by-event simulation, and the numerical solutions to Maxwell's equations
for the two-slit geometry shown in Fig.~\ref{YoungFig3}.
The three distinct approaches yield results in excellent agreement.
The command used to generate this plot was
{\tt py run\_youngs\_interference\_experiment.py -TDME}.
}
\label{YoungFig3b}
\end{figure}

If light is composed of discrete photons, it is natural to ask whether the interference patterns observed
by Young would still emerge if these photons arrived at the detector one by one.
As emphasized repeatedly, this question cannot be addressed within the framework of a traditional wave theory,
be it Maxwell's equations or standard quantum mechanics,
which primarily describes the evolution of fields or probability amplitudes rather than individual detection events.
To resolve this, it is productive to move beyond specifically optical phenomena and adopt a more generalized perspective,
e.g., the logic employed in the ``YOUNG game'' involving human participants.

Figure~\ref{YoungFig3b} illustrates the far-field intensity distribution as obtained through three distinct approaches:
\begin{enumerate}
\item
generated by the Event-Based Ensemble Simulation (EBES),
\item
extracted from the numerical solution of Maxwell's equations (cf. Fig.~\ref{YoungFig3}),
\item
predicted by the analytical approximation derived from the geometry in Fig.~\ref{YoungFig2}.
\end{enumerate}
Evidently, the agreement between the three approaches is excellent.

It is worth noting the significant disparity in computational requirements between the two methods.
Solving Maxwell's equations to produce Fig.~\ref{YoungFig3} (including the data in Fig.~\ref{YoungFig3b})
required 64 Intel Xeon Platinum 8168 CPUs (totaling 1,536 cores) running at 2.7 GHz for approximately 110 minutes.
In sharp contrast, the Python implementation of the event-by-event simulation generates the same intensity distributions
on a single PC core in a matter of minutes.

This orders-of-magnitude difference in effort stems from the nature of the algorithms.
The Maxwell solver must update over 174 million field values at each of the 21,360 time steps.
The vast majority of these calculations are dedicated to the propagation of the partial waves
from the slits to the detection region.
Conversely, in the event-by-event simulation, propagation through free space is computationally trivial,
as the algorithm does not need to resolve the intermediate field at every point in the intervening vacuum.

\section{Summary}\label{CONC}

The event-by-event simulation of the YOUNG game,
conducted with human participants,
reproduces the exact interference pattern obtained through the standard,
approximate wave-mechanical treatment of Young's two-beam interference experiment.

The simulation software demonstrates that the event-by-event emergence of the interference pattern
can be fully explained without invoking wave theory, quantum theory, or any kind of mysticism.
There is nothing mysterious at all, only a shift in perspective toward a message-passing framework.

In a pictorial description, one may imagine traveling agents (``particles'')
equipped with an internal clock, and stationary observing agents (``detectors'')
capable of reading that clock and performing a simple calculation.
Agents representing particles (and those representing detectors)
never communicate among themselves. Interaction occurs only when a particle-agent encounters a detector-agent;
at that moment, the detector receives the data carried by the particle.

The rules governing how a traveling agent modifies the state of an observing agent are precisely
what generate the interference pattern, event by event.
Even in the wave-mechanical or quantum-theoretical description,
where the event-by-event aspect is absent,
it is ultimately the interaction between the wave field and the detector material
(such as a photosensitive plate), rather than the waves themselves,
that produces the interference pattern.

\ifFULLBOOK
\begin{subappendices}

\section{Simplified wave theory}\label{secYOUNG1}

The wave mechanical analysis of Young's experiment
is based on the model shown in Fig.~\ref{YoungFig2}
and uses some simplifying approximations~\cite{BORN64}.

By assumption, waves leaving the sources initially have the same amplitude and the same phase.
Their time-dependent, oscillatory behavior is described by
\begin{equation}
A_{\pm} = A \sin 2\pi f t
\;,
\label{YOUNG0}
\end{equation}
where $A$ is the amplitude of the wave,
$f$ denotes the frequency by which the wave amplitude goes up and down, and $t$ is the time variable.
In this subsection, the subscript ``$+$'' (``$-$'') refers to pinhole 1(2) from which the wave emerges, see
Fig.~\ref{YoungFig1}.
Without missing much of the essence of the matter, the mathematics can be simplified considerably
by adopting the approximation that waves travel in straight lines,
in concert with laws of geometrical optics~\cite{BORN64}.
Let $v$ denote the velocity by which the wave crest moves ($v$ is the speed of light in Young's experiment).
The frequency $f$ is measured in one over seconds ($1/\mathrm{s}$)
and the speed $v$ is expressed in meters per second ($\mathrm{m}/\mathrm{s}$).

Next, pick a point on the detector screen and label this point by the coordinates $(x=0,y)$ (see Fig.~\ref{YoungFig2}).
By Pythagoras' rule, the squares of the shortest distances between this point of observation
and the virtual sources are given by
\begin{equation}
s^2_{\pm} = R^2 + \left( y\pm d/2 \right)^2
\label{YOUNG1a}
\;,
\end{equation}
where the ``$+$'' (``$-$'') sign refers to the upper (lower) virtual source in Fig.~\ref{YoungFig2}.

According to wave mechanics (and Eq.~(\ref{YOUNG0})), it takes the wave crest a time $\tau_{\pm}=s_{\pm}/v$
to travel the distance $s_{\pm}$.
Thus, the amplitudes of the waves merging at $(x=0,y)$ are given by
\begin{equation}
A_{\pm} = A \sin 2\pi f (t + \tau_{\pm})
\;.
\label{YOUNG2}
\end{equation}

The key feature of wave theory is that in order to compute the intensity observed at a particular point in space
the amplitudes of all the waves merging at that point have to be added, and their sum squared~\cite{BORN64}.
In the case at hand, this yields
\begin{eqnarray}
I(y,t) &=& \left(A_{+}+A_{-}\right)^2 = A^2 \left[\sin 2\pi f (t + \tau_{+}) + \sin 2\pi f (t + \tau_{-}) \right]
\nonumber \\
&=&4A^2\sin^2 \pi f (2t + \tau_{+} + \tau_{-})\; \cos^2 \pi f (\tau_{+} - \tau_{-})
\;.
\label{YOUNG3}
\end{eqnarray}
Frequencies of visible light are of the order of $10^{14}\,\mathrm{Hz}$ and
typical light detectors report the average over many periods of the rapidly oscillating intensity $I(y,t)$.
Integrating the time-dependent factor in Eq.~(\ref{YOUNG3}) over a time $T$ that covers many of these oscillations
gives
\begin{eqnarray}
&&\frac{1}{T} \int_{0}^{T} \sin^2 \pi f (2t + \tau_{+} + \tau_{-})\;dt
=\frac{1}{2T} \int_{0}^{T} \left[1 - \cos 2\pi f (2t + \tau_{+} + \tau_{-})\right]\;dt
\nonumber \\
&=&\frac{1}{2} - \frac{1}{8\pi fT} [\sin  2\pi f (2T + \tau_{+} + \tau_{-})-\sin  2\pi f (\tau_{+} + \tau_{-})]
\xrightarrow{T\to\infty}\frac{1}{2}
\label{YOUNG4a}
\;,
\end{eqnarray}
and for sufficiently large time $T$, Eq.~(\ref{YOUNG3}) becomes
\begin{eqnarray}
I(y) =\frac{1}{T} \int_{0}^{T} I(y,t)\;dt \approx 2A^2\cos^2 \pi f (\tau_{+} - \tau_{-})
\label{YOUNG4}
\;.
\end{eqnarray}
From Eq.~(\ref{YOUNG1a}) and with the help of some elementary algebra, it follows that
\begin{eqnarray}
f (\tau_{+} - \tau_{-})=
\frac{f}{v}\frac{2yd}{\sqrt{R^2 + \left( y+d/2 \right)^2}+\sqrt{R^2 + \left( y-d/2 \right)^2}}
\xrightarrow{R\gg|y\pm d/2|} \frac{f}{v}\frac{yd}{R}
\label{YOUNG5a}
\;.
\end{eqnarray}
Therefore, if the condition $R\gg|y\pm d/2|$ is satisfied,
the approximate wave mechanical treatment predicts that the intensity observed at the screen
is given by~\cite{BORN64}
\begin{eqnarray}
I(y) &=&2A^2\cos^2 \frac{\pi ydf}{vR} 
\label{YOUNG6a}
\;.
\end{eqnarray}
Equation~(\ref{YOUNG6a}) tells us that as a function of $y$, the intensity changes from
its maximum ($2A^2$) to zero in periodic manner, the distance between two successive
maxima being $2fR/vd$.
Qualitatively, Eq.~(\ref{YOUNG6a}) captures the essential features of the intensity pattern observed
in Young's two-beam interference experiment.

\subsection{Conditions to observe a nice interference pattern}\label{NOSHOW}

As is clear from the simplified wave mechanical treatment of section~\ref{secYOUNG1},
in order to obtain the simple sinusoidal interference pattern Eq.~(\ref{YOUNG6a}),
the value of $v/f$ has to be chosen properly in relation to the distances $y$, $R$, and $d$.

The first condition is that the distance $d$ between the pinholes
is much smaller than the size $Y$ of the detector array, that is $d\ll Y$, see Fig.~\ref{YoungFig2}.
The second condition is that $Y$ is much smaller than
the shortest distance $R$ between pinholes and detectors, that is $Y\ll R$, see Fig.~\ref{YoungFig2}.

When {\tt run\_youngs\_interference\_experiment.py} starts, the settings appropriate for humans are selected,
that is $v=1\,\mathrm{m/s}$ ($=3.6\,\mathrm{km/hour}$), $f=1 \,\mathrm{Hz}$,
$d=50 \,\mathrm{m} \ll Y= 500 \,\mathrm{m}\ll R= 5000 \,\mathrm{m}$.
Clearly, the required conditions are, approximately, satisfied.

When {\tt run\_youngs\_interference\_experiment.py -LIGHT}
or\hfil\break {\tt run\_youngs\_interference\_experiment.py -TDME}
starts, the settings appropriate for green light are chosen.
For instance, in the case that the software compares with the data obtained by solving Maxwell's equation (option -TDME),
that is $v=1\,\mathrm{m/s}$ (in units of the speed of light), $f=5.45\times10^{14}\,\mathrm{Hz}$,
$d=24 \,\mathrm{nm} \ll Y= 2000 \,\mathrm{nm}\ll R=100000 \,\mathrm{nm}$ so that
the required conditions are satisfied.

\section{Over interpretation of pictures}\label{PICTURES}
The \hyperlink{WARNING}{WARNING},
made in point 4 of the description of the game deserves further discussion.
Imagine that instead of a clock, the walker carries a device to measure the distance $D$ traveled
and a counter $C$ that counts the number of the steps made by the walker.
The shortest distances between the virtual sources and the detector at $y$ are given by Eq.~(\ref{YOUNG1a}).
Thus, for a fixed ``$+$'' path, $D=s_{+}$. For a fixed ``$-$'' path, $D=s_{-}$.
The observer now reads off the distance $D$ and divides $D$ by the count $C$
to obtain $\lambda=D/C$ (units of metres).
Clearly, setting the ratio $v/f$ equal to $\lambda$, the interference pattern
Eq.~(\ref{YOUNG6a}) does not change.
In other words, to obtain latter,
it does not matter what kind of picture is attached to the message carried
by the walker, as long as it is possible to infer a length scale (e.g., $v/f$ or $D/C$) from it.
In wave mechanics $\lambda$ is called the wavelength.
The wavelength $\lambda$ is the distance between two successive crests (or troughs) of the wave
and determines the characteristic length scale of the wave motion.
In the event-by-event approach, $\lambda$ is a characteristic length scale
of the experiment under consideration but since there are no waves in this approach,
it also does not warrant interpreting $\lambda$ as a wavelength.

\section{Details of the EBES of the YOUNG game}\label{MATHGAME}

Let us focus on one particular observer and ask ourselves how to
encode and process the stopwatch data which the walkers, arriving
it this particular observer, carry with them.

Although not essential (see section~\ref{PICTURES}), it is computationally convenient to encode
the position of the hand of the stopwatch by a pair of numbers, namely
$c=\cos2\pi f\tau$ and $s=\sin2\pi f\tau$ where $\tau$ is the time
it took the walker to reach the chosen observer.

The frequency $f$ is defined as $f=1/P$ where $P$ is the time it takes the hand
of the stopwatch to complete one full turn.
In one full rotation of the hand, taking a time $P$, the walker marches a distance $\lambda=vP=v/f$
where $v$ denotes the constant walking velocity.

In the case of monochromatic light, the most obvious choice for $f$ is to take the frequency
(corresponding to the color) of the light.
For instance, for green light $f\approx 5.5\times 10^{15}\,\mathrm{Hz}$.
However, in the case that the YOUNG game is played with people,
the value of $f$ will be different, as discussed earlier.

The internal state of the processing device is encoded in two numbers $-1\le x_0\le 1$ and $-1\le x_1\le 1$.
The following, very simple update rules~\cite{JIN10b} play the major part in all event-by-event simulations:
\begin{subequations}
\label{UPDATErule0}
\begin{eqnarray}
x_0 &\leftarrow&\gamma x_0 + (1-\gamma) c,
\label{UPDATErule0a}
\\
x_1 &\leftarrow&\gamma x_1 + (1-\gamma) s
\label{UPDATErule0b}
\;,
\end{eqnarray}
\end{subequations}
where $0\le\gamma<1$ is a free parameter by which controls
the pace with which the processing device adapts to new data.

The left arrow indicates that the current values of $x_0$ and $x_1$ are to be replaced by their respective values
given by the formulas of the right of the left arrow.
The main reason for writing the update rules in this form is
that this is exactly the operation which a digital computer actually performs when executing the Python program.
Note that the internal state requires storage for only two floating point numbers

Obviously, if $\gamma=1$, the state of the device is frozen, making
the device useless and if $\gamma=0$,
the device simply copies the incoming data to the internal state.
For the present purposes, the most interesting regime is when
$\gamma$ is close to but less than one.
Then, one can show that the value of $x_0$ ($x_1$) approaches the average of many values
of $c$ ($s$) that have been supplied as input~\cite{JIN10b}, see also appendix~\ref{WHYTHERULES}.
It is in this regime that the event-by-event simulations can produce
the results, that is the distributions of events, that are characteristic of quantum physics experiments.

Focusing on the same observer as before,
the times $\tau$ that this observer will read off from the stopwatches of the walkers
can at most take two different values, depending on the door through which a walker left.
Let us denote these times by $\tau_{\pm}$
where the ``$+$'' (``$-$'') sign refers to the upper (lower) door in Fig.~\ref{YoungFig4}.
The rules of the game imply that the number of times the observer reads off $\tau_{+}$ or $\tau_{-}$
is about the same.
As mentioned above, if $\gamma$ is close to but less than one,
$x_0\approx (\cos2\pi f\tau_{+}+\cos2\pi f\tau_{-})/2$ and
$x_1\approx (\sin2\pi f\tau_{+}+\sin2\pi f\tau_{-})/2$.

To complete the construction of the processing device, a rule
by which the observer send the signal to his/her companion needs to be defined.
Note that by elementary trigonometry
\begin{eqnarray}
x_0^2+x_1^2
&\approx&\frac{1}{4}(\cos2\pi f\tau_{+}+\cos2\pi f\tau_{-})^2+(\sin2\pi f\tau_{+}+\sin2\pi f\tau_{-})^2
\nonumber \\
&&= \frac{1}{2} + \frac{1}{2}(\cos2\pi f\tau_{+}\cos2\pi f\tau_{-})+(\sin2\pi f\tau_{+}\sin2\pi f\tau_{-})
\nonumber \\
&&=\frac{1}{2} + \frac{1}{2}[\cos2\pi f(\tau_{+}- \tau_{-})]
= \cos^2\pi f(\tau_{+}- \tau_{-})
\label{UPDATErule1}
\;,
\end{eqnarray}
showing that the final expression in Eq.~(\ref{UPDATErule1}) only depends of the
difference $\tau_{+}- \tau_{-}$ of the time of flights.

As Eq.~(\ref{UPDATErule1}) never takes negative values, the observer can use this
expression as a weight to decide whether or not to send a signal to his/her companion
The rate at which these signals are sent is then given by $x_0^2+x_1^2$.
Technically, for each walker arriving at the observer,
a uniform random number $0< r< 1$ is generated for use in the application of the rule
\begin{eqnarray}
r < x_0^2 + x_1^2\;\;\implies\;\;\hbox{send the signal to the companion}
\label{UPDATErule2}
\;.
\end{eqnarray}

According to the rules of game, the companion simply counts the number of signal received.
Consequently, after many walkers have been processed,
the companion's count will be proportional to $\cos^2\pi f(\tau_{+}- \tau_{-})$.

Note that an arbitrarily chosen companion was chosen to discuss the construction of the update rules.
Consistency of the approach requires that the update rules are the same for all companions.

To compute $\tau_{\pm}$ by Euclidean geometry,
it is expedient to use Fig.~\ref{YoungFig2}, a slightly more abstract version
of Fig.~\ref{YoungFig4}.
We pick a point on the detector screen and label this point by the coordinates $(x=0,y)$ (see Fig.~\ref{YoungFig2}).
By Pythagoras' rule, the squares of the shortest distances between this point of observation
and the virtual sources are given by
\begin{equation}
s^2_{\pm} = R^2 + \left( y\pm d/2 \right)^2
\label{YOUNG1}
\;,
\end{equation}
where the $+$ ($-$) sign refers to the upper (lower) virtual source in Fig.~\ref{YoungFig2}.

Using Eq.~(\ref{YOUNG1}) and some elementary algebra yields
\begin{eqnarray}
\tau_{+} - \tau_{-}=(s_{+} - s_{-})/v
=\frac{2yd/v}{\sqrt{R^2 + \left( y+d/2 \right)^2}+\sqrt{R^2 + \left( y-d/2 \right)^2}}
\;,
\nonumber \\
\label{YOUNG5}
\end{eqnarray}
where $v$ denotes the constant velocity of the walkers.

Eq.~(\ref{YOUNG5}) simplifies considerably if
the distance between the sources and the range $Y$ of $y$-values
of the points of observation are small compared to the distance $R$ between the sources
and the array of detectors.
Expressed in symbols, if $|y|\le Y \ll R$ and $d \ll R$,
the square roots in Eq.~(\ref{YOUNG5}) may be, to a good approximation, replaced by $R$.
Therefore,
\begin{eqnarray}
\tau_{+} - \tau_{-}&=&\frac{yd}{vR}
\label{YOUNG6}
\;,
\end{eqnarray}
if $Y \ll R$ and $d \ll R$, simple expressions indeed.

Putting all this together, the conclusion is that if the condition $R\gg|y\pm d/2|$ is satisfied,
the YOUNG game (experiment) with people (photons) yields a distribution of counts which is given by
\begin{eqnarray}
\mathrm{Counts}(y) &\approx&2A^2\cos^2 \frac{\pi ydf}{vR}=A^2\left(1+\cos\frac{2\pi ydf}{vR}\right)
\label{YOUNG7}
\label{YOUNGTHEORY}
\;,
\end{eqnarray}
where $A$ is a constant of proportionality and
the approximate sign ``$\approx$'' indicates that right hand side has been obtained
under the assumptions that $\gamma$ is close to but smaller than one and that the number of walkers is very large.
Obviously, Eq.~(\ref{YOUNG7}) is identical to the intensity pattern obtained by
the approximate wave mechanical description of Young's two-beam interference experiment, see Eq.~(\ref{YOUNG6a}).

Equation~(\ref{YOUNG7}) tells us that as a function of $y$, the count varies between its
maximum value ($2A^2$) and zero in periodic manner.
The counts reach their maximum at $y=nvR/df$ and their minimum at  $y=(n+1/2)vR/df$
where $n=0,\pm1,\pm2,\ldots$.
The periodic variation of the counts as function of the position of observation
is ``the'' key feature of an interference pattern.
However, as explained in section~\ref{NOSHOW}, to actually ``see''
a nice interference pattern in a realistic setting,
the value of $v/f$ has to be chosen properly in relation to the distances $y$, $R$, and $d$.

It is important to mention here that there is no relation between the appearance of the cosine in Eq.~(\ref{YOUNG7})
and the encoding of the stopwatch hand in terms of a cosine and a sine.
For instance, the event-by-event simulation of Fraunhofer diffraction employs
the same encoding but the interference pattern cannot be represented by
the simple expression Eq.~(\ref{YOUNG7}).

\begin{center}
\framebox{
\parbox[t]{0.9\hsize}{%
In summary,
using the update rules Eqs.~(\ref{UPDATErule0a}),~(\ref{UPDATErule0b}), and~(\ref{UPDATErule2})
for playing the YOUNG game with a lot of people and
with enough (observer,companion) pairs to cover the line of observation with high resolution
will produce an interference pattern.
Moreover, playing the YOUNG game with people yields the same interference pattern Eq.~(\ref{YOUNG7})
as the one obtained by the standard, approximate wave mechanical treatment Eq.~(\ref{YOUNG6a})
of Young's two-beam interference experiment.
}}%
\end{center}

\subsection{What happens if one door is closed?}
Imagine that instead of having two open doors, see Fig.~\ref{YoungFig4}, the game is played with only one door open.
If the value of $\gamma$ is close to but less than one, the two-door game produces an interference pattern.
However, for the one-door game, the distribution of counts will be flat, that is no interference pattern
will be observed.

To see how this come about (without changing the rules of the game!), it is sufficient to recall that the
update rules Eqs.~(\ref{UPDATErule0a}) and ~(\ref{UPDATErule0b})
compute the averages of the $c$'s and $s$'s that the observer obtains by reading off the stopwatch time.

Say that the upper door (corresponding to pinhole 1) is closed and the game is played according to the rules given above.
Then, when a walker meets an observer, the stopwatch of this walker will {\bf always} show a time $\tau_{-}$
corresponding to the distance travelled.
Thus, after some walkers arrived at a particular observer,
$x_0\approx\cos2\pi f\tau_{-}$, $x_1\approx\sin2\pi f\tau_{-}$, and $x_0^2 + x_1^2\approx1$ (here the $\approx$ sign
has the same meaning as above).

The rule Eq.~(\ref{UPDATErule2}), the fact that the random number $r<1$,
together with $x_0^2 + x_1^2\approx1$ imply
that the observer will send a signal to the companion, almost every time a walker meets the observer.
Disregarding statistical fluctuations, the distribution of counts is flat, that is there is no interference pattern.

\begin{center}
\framebox{
\parbox[t]{0.9\hsize}{%
In summary, if $\gamma$ is larger than zero but smaller than one, playing the YOUNG game
will produce an interference pattern if walkers are allowed to leave through the two doors
but will not produce an interference pattern if one of the two doors is closed,
and this without having to modify the rules of the game!
Even though there are no waves in play, the outcome of the game in terms of open/closed doors is
in complete concert with the one obtained from a wave mechanical description of experiments
with one and two beams of light (or one and two sources of water waves).
All these facts are most easily verified by simulation of the YOUNG game.
}}%
\end{center}

\subsection{Role of the control parameter} 

While interference patterns are often associated with waves, the analysis above shows that
they can just as easily arise from event-by-event processes.
By applying the update rules in Eqs. (\ref{UPDATErule0a}) and (\ref{UPDATErule2}),
we can explicitly model a discrete process that culminates in a standard interference pattern.

Section~\ref{YOUNGGAME} only briefly touched upon the role of the free parameter $\gamma$
in Eqs.~(\ref{UPDATErule0a}) and~(\ref{UPDATErule0b}).
For a deeper understanding of the working of these rules, it is necessary to examine
the effect of $\gamma$ on the interference pattern in more detail.

As mentioned earlier, the case $\gamma=1$ can be excluded from further consideration
because then the internal state of the processing device is frozen.
The opposite case, $\gamma=0$ is a little more interesting.
From Eqs.~(\ref{UPDATErule0a}) and~(\ref{UPDATErule0b}) with $\gamma=0$,
it follows immediately that
$x_0=\cos2\pi f\tau_{\pm}$ and $x_1=\sin2\pi f\tau_{\pm}$,
where the ``$+$'' (``$-$'') sign indicates that the walker left through to the upper (lower) door
(see Fig.~\ref{YoungFig4}).
But in this case $u^2+v^2=1$ so that by the rule Eq.~(\ref{UPDATErule2})
and the fact that the random number $r<1$,
the observer will send a signal to the companion, every time a walker meets the observer.
In terms of detector performance, this means that the detection efficiency is 100\%.
In this case, after many walkers have left the room,
all companions will count (approximately) the same number of walkers.
In other words, the distribution of counts will be approximately flat, and there is no resemblance
to an interference pattern at all.

For intermediate values of $\gamma$, e.g. $\gamma=1/2$, the distribution of counts
exhibits periodic oscillations with an amplitude that is (much) less than if $\gamma$ is close to one.
This amplitude is a monotonically increasing function of $\gamma$.
Its actual value is most easily determined by simulation.
In wave of quantum theory parlance, changing the parameter $\gamma$ mimics the change of coherence of the wave.

\subsection{Why the rules Eqs.~(\ref{UPDATErule0}) and~(\ref{UPDATErule2})?}\label{WHYTHERULES}

As previously noted, the update rule in Eq. (\ref{UPDATErule0a})
is used to compute the average of the input sequence $\{c_1,c_2,\ldots,c_N\}$ to the data processor.
This raises a natural question: why use Eq. (\ref{UPDATErule0a}) instead of the standard arithmetic mean,
$(c_1+c_2+\ldots+c_N)/N$? At first glance, the latter appears to be a much simpler approach.

Formalizing this idea in terms of rules that the data processor has to carry out gives
\begin{subequations}
\label{UPDATErule3}
\begin{eqnarray}
N &\leftarrow&N+1
\label{UPDATErule3a}
\\
w &\leftarrow&w + c
\label{UPDATErule3b}
\\
u' &\leftarrow&w/N
\label{UPDATErule3c}
\;,
\end{eqnarray}
\end{subequations}
where initially, it is necessary to set $N=0$ and $w=0$
in order that the value of $u'$ agrees with the arithmetic average.
Obviously, these rules compute the exact average of the sequence $(c_1+c_2+\ldots+c_N)/N$.
However, that is not the only requirement for the data processor to be able to play the YOUNG game properly.

For instance, consider the case that the game is played with $N_1$ walkers and with the upper door closed
(showing no interference pattern). Then suddenly open the lower door and send $N_2$ walkers
to one of both doors (in which case an interference pattern is expect to appear).
According to Eq.~(\ref{UPDATErule3}), the arithmetic average is given by
\begin{eqnarray}
u' = \frac{N_1 c_{-} + N_2 (c_{-}+c_{+})}{N_1+N_2}
\label{UPDATErule4}
\;,
\end{eqnarray}
where $c_{\pm}=\cos 2\pi f\tau_{\pm}$.
However, if $N_1$ is much, much larger than $N_2$, at the end of the game, $u' \sim c_{-}$
yielding no interference pattern.
But in order to observe an interference pattern, it is necessary that $u' \sim c_{-}+c_{+}$.
Clearly, a data processor that operates according to Eq.~(\ref{UPDATErule3}) is pretty useless for
EBES purposes.

In contrast, update rule Eq.~(\ref{UPDATErule0a}) computes a good approximation to average
of the sequence $(c_1+c_2+\ldots+c_N)/N$ {\bf and} has the virtue that the value of $\gamma$ rather than $N$
determines the accuracy of the estimate.
To see how this works, consider the explicit expression for $u$ after processing the input
sequence $\{c_1,c_2,\ldots,c_N\}$ obtained by starting from Eq.~(\ref{UPDATErule0a})
and performing some elementary algebraic manipulations yielding
\begin{eqnarray}
u_N = \gamma^N u_0 + (1-\gamma)\left(c_N+\gamma c_{N-1} + \ldots \gamma^{N-1} c_1\right)
\label{UPDATErule5}
\;.
\end{eqnarray}
There are two aspects of this formula that deserve attention.

First, because $\gamma<1$, $\gamma^N $ vanishes exponentially with increasing $N$.
This implies that, independent of the initial value of $u_0$, the first term does not significantly
contribute to the average if $N$ is large.
Recall that for Eq.~(\ref{UPDATErule4}) to be useful, it is essential that initially $N=0$ and $w=0$.
Using Eq.~(\ref{UPDATErule0a}), the initial value $u_0$ is irrelevant (in practice).

Second, the contribution of the $c$'s to the average are weighted by a power of $\gamma$.
The most recent value of $c$ ($c_N$) appears with weight equal to one
whereas the first $c$ ($c_1$) contributes with weight $\gamma^{N-1}$, a small number.
In plain words, Eq.~(\ref{UPDATErule0a})
defines a data processor with the capacity to ``learn'' and to ``forget''.
Such a data processor is a primitive adaptive machine.

Finally, we consider Eq. (\ref{UPDATErule2}).
The sole purpose of employing a random number to gate the transmission of signals is to ensure that,
to an uninformed observer, the signal sequence appears unpredictable.
If this stochasticity is not a requirement, alternative adaptive processors,
distinct from the one defined by Eq. (\ref{UPDATErule0}),
can generate regular signal series that yield nearly identical interference patterns \cite{RAED05b}.
Note that the current simulation software strictly implements the random-number approach.

\subsection{How ``realistic'' is the detector model?}

Equations~(\ref{UPDATErule0}) and~(\ref{UPDATErule2})
describe the operation of a device that has the functionality of the
detector. If $\gamma=0$, the device clicks with every walker meeting an observer.
If $0<\gamma<1$, the number of clicks depends on the sequence of stopwatch times
that on observer reads off from the stopwatches of the incoming walkers.

Clearly, this ``detection process'' does not attempt to account for the intricate physical mechanisms within a detector,
such as the cascade of events in a photomultiplier or the nuclear reactions triggered by neutron capture.
While the simulation's implementation is a significant simplification of laboratory hardware,
its utility lies precisely in this abstraction. The following analogy illustrates this point.

Ohm's law gives the relation between the electrical current $I$ flowing through a conductor
and the constant potential $V$ applied across it.
In its fundamental form, $V=RI$, where $R$ represents the resistance.
For the engineering of electrical and electronic devices, Ohm's law is indispensable.
However, it provides no insight into the microscopic dynamics of charge carriers;
describing the flow of electrons in detail requires extensive knowledge of solid-state and statistical physics.
While such a detailed description is scientifically more rigorous,
it adds very little to the practical utility of Ohm's law in most applications.

Viewed through this lens, the processing units in our event-by-event simulations should be regarded as
minimalist models of device behavior.
They are designed in the spirit of Occam's razor:
stripped of unnecessary complexity, yet sufficiently sophisticated to reproduce the phenomena of interest.

\subsection{Elapsed versus simulation time}

As discussed in Section~\ref{sectionEBES}, the state of an EBES system remains static between consecutive events.
In this framework, ``time'' is discrete, jumping directly from one event to the next.
Consequently, EBES time differs fundamentally from the continuous time variable found in contemporary ``laws of physics''.

The rules of the game do not dictate exactly when a walker should depart,
only that they must do so sequentially:
a walker cannot leave until the previous one has reached an observer.
For instance, if any walker takes less than ten minutes to reach an observer,
then as long as the departure interval exceeds ten minutes,
the final distribution of counts remains invariant.
In this sense, the game is agnostic toward ``wall-clock time''.

Furthermore, the observer+companion unit records arrivals independently of the real-time interval between them.
However, the update rules in Eq.~(\ref{UPDATErule0}) imply that these units possess a primitive
form of memory that gradually ``forgets'' earlier inputs.
Here, ``earlier'' refers to the relative position in the sequence of walkers rather than elapsed physical time.
If the unit ceases to receive new input, its internal state remains unchanged indefinitely.

Admittedly, this is an idealization; in physical reality, memory decay is typically tied to elapsed time.
If this temporal decay is significant for a specific experiment,
it can be integrated into the EBES almost trivially.
This is achieved by introducing a clock generator to synchronize the steps of the game.
After each application of the update rules (Eq.~\ref{UPDATErule0}), the state variables are scaled:
$x_0\leftarrow\kappa x_0$ and $x_1\leftarrow\kappa x_1$.
The parameter $0<\kappa=1$ determines the rate at which the unit loses
memory relative to the ``elapsed time'' defined by the clock.
When $\kappa=1$, memory is preserved perfectly regardless of the time elapsed between events.

\subsection{Are the rules Eqs.~(\ref{UPDATErule0}) and~(\ref{UPDATErule2}) special?}

The rules defined by Eqs.~(\ref{UPDATErule0}) and~(\ref{UPDATErule2}),
along with their straightforward extensions,
have been utilized to perform event-by-event simulations of a wide array of quantum physics experiments.
To the best of our knowledge, there is currently no single- or two-particle laboratory experiment
that cannot be successfully modeled using the EBES approach.
Notably, the EBES provides a level of granularity, specifically, the description of individual detection events,
that is fundamentally inaccessible to standard quantum theory.
Given its success across such a broad class of experiments,
it is reasonable to consider the EBES approach as a legitimate alternative for theory-building in fundamental quantum physics.

This proposition naturally raises the question of uniqueness:
is this the only such theory?
The answer is clearly no.
The rules established in Eqs.~(\ref{UPDATErule0}) and~(\ref{UPDATErule2})
are not the exclusive path to the count distribution predicted in Eq.~(\ref{YOUNG7}).
Indeed, the earliest event-by-event simulations utilized more sophisticated adaptive machines~\cite{RAED05b}.
While those machines could be optimized for faster adaptation or deterministic output,
the underlying conceptual foundations remain identical to the rules presented here. Ultimately,
there is no a priori reason why a single, unique theoretical framework must describe a given class of phenomena,
a point the EBES approach effectively demonstrates for the field of quantum optics.

\end{subappendices}
\fi


\chapter{Neutron interferometry experiment}\label{NEUTRONINTRO}
\begin{figure}[!bp]
\centering
\includegraphics[width=0.9\hsize]{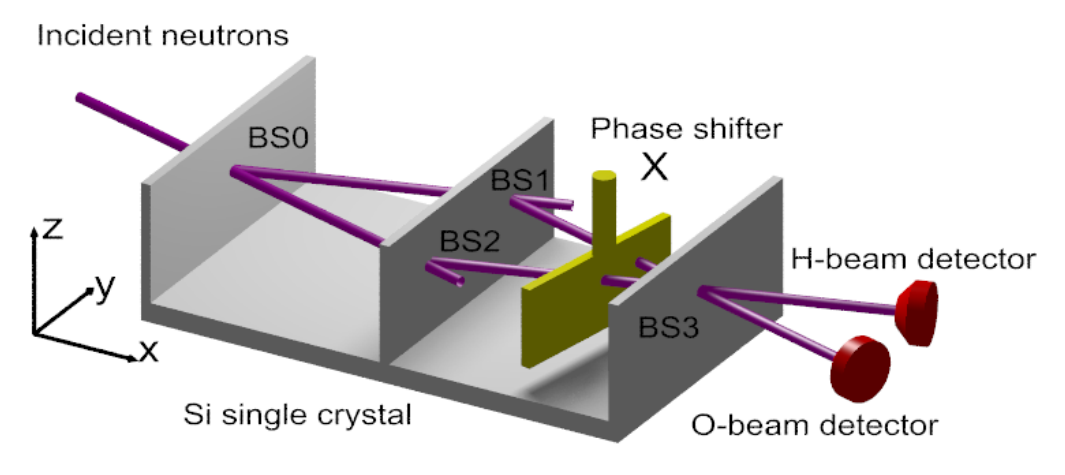}
\caption{
Layout of the perfect monolithic silicon crystal neutron interferometer~\cite{RAUC74a,RAUC15}.
BS0, ..., BS3: silicon plates acting as beam splitters;
phase shifter: aluminum plate tilted by an angle (controlled by the setting $X$ of a stepper motor)
with respect to the $y$ axis.
Detectors at the far right count the number of neutrons in the O- and H-beams.
Neutrons that are transmitted (or refracted) by BS1 or BS2 leave the interferometer
and do not contribute to the interference signal.
}
\label{NeutronFig1}
\end{figure}

Figure~\ref{NeutronFig1} illustrates the primary components of a laboratory neutron interferometry experiment~\cite{RAUC15}.
For brevity, the apparatus, excluding the detectors,
is hereafter referred to as the ``interferometer''.
Incident neutrons, selected for specific energy and direction following nuclear reactions,
form the beam entering the device~\cite{RAUC15}.
These neutrons traverse a near-perfect, monolithic silicon crystal.
Under specific Bragg conditions~\cite{RAUC15} which, while complex, are not central to the present discussion,
the neutrons follow the paths indicated by the solid beams in Fig.~\ref{NeutronFig1}.
Upon exiting the final plate of the crystal, neutrons in the O- and H-beams (this terminology is standard for
neutron interference experiments~\cite{RAUC15})
trigger detectors with efficiencies exceeding 99\%.
Notably, once a neutron interacts with a detector, it is absorbed and ceases to exist~\cite{RAUC15}.

\begin{figure}[!tp]
\centering
\includegraphics[width=0.85\hsize]{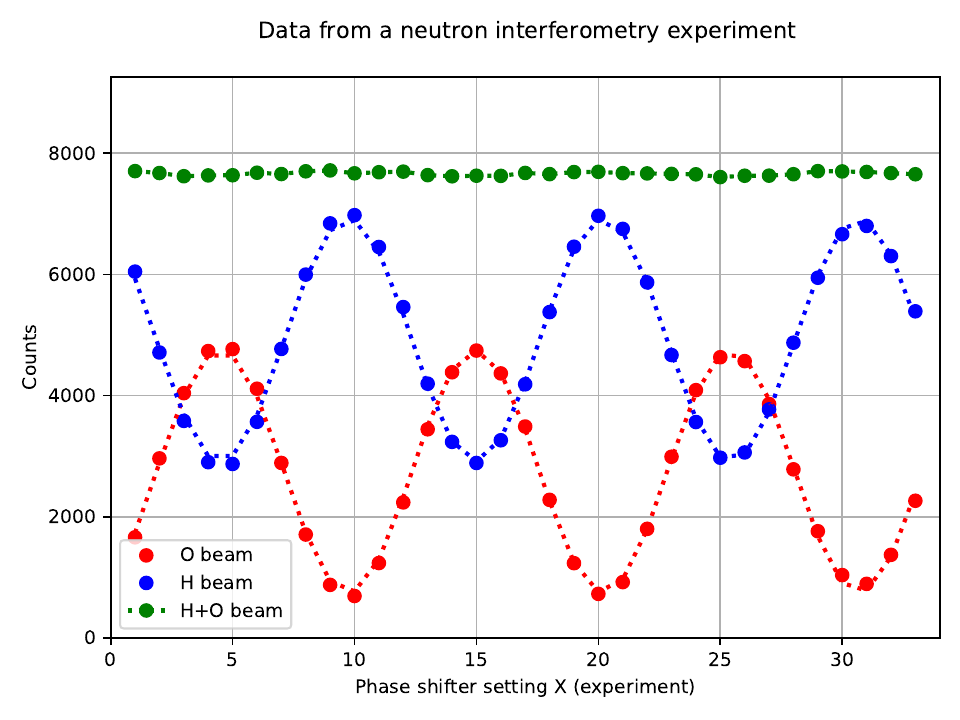}
\caption{
Neutron counts as a function of the phase shifter setting $X$
as obtained from the data files of a real experiment, reported on in Ref.~\onlinecite{WILL20c}.
Red circles: O-beam count;
blue circles: H-beam count;
green circles: H-beam count + O-beam count.
The dotted lines are guide to the eye.
}
\label{NeutronFig2}
\end{figure}

\begin{center}
\framebox{
\parbox[t]{0.9\hsize}{%
A neutron triggering a detector, entering or leaving one of the silicon plates, or passing through
the phase shifter is regarded as an {\bf event}.
}}%
\end{center}

\begin{center}
\framebox{
\parbox[t]{0.9\hsize}{%
The relative frequency of an event, for instance, a neutron exiting a silicon plate via one of two possible paths,
is defined as the ratio of the number of such occurrences to the total number of events.
In this context, the total is represented by the aggregate number of neutrons exiting that same silicon plate.
}}%
\end{center}

The central feature of this experiment is that the detection counts in each outgoing beam exhibit
a near-periodic variation as a function of the phase shifter's orientation, $X$ (see Fig.~\ref{NeutronFig2}).
Crucially, while the individual counts fluctuate,
the total number of neutrons across both beams remains essentially constant.

The experimental data shown in Fig.~\ref{NeutronFig2} are represented rather well by the simple functions
\begin{subequations}
\label{app3}
\begin{eqnarray}
N_\mathrm{O}&\approx& 
A_\mathrm{O}\left[ 1+ B_\mathrm{O}\cos\varphi  \right]
\;,
\label{app3o}
\\ 
N_\mathrm{H}&\approx& 
A_\mathrm{H}\left[1 - B_\mathrm{H}\cos\varphi \right]
\;,
\label{app3h}
\end{eqnarray}
\end{subequations}
with $\varphi=\Omega X + \phi_\mathrm{O}$ and
$\Omega\approx0.60$,
$\phi_\mathrm{O}\approx-2.75$,
$A_\mathrm{O}\approx 2781$,
$A_\mathrm{H}\approx 4951$,
$B_\mathrm{O}\approx 0.74$,
$B_\mathrm{H}\approx 0.42$,
and the subscripts referring to a particular beam~\cite{WILL20c}.
Evidently, the distribution of neutrons between the two beams varies with the phase shifter setting $X$ (or angle $\varphi$),
while the total intensity remains approximately constant.
To streamline the subsequent discussion, we shall adopt Eq. (\ref{app3})
as the definitive representation of the neutron interferometry results.

While the periodic variation of counts relative to $\varphi$ is termed an ``interference pattern''
and cited as proof of wave-like behavior~\cite{RAUC15}, this interpretation is incomplete.
As shown by the EBES discussed in this text, an interference pattern,
specifically one following Eq.~(\ref{app3}),
does not uniquely imply a wave character, proving that the traditional conclusion is misplaced.

\begin{center}
\framebox{
\parbox[t]{0.9\hsize}{%
Conceptually, the neutron interferometry experiment closely parallels Young's double-slit interference experiment.
The beam splitters BS0, BS1, and BS2 serve to prepare well-defined neutron beams that subsequently
``interfere' at BS3.
The phase shifter plate modulates the time-of-flight along the paths BS1$\to$BS3 and BS2$\to$BS3;
this is physically equivalent to varying the optical path length from point sources to a detector
by shifting the detector's position
}}%
\end{center}

In a typical neutron interferometry experiment,
approximately one neutron enters the interferometer every $10\,\mathrm{ms}$~\cite{WILL20c},
while each neutron resides in the silicon crystal region for only about $50\,\mu\mathrm{s}$.
Furthermore, experimental data indicate that the probability of two detectors firing
within the average interval between trigger events is negligible~\cite{WILL20c}.
Consequently, it is highly reasonable to assume that neutrons pass through the interferometer independently,
without direct communication with one another.

The core arguments are summarized below:
\begin{enumerate}
\item
Neutron experiments demonstrate that massive particles generate interference patterns even under
one-by-one detection conditions~\cite{RAUC74a,RAUC15}.
\item
Quantum theory fails to explain the event-by-event formation of that pattern.
Because quantum theory lacks a formal representation of an individual event,
it cannot account for the trajectory or behavior of a single neutron as it contributes to the emerging whole.
\end{enumerate}

\begin{center}
\framebox{
\parbox[t]{0.9\hsize}{%
Contemporary physics ``explains'' the emergence of interference patterns in one-by-one detection experiments
by invoking abstract concepts like ``particle-wave duality'' and the elusive ``wave function collapse''.
However, because an EBES can replicate the experimental data without recourse to such esoteric constructs,
it provides a more transparent, common-sense alternative.
}}%
\end{center}

Although it may be considered ``heretical'' from the standpoint of orthodox quantum theory,
it is entirely natural to ask how neutrons, passing through the interferometer one-by-one, never communicating,
and traversing only a single path at a time, can produce a periodic interference pattern.

In many respects, this inquiry mirrors the one posed by Young's two-beam experiment when performed photon-by-photon.
In that context, the interference pattern emerges as a function of the detector coordinate $y$ (see Fig.~\ref{YoungFig4});
in the neutron experiment, the phase shifter setting $X$ plays the exact same role as $y$.

To address this question, we find it productive to move beyond the language of wave mechanics.
Instead, as we did with the Young's experiment in Section~\ref{YOUNGGAME},
we adopt a broader perspective by framing the process as a game.
For reasons that will become clear, this will be referred to as the ``NEUTRON'' game.

\section{A game with people}\label{NEUTRONGAME}

\begin{figure}[!bp]
\centering
\includegraphics[width=0.90\hsize]{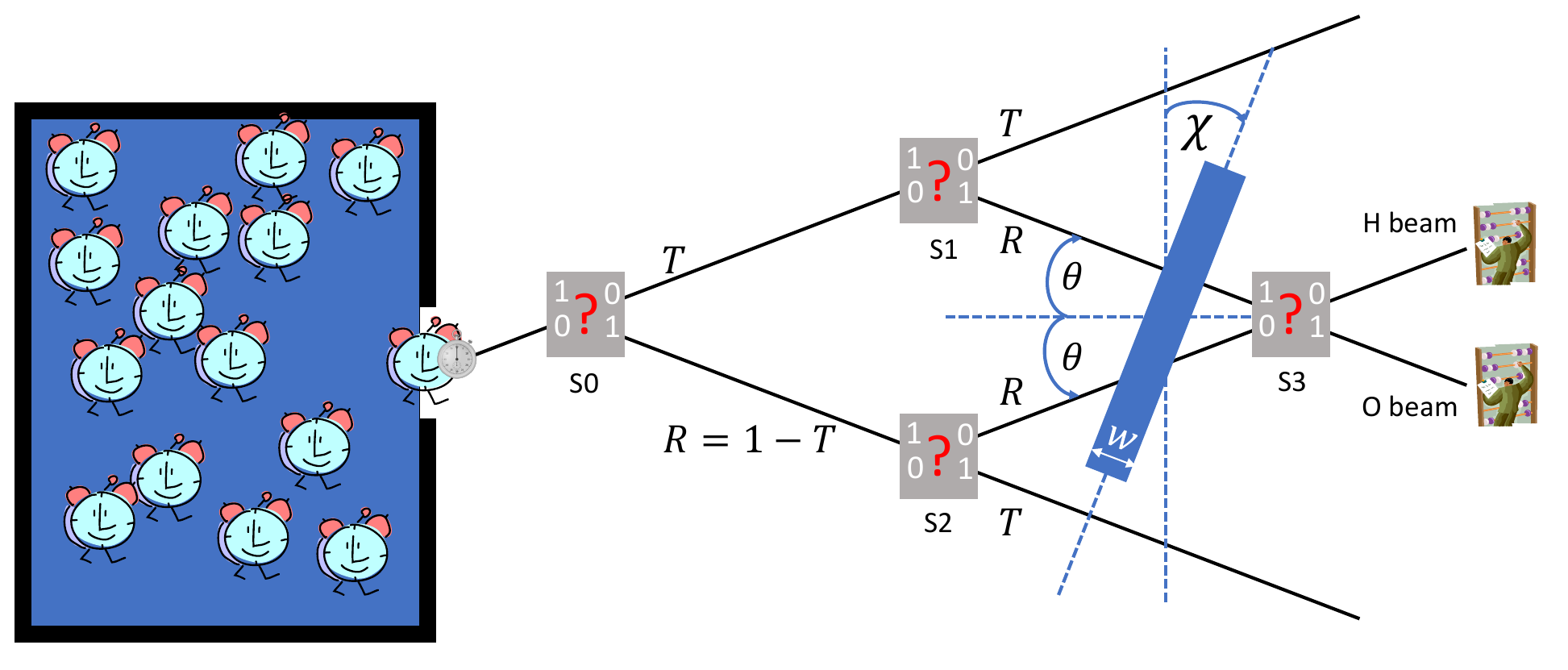}
\caption{
Picture of a NEUTRON game played with people, designed to demonstrate that one-by-one, people
can build up an interference pattern without ever communicating with each other.
The zero's and one's in the gray boxes labeled S0,...,S3 indicate the port number (0 or 1) through
which the walkers enter and leave.
}
\label{NeutronFig4}
\end{figure}

Imagine playing the NEUTRON game, sketched in Fig.~\ref{NeutronFig4}.
Not surprisingly, this game involves more elements than just sources and detectors
that in the case of the Young game. Consequently, its description is also longer.
With the help of little imagination and Figs.~\ref{NeutronFig1} and~\ref{NeutronFig4}, it is easy to
map the NEUTRON game onto the real neutron interferometry experiment.
The elements and rules of the NEUTRON game are as follows:

\begin{enumerate}
\item
On the far left there is a room filled with many walkers.
\item
There are four ``dark'' places (the gray objects with a question mark inside), labeled S0, S1, S2, and S3.
\item
The shortest distances between S0 and S1 and S0 and S2 are the same.
Also the shortest distances between S1 and S3 and S2 and S3 are the same.
\item
On the far right, two bookkeepers use their abacus to keep track of the numbers of walkers that have passed by.
When a walker has been registered by a bookkeeper, the walker leaves the game.
\item
Walkers leave the room one-by-one at arbitrary intervals but not before the previous walker has left the game.
\item
At the door opening the walker picks up a stopwatch, starts it and walks toward the dark place S0.
All walkers always move with exactly the same velocity, to be denoted by $v$.
\item
On their way between two splitters, walkers always take the shortest path, see Fig.~\ref{NeutronFig4}.
\item
Inside each dark place sits a person, referred to as a splitter.
Splitters have no idea of what is going on outside their own dark place.
Each dark place has two input ports and two output ports which are
labeled by an integer, taking the value zero or one, as indicated by the white ``0'''s and ``1'''s
in each of the dark places in Fig.~\ref{NeutronFig4}.
\item
When a walker enters a dark place, the splitter makes a note of the port through which
the walker entered, reads off the position of the stopwatch hand
and writes it down on a piece of paper.
The port number and the position of the hand are
the only piece of information that the splitter acquires from the walker.
\item
Based on the data acquired, the splitter instructs walkers to leave the dark place by continuing along their
path or by taking the alternative route.
All splitters use the same procedure, represented the question marks in Fig.~\ref{NeutronFig4},
\item
Walkers always arrive at  S0 through the same path.
The same holds for arrivals at S1 or S2.
In the case that walkers enter a dark place in one direction only,
the relative frequency by which walkers are instructed
to continue along their path is given by $T$ (transmission, a number between zero and one)
The number $R=1-T$ (refracted) is the relative frequency
for a walker to be instructed to take the alternative route.
The value of $T$ (and $R$) are fixed during the course of the game.
\item
Walkers arriving at S1 or S2 who are instructed to continue along their paths leave the game.
\item
Walkers arriving at S1 or S2 and are being refracted (change direction) will move towards splitter S3.
\item
On their way to S3, walkers have to pass a stretch of muddy ground (indicated by the blue, tilted rectangle).
\item
In the muddy stretch, the walking velocity (denoted by $v_{\mathrm{mud}}$) is reduced but still constant.
The time it takes the walker to leave the muddy ground depends on the tilt angle $\chi$.
For instance, for the tilt angle $\chi$ shown in Fig.~\ref{NeutronFig4},
it will take a walker less time to go from S1 to S3 than to go from S2 to S3
because in the former case, the path through the muddy ground is shorter than in the latter case.
\item
Walkers can arrive at S3 by two different paths, starting from either S1 or from S2.
They also (only) leave via one of the two different paths, traditionally called O- and H-beam~\cite{RAUC15}.
\item
The splitter in S3 uses the same procedure as S0, S1, and S2 to send the walker along either the O- or H-beam.
\item
Blocking one of the two paths to S3, the data acquired by splitter in S3
should automatically force this splitter to operate in exactly the same manner as the splitters in S0 (and S1 and S2).
\end{enumerate}

\begin{center}
\framebox{
\parbox[t]{0.9\hsize}{%
The winner of the NEUTRON game is the one who can replace all question marks
in Fig.~\ref{NeutronFig4} by one and the same procedure
that yields the experimentally observed distribution of counts, shown in Fig.~\ref{NeutronFig2}.
}}%
\end{center}

\section{Event-by-event simulation of the NEUTRON game}\label{NEUTRONRESULTS}

Clearly, playing the NEUTRON game with human participants is impractical.
Instead, it is far more efficient and economical to implement the rules in computer code,
as we will discuss next.
Attempting to solve the NEUTRON game using the traditional methods of theoretical physics
is an almost insurmountable task.
In contrast, as the code snippets in Section~\ref{COMPUTERCODE.N} illustrate,
simulating the game requires relatively little programming effort.
What follows is a brief analysis of representative simulation results,
intended to assist readers in performing their own experiments.

\begin{figure}[!htp]
\centering
\includegraphics[width=0.90\hsize]{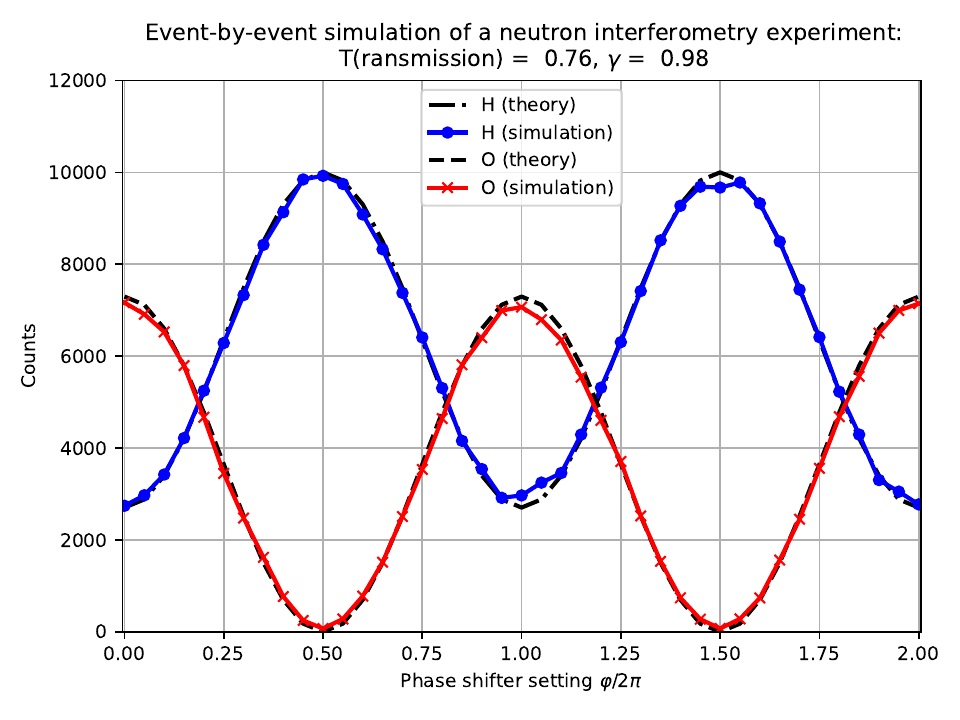}
\caption{
The theoretically expected result (dashed lines) and the final O and H beam counts (makers)
obtained by the EBES of the game with people.
The value of parameter $\gamma=0.98$
}
\label{neutronresults2}
\end{figure}

Assuming that Python3 and a few additional packages are installed and operational (see appendix~A for more details)
the procedure to run a simulation is simple.
\begin{enumerate}
\item
Open a command window on your computer.
\item
Windows users type
``{\tt py run\_neutron\_interference\_experiment.py}''
\hfill\break
(adding the {\tt -w} option activates noninteractive mode).
Linux or Mac users type {\tt python3} instead of {\tt py}.
The window with default parameter settings, shown in Fig.~\ref{NeutronFig5}, appears.
\item
Press RETURN or click on the START button.

\begin{figure}[!tp]
\centering
\includegraphics[width=0.90\hsize]{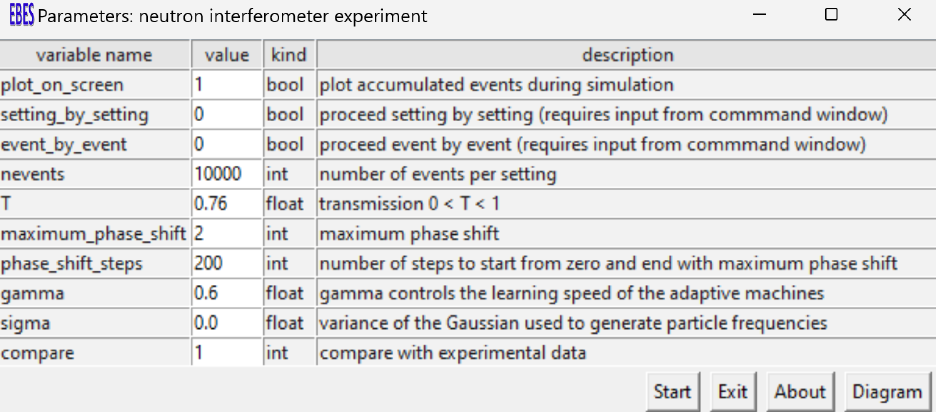}
\caption{
Graphical interface to control the EBES of the game with people
shown in Fig.~\ref{NeutronFig4} or, equivalently, of a real neutron interferometry experiment~\cite{WILL20c}.
}
\label{NeutronFig5}
\end{figure}

\begin{figure}[!bp]
\centering
\includegraphics[width=0.90\hsize]{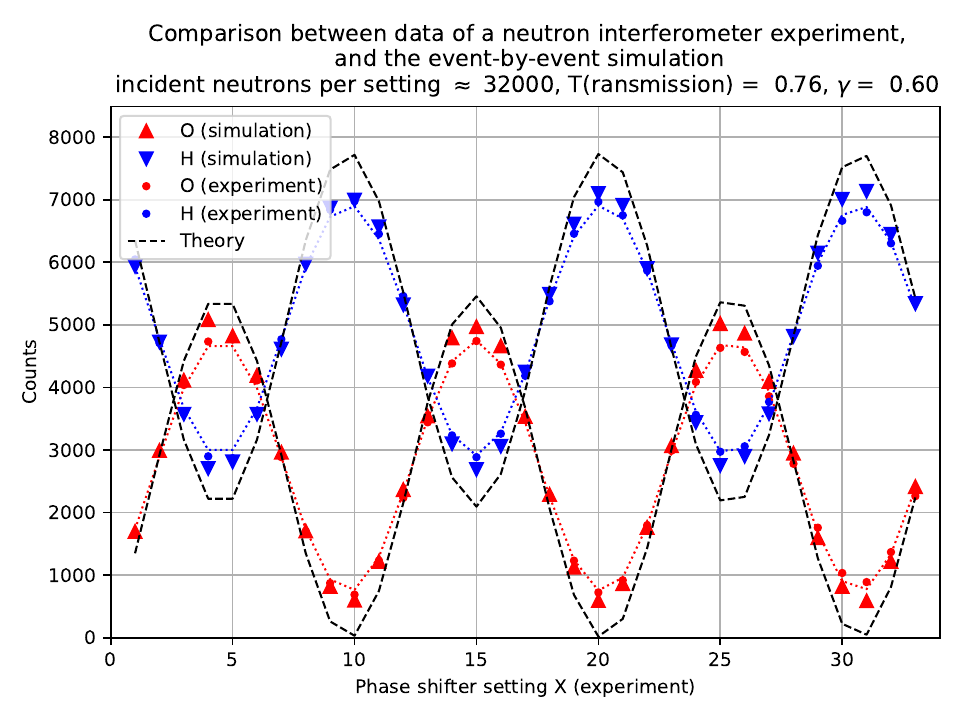}
\caption{
Neutron counts as a function of the phase shifter setting $X$
as obtained from the data files of a real experiment, reported on in Ref.~\onlinecite{WILL20c}.
Red circles: O-beam count;
blue circles: H-beam count;
green circles: H-beam count + O-beam count.
The dotted lines are guide to the eye.
Triangles: results obtained by the EBES
of the game with people with $\gamma=0.6$ and $T=0.78$~\cite{WILL20c}.
The dashed lines show the predictions of quantum theory.
}
\label{neutronresults3}
\end{figure}

\item
Another window appears, showing the data~\cite{WILL20c} presented in Fig.~\ref{NeutronFig1}.
The simulation starts with the phase shifter setting $X=1$ and after a few seconds
a red (O beam count) and blue (H beam count) triangle is drawn on top of the experimental data.
The simulation continue with $X=2,3,\ldots$, adding triangles to the picture.
The parameter $\gamma$, the number of incident neutrons and the value of the
transmission $T$ are preset such that the agreement between experiments and simulation is excellent,
see Fig.~\ref{neutronresults3}.
This agreement justifies the claim made earlier that the EBES
can reproduce the experimental neutron interferometry data~\cite{WILL20c}.
Also shown (by the black dashed line) is the prediction of quantum theory (see below).
\item
Click on the small window asking you to continue.
\item
By changing the value of the parameter {\tt compare} from 1 to 0, the code will no longer
make the comparison with experimental data but perform a comparison with
quantum theory (see below).
Also change the value of {\tt gamma} to 0.98 and click on start (or press ENTER).

Another window appears, showing the O beam counts generated by the observers during the simulation.
The window is updated each time 1000 (= nevents / 10) walkers have met an observer.
After all 10000 (= variable nevents) walkers per phase shifter setting have been processed, the simulation
stop and draws a solid line (= prediction of quantum theory, see below) on top of the histogram,
see Fig.~\ref{neutronresults1}.
There is no fitting involved in drawing this solid line.

\begin{figure}[!htp]
\centering
\includegraphics[width=0.90\hsize]{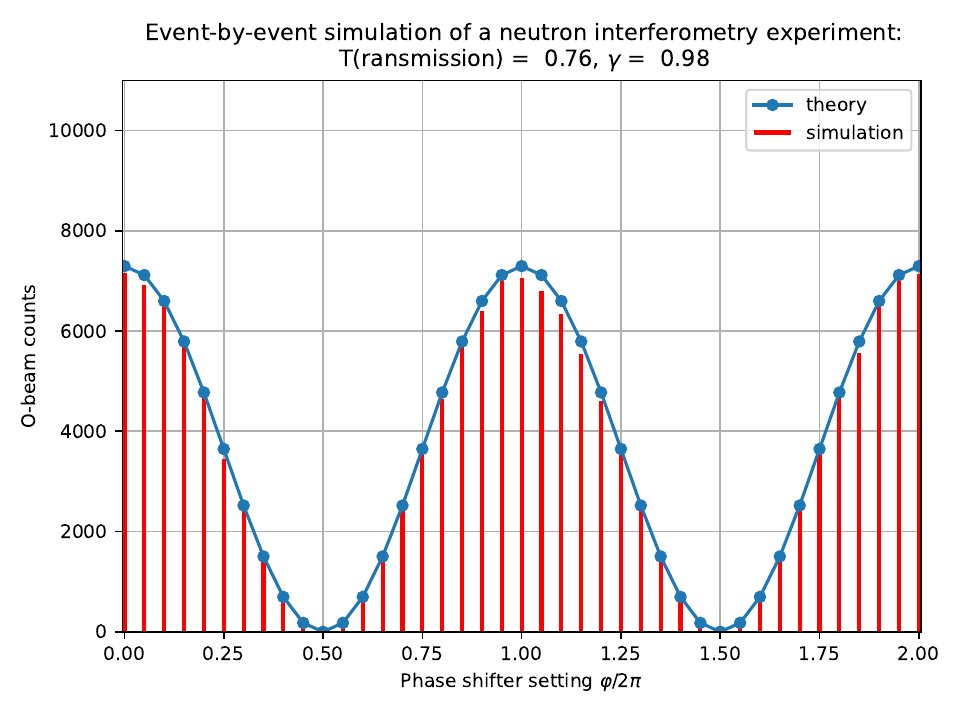}
\caption{
The theoretically expected result (solid line) and the accumulated O-beam counts (red bars)
obtained by the EBES of the NEUTRON game with people.
The value of parameter $\gamma=0.98$
}
\label{neutronresults1}
\end{figure}

\item
Click on the small window asking you to continue.
\item
Figure~\ref{neutronresults2} appears, showing the final O and H beam counts (markers)
and the results (dashed lines) predicted by quantum theory.
Clearly, for $\gamma=0.98$, the EBES reproduces the theoretically expected
results quite well.
\end{enumerate}

\begin{center}
\framebox{
\parbox[t]{0.9\hsize}{%
In summary:  the EBES of the NEUTRON game with people
yields the same interference pattern as experimentally observed and
can also produce results that are excellent agreement with
quantum theoretical results for the neutron interferometer.
}}%
\end{center}


\section{Python code for simulating the NEUTRON game}\label{COMPUTERCODE.N}

%
%
The level of abstraction required to translate the NEUTRON game and its operational rules (Section~\ref{NEUTRONGAME})
into computer code is remarkably low.
The core elements of the simulation are defined simply as objects and their respective interactions.
For convenience, Table~\ref{NeutronTab1} provides a one-to-one mapping between these computational objects,
the actors and actions defined in the human-based game, and the physical components of the neutron interferometry experiment.

\begin{table}[!htp]
\centering
\caption{%
Correspondence between the actors in the game shown in Fig.~\ref{NeutronFig4},
the objects used in the Python code,
and the concepts used to describe the neutron interferometry experiment.
}
\begin{tabular}{ccc}
\noalign{\medskip}
\hline\hline\noalign{\smallskip}
game with people       &  simulation code &  experiment           \\
\hline\noalign{\smallskip}
walker                 & messenger/particle        &  neutron              \\
door                   & virtual source            &  nuclear reactor      \\
velocity               & velocity                  &  velocity             \\
stopwatch              & internal clock            &  ???                  \\
stopwatch period       & 1 / frequency             &  ???                  \\
stopwatch time         & message                   &  time of flight       \\
distance travelled     & distance travelled        &  distance travelled   \\
splitter               & beam splitter             &  beam splitter        \\
muddy stretch          & time delay units          &  phase shifter        \\
person with abacus     & detector                  &  neutron detector     \\
distribution of counts & graphical output          &  graphical output     \\
\hline\noalign{\smallskip}
\end{tabular}
\label{NeutronTab1}
\end{table}

Having described the operation of the walker, the splitter in S3, and the effect of walking
through the stretch of mud, the remaining task is to arrange the splitters, muddy stretch
and bookkeepers according to the layout shown in Fig.~\ref{NeutronFig4}.
Thereby, for the consistency of the event-by-event approach,
it is essential that the splitters in S0, S1, S2 use {\bf exactly} the same rules
of operation as the splitter in S3.


As throughout this book, terms such as ``particle'' serve merely as convenient labels
for a data object or class within the code;
they do not refer to the everyday or classical concept of a physical particle.
However, within the specific context of the NEUTRON game, it remains perfectly appropriate to conceptually associate this
``particle'' with an actual neutron.

\renewcommand\code{{\bf experiment/neutron\_interference\_experiment.py}}
Listing~\ref{Lmain.neutrons} presents the code snippet (slightly modified from \code) that performs the simulation
in the NEUTRON game with people (see section~\ref{NEUTRONGAME}), or, equivalently
the neutron interferometry experiment (see section~\ref{NEUTRONINTRO}).
The structure of this snippet is very similar to the corresponding
snippet Listing~\ref{Lmain.neutrons} of the simulation code for Young's two-beam interference experiment.

Lines 2 to 9 show how the static elements in the game are represented by objects
created by reference to the classes of detectors, time delays units, and beam splitters,
each of which discuss in more detail later.

The statement in line number 12 initiates a loop over {\tt nevents} events.
The value of {\tt nevents} is taken from the GUI parameter window (see Fig.~\ref{NeutronFig5}).

The object named {\tt p} which is an instance of the class {\tt Particle} is created in line 14.
There is no need to discuss this class here because it is exactly the same as the one
used for playing the two-beam game, see section~\ref{COMPUTERCODE.N}.
In the NEUTRON game simulation, the actual values of particle velocity and frequency
have been absorbed in the definition of $\varphi$ 
and therefore, the particle object is created with default settings {\tt f=1} and {\tt v=1}.

For the logic of the code, it is important to recall that (i) at the end of the loop,
the object {\tt p} is destroyed (a Python feature),
(ii) with each iteration, a new object {\tt p} is being created, and (iii) upon creation all objects {\tt p}
the
have their clock time $t$, frequency $f$ and velocity $v$ set to zero, {\tt f=1}, and {\tt v=1}, respectively.

{\scriptsize
\lstset{style=mystyle,linewidth=0.95\textwidth,caption={Python code for simulating the game with people, described in section~\ref{NEUTRONGAME}.},label={Lmain.neutrons}}
\hbox{\hbox to 0.3cm{}
\begin{lstlisting}
# create an arrays of two objects of class ParticleCounter <=> simple counters
det = [ParticleCounter(), ParticleCounter()]  # create and reset two detectors, count after BS3
# create two delay units, one with delay = 0 and one with variable delay = dt * time_delay
delay = [TimeDelay(delay= -dt * time_delay / 2 ), TimeDelay(delay= + dt * time_delay/2)]
# Create an array with four beam splitters objects, set gamma, T and the random number generator
BS = [BeamSplitter(port=0, gamma=self.settings.options.gamma, T=self.settings.options.T, rng=self.rng),
      BeamSplitter(port=0, gamma=self.settings.options.gamma, T=self.settings.options.T, rng=self.rng),
      BeamSplitter(port=0, gamma=self.settings.options.gamma, T=self.settings.options.T, rng=self.rng),
      BeamSplitter(port=0, gamma=self.settings.options.gamma, T=self.settings.options.T, rng=self.rng)]

# loop over all events
for i in range(nevents):
# Create the object p, an instance of the class Particle, with defaults settings
    p = Particle(t=0)
# Construct the game/experimental setup, particle enters at port 0 of BS[0], and  leaves at port=0,1
    port = BS[0].process(port=0,particle=p)  # particle leaves BS[0] through port 0 or 1
# follow the particle moving on one of the two paths
    if port == 0: # go to BS1
        port1 = BS[1].process(port=port,particle=p)
    else:  # go to BS2
        port1 = BS[2].process(port=port,particle=p)
# if port1 == port, the particle leaves the interferometer --> next event
    if not port1 == port:
# delay the particle traveling along path "port"
        delay[port].process(particle=p)
# particle enters BS3 and through port 0 or 1
        port=BS[3].process(port=port,particle=p)
# count the particles leaving BS3 through port 0 (H beam) or 1 (O beam)
        det[port].process(p)
\end{lstlisting}
}
}
\medskip


Lines 16 to 29 are just the step-by-step implementation of the NEUTRON game with people.
The particle enters the beam splitter BS0 via port 0 (using port 1 only interchanges the roles
of H and O beam).
The object BS0 returns the number of the output port (0 or 1) and also changes the message
carried by the particle (more about this later).
The output port number returned by BS0 determines whether the particle proceed to
BS1 or BS2, exactly as in the NEUTRON game.

Line 23 checks if the particle leaving BS1 or BS2 has been refracted or not.
If not, according to the rules of the NEUTRON game, the particle leaves the game
and a new one will be created (unless the specified number of events has been exhausted).

The time delay incurred by passing the stretch of mud is implemented
as a call to the function {\tt process(particle=p)} of the delay object (to be discussed later).
The {\tt delay[port]} construct selects the proper delay unit:
if {\tt port = 0 (1)}, the particle travels from BS1 (BS2) to BS3.

In essence, line 27 is the same as line 16, 19 or 21, except that it refers to BS3 instead
of to BS0, BS1, or BS2, respectively.

Finally, line 29 shows how the detector objects are used to simply count the number of particles that leave BS3
via port 0 or 1.

\renewcommand\code{{\bf dlm/time\_delay.py}}

{\scriptsize
\lstset{style=mystyle,linewidth=0.95\textwidth,caption={Python class definition of a delay object.},label={Ldelay}}
\hbox{\hbox to 0.3cm{}
\begin{lstlisting}
class TimeDelay():
    def __init__(self, delay=0):  # create a delay unit and set the delay for later use
        self.delay = delay

    def set(self, delay=0):  # set the delay for later use
        self.delay = delay

    def process(self, particle):  # plane rotation of message vector
        t = twopi * self.delay * particle.frequency
        c, s = cos(t), sin(t)
        x, y = particle.message[0], particle.message[1]
        particle.message = [ c * x - s * y, s * x + c * y ]  # replace the message
\end{lstlisting}\label{LISTING\thelstlisting}
}
}
\medskip

Listing~\ref{Ldelay} shows the code snippet (taken from \code)
to create a delay object ({\tt \_\_init\_\_(...)}) and to ``process'' a particle.
The former simply takes the provided values of the variable {\tt delay} and stores
the value in the object being created.
The latter extracts the message (the hand of the stopwatch)
from the particle object and rotates the hand by the amount corresponding to the delay.
\ifFULLBOOK In other words, the function {\tt process} implements Eq.~(\ref{UPDATErule3.N}).\fi

\renewcommand\code{{\bf dlm/detectors.py}}
Listing~\ref{Ldet.neutrons} shows the code (taken from \code) to create the most simple detector object
and process a particle by incrementing a counter.
The ``class'' {\tt ParticleCounter} refers to another, more general class
{\tt Detectors} that contains code that is common to three kinds of detectors available in the simulation software.

{\scriptsize
\lstset{style=mystyle,linewidth=0.95\textwidth,caption={Python class definition of the most simple detector object.},label={Ldet.neutrons}}
\hbox{\hbox to 0.3cm{}
\begin{lstlisting}
class ParticleCounter(Detectors):  # detector object: simple counting detector
    def reset(self, _):
        self.count = 0

    def process(self, _):  # increments counter
        self.count += 1
\end{lstlisting}\label{LISTING\thelstlisting}
}
}
\medskip

\renewcommand\code{{\bf dlm/beam\_splitter.py}}
Listing~\ref{Lbs} shows the relevant parts of the code (taken from \code)
to create a beam splitter object and to process a particle entering the beam splitter through
the specified port {\tt port}.
Obviously, the code for the {\tt BeamSplitter} class is a little longer than for the other classes
encountered so far but with the experience of reading previous code snippets
the reader should not encounter major problems identifying
the steps in the operation of the splitter and the lines of computer code.

{\scriptsize
\lstset{style=mystyle,linewidth=0.95\textwidth,caption={Python class definition of a beam splitter object.},label={Lbs}}
\hbox{\hbox to 0.3cm{}
\begin{lstlisting}
class BeamSplitter():
    def __init__(self, port, gamma, T=0.5, rng=None):  # creates an instance of a beam splitter
        self.x = np.zeros(2)                  # internal x-vector, 2 elements
        self.y = np.zeros(4)                  # internal messages vector, 4 elements
        self.reset(port, gamma)
        t = sqrt(T)
        r = sqrt(1 - T)
    # beam splitter matrix according to optics / quantum theory
        self.BS = np.array([[t , 0, 0, -r],[0, t, r, 0],[0 , -r, t, 0],[r, 0, 0, t]])
        if rng is None:
            self.rng = np.random.default_rng()
        else:
            self.rng = rng

    def reset(self, port, gamma):  # reset internal vectors
        self.x[port] = 1
        self.x[1-port] = 1 - self.x[port]  # sum(x) = 1
        self.y[0] = 1
        self.y[1] = 0
        self.y[2] = 1
        self.y[3] = 0
        self.gamma = gamma

    def process(self, port, particle):  # process incoming particle
    # input stage:
        self.x[port] = self.gamma * self.x[port] + (1 - self.gamma)  # adaptive machine
        self.x[1 - port] = 1 - self.x[port]
        self.y[port + port] = particle.message[0]  # store message in registers
        self.y[port + port + 1] = particle.message[1]

    # transformation stage:
      # construct intermediate state vector
        z = [sqrt(self.x[0]),0,sqrt(self.x[1]),0]
        z[1], z[3] = z[0], z[2]
        z[0:4] = z[0:4] * self.y[0:4]  # sqrt{internal_x} . internal_message
      # change the intermediate state vector to match beam splitter functionality
        z = self.BS @ z  #  = matrix_BS . sqrt{internal_x} . internal_message
    # output stage:
        r = z[0]**2 + z[1]**2  # determine output port
        if self.rng.random() < r:  # use the random number generator to mimic "randomness"
            r = sqrt(r)
            particle.message[0], particle.message[1] = z[0] / r, z[1] / r  # compute new message
            return 0  # particle leaves via port 0
        else:
            r = sqrt(1 - r)
            particle.message[0], particle.message[1] = z[2] / r, z[3] / r  # compute new message
            return 1  # particle leaves via port 1
\end{lstlisting}\label{LISTING\thelstlisting}
}
}
\medskip

\section{Summary}\label{CONC.N}


The EBES of the NEUTRON game reproduces, in remarkable detail (see Fig.~\ref{NeutronFig2}),
the data from single-neutron interference experiments performed in the laboratory.
The same algorithm, extended to incorporate the neutron's magnetic moment,
has also been used to successfully model the results of a wide range of
single-neutron interference experiments~\cite{RAED12b,RAED14a}


In a pictorial description of the simulation, one may imagine traveling agents (neutrons)
equipped with an internal clock, and observing agents (splitters) that can read this clock,
perform a simple calculation, and instruct the traveling agents which of the two paths to follow.
After a neutron-agent encounters the fourth observing agent (S3 in Fig.~\ref{NEUTRONGAME}),
the counts of agents in each path, plotted as a function of the tilt angle of the muddy stretch,
produce the familiar sinusoidal interference patterns.
Agents representing neutrons (and those representing splitters) never communicate among themselves.
The rules governing how a traveling agent updates the state of an observing agent are what generate the interference pattern,
event by event.


The astute reader may wonder how an EBES can generate interference patterns that,
in the limit $\gamma\rightarrow1$, coincide with those predicted by quantum theory,
yet without invoking a single quantum-theoretical concept.
One way or another, the simulation must construct, event by event,
the equivalent of the probability distribution obtained in quantum theory.
This is precisely what happens. After many events have been processed and the simulated system has reached a stationary state,
the internal state of splitter S3 (see Fig.~\ref{NEUTRONGAME}),
represented by four floating-point numbers,
assumes the role played by the wave function in the quantum-theoretical description.

There is, however, a crucial conceptual and operational distinction.
In quantum theory, the wave function is associated with an ensemble of neutrons,
each one moving through space.
In the event-by-event approach, by contrast,
the internal state (or ``wave function'') belongs to the splitter itself and remains localized there.


In the event-by-event approach, the neutron is not represented by a wave function but behaves like a postman carrying a message.
The internal states of the splitters adapt to the number of postmen arriving at each of their two inputs
and to the messages they deliver,
and on that basis direct the postmen along the appropriate path.
This shift in perspective, where the wave function, or in event-by-event terms the internal state,
is not attached to the particle but to the material with which it interacts,
effectively removes the mysteries traditionally associated with particle-wave duality in quantum theory,
a point recognized for at least a century~\cite{DUAN23,LAND65,LAND15}.

\ifFULLBOOK
\begin{subappendices}
\section{Winning the NEUTRON game}\label{NEUTRONWIN}


%

To win the NEUTRON game, one must identify the operational rules for walkers and splitters
that replicate the experimental count distributions shown in Fig.~\ref{NeutronFig2} and described by Eq.~(\ref{app3}).
Logically, the splitter at S3 faces the most complex challenge,
as it must process walkers entering from two ports rather than one.
It is therefore expedient to first derive the rules for S3;
we can then demonstrate that applying these identical rules to S0, S1, and S2 is sufficient to win the game.

\subsection{Rules of operation for a walker}
As in the YOUNG game (see section~\ref{YOUNGGAME}),
walkers encode the position of the hand of their stopwatch as
$c=\cos2\pi f t$ and $s=\sin2\pi f t$ where $t$ is the time indicated by the stopwatch.
This encoding offers a convenient picture but is not unique nor essential, see
\hyperlink{WARNING}{{\color{red}WARNING}} and section~\ref{PICTURES}.
Recall that according to point 6 of the game description, when a walker leaves the room, $t=0$.

The frequency $f$ is defined as $f=1/P$ where $P$ is the time it takes the hand
of the stopwatch to complete one full turn.
In one full rotation of the hand, taking a time $P$, the walker marches a distance $\lambda=vP=v/f$
where $v$ denotes the constant walking velocity.
The choice of the value of $f$ is discussed later.

If a walker moves for a time $\tau$, the encoded values change according to
\begin{subequations}
\label{UPDATErule4.N}
\begin{eqnarray}
c &\leftarrow&c\;c_{\tau}- s\; s_{\tau}
\;,
\label{UPDATErule4.Na}
\\
s &\leftarrow&s\; c_{\tau} + c\; s_{\tau}
\label{UPDATErule4b.N}
\;,
\end{eqnarray}
\end{subequations}
$c_{\tau}=\cos2\pi f\tau$ and $s_{\tau}=\sin2\pi f\tau$.
Geometrically, the update rule Eq.~(\ref{UPDATErule4.N}) implements a plane rotation
of the vector $(c,s)$ about the angle $2\pi f\tau$ or, equivalently,
performs a rotation of the hand of the stopwatch by $2\pi f\tau$.

Because of point 3 in the description of the NEUTRON game, in the absence of a muddy stretch,
walkers arriving at S3 will have their stopwatches show exactly the same time.
Furthermore, the splitter in S3 can only learn about the presence of the muddy stretch and its
tilt angle $\chi$ by reading off the stopwatch times of the walkers
(see point 9 of the description of the NEUTRON game).
Clearly, if $f\tau$ is to contain information about the time delay caused by the presence
of the muddy stretch, $f\tau$ should depend on angle $\chi$.

\subsection{Rule governing the passage through the stretch of mud}\label{MUD}
Consider a walker taking the path from S1 to S3 (the same arguments hold
for a walker moving from S2 to S3, with  $\chi$ replaced by $-\chi$) through the stretch of mud.
Using elementary geometry, it is easy to show that the shortest distance
through the mud has length $l=w/\cos(\theta-\chi)$, see Fig.~\ref{NeutronFig4} for the meaning of the symbols.
Therefore, $\tau=w/v_{\mathrm{mud}}\cos(\theta-\chi)$
is the corresponding time for a walker to pass the stretch of mud of width $w$.

An arbitrary choice of $f$, $v_{\mathrm{mud}}$, and $w$ will, in general, not
yield the nice, sinusoidal interference patterns of the form given by Eq.~(\ref{app3}).
Fortunately, it is straightforward to show (see appendix~\ref{APP1})
that if $fw/v_{\mathrm{mud}}$ is a very large (dimensionless) number and if $\chi$ is restricted to small angles,
the time spent in the muddy stretch is, to a very good approximation, a linear function of the angle $\chi$.
It is under exactly these conditions that the quantum theoretical description
of the neutron interferometry experiment~\cite{RAUC15} and the EBES of the NEUTRON game
yield interference patterns of the simple form Eq.~(\ref{app3}).
This seems to be a generic feature: nice simple patterns only appear under certain
well-chosen conditions.

Under the mentioned conditions, $f\tau\approx a\chi$ where $a$ is a (very) large, unknown, dimensionless constant.
Therefore, writing $2\pi f\tau=\phi$ where the angle $\phi$ takes ``reasonable'' values,
e.g. cover the interval $[-2\pi,2\pi]$, simplifies matters a lot.

The rule governing the passage through the stretch of mud: increase the walking time by
the amount of time (depending on the tilt angle $\chi$) it took the walker to pass through the stretch.
Operationally, this means updating the position of the hand of the stopwatch according to:
\begin{subequations}
\label{UPDATErule3.N}
\begin{eqnarray}
c &\leftarrow&c\; c_{\mathrm{mud}} - s\; s_{\mathrm{mud}}
\;,
\label{UPDATErule3.Na}
\\
s &\leftarrow&s\; c_{\mathrm{mud}} + c\; s_{\mathrm{mud}}
\label{UPDATErule3b.N}
\;,
\end{eqnarray}
\end{subequations}
where $c_{\mathrm{mud}}=\cos\varphi/2$, $s_{\mathrm{mud}}=\sin\varphi/2$.
The angle $\varphi$ corresponds (in a complicated manner) to the phase shifter position $X$ in the real
experiment and to the tilt angle $\chi$ in the NEUTRON game.
Recall that the Eq.~(\ref{UPDATErule3.N}) holds for walkers moving from
S1 to S3 and with $\varphi\to-\varphi$ for walkers moving from S2 to S3.
The reason for introducing the factor $1/2$ is that the difference
of the delay times in paths S1$\to$S3 and S2$\to$S3 is $\varphi$,
the variable which appears in Eq.~(\ref{app3}).
In summary, Eq.~(\ref{UPDATErule3.N}) models what happens if the walker passes the stretch of mud.

\subsection{Operational rules for the splitters}\label{SPLITTER}
As might be expected, the operation of the splitter is more complicated than the operation
of the other objects encountered thus far.
Focusing on splitter S3, from the description of the NEUTRON game it follows that the number of walkers
arriving at S3 via S1 is not necessarily the same as the one for arriving via S2.
Indeed, the ratio of these numbers depends on the value of the transmission $T$ (and refraction $R=1-T$).

Although the value of $T$ is fixed before the game is started,
the value of $T$ can be any number between zero and one and
the rules of operation should not dependent on the particular the value of $T$.

In order to mimic the operation of an actual beam splitter (see later), it is necessary that
a splitter should be able to estimate the number of walkers that arrived at S3 via S1
and the numbers of walkers that arrived via S2.

As explained in section~\ref{WHYTHERULES}, simply counting
the numbers of walkers arriving via S1 and S2 is not going to work in all cases.
Instead, the splitter should estimate the relative frequencies of arrivals
through ports 0 and 1.
Of course, the splitter should also process the stopwatch data.

It is most convenient, also with the eye on later applications,
to break up the operation of a splitter into an input, transformation and output stage.

{\bf Input stage:\ }
The relative frequencies $x_0$ and $x_1$ of walkers who arrive at port 0 or 1, respectively,
can be estimated by applying the following, very simple update rules~\cite{RAED05d}:
\begin{subequations}
\label{UPDATErule0.N}
\begin{eqnarray}
x_0 &\leftarrow&\gamma x_0 + (1-\gamma) (1-p)
\label{UPDATErule0a.N}
\\
x_1 &\leftarrow&\gamma x_1 + (1-\gamma) p
\label{UPDATErule0b.N}
\;,
\end{eqnarray}
\end{subequations}
where the variable $p$ takes the value zero (one) if the walker enters the dark room through port zero (one)
and $0\le\gamma<1$ is a free parameter which controls
the pace with which the processing device adapts to new data.
Note that if $x_0+x_1=1$ before applying the update rule Eq.~(\ref{UPDATErule0.N}),
application of the update rule yields $x_0+x_1=1$.
Thus, if $x_0$ and $x_1$ are given an interpretation in terms of the probability to
arrive at port $p=0$ and $p=1$, respectively, applying Eq.~(\ref{UPDATErule0.N}) keeps this interpretation intact.

From the viewpoint of general applicability, it is important to recognize that the update rule Eq.~(\ref{UPDATErule0.N})
are identical to the one used to play the Young game, see Eq.~(\ref{UPDATErule0}).
The latter reduces to Eq.~(\ref{UPDATErule0.N}) by noting that
$(c=1,s=0)$ corresponds to $p=0$ and $(c=0,s=1)$ corresponds to $p=1$.

If $\gamma=1$, the state of the unit is frozen, making the model useless and if $\gamma=0$,
the internal state of the unit simply reflects the port through which the last walker entered the dark place.
The interesting regime is when $\gamma$ is close to but less than one.
In this case, it can be shown that the value of $x_0$ and $x_1$
approach the relative frequencies of walkers arriving at port 0 or 1, respectively~\cite{MICH11a}.
It is in this regime, referred to as {\bf stationary regime} in the following,
that the EBES can produce the results,
that is the distributions of events, of the neutron interferometry (and many other) experiments.

Next, the stopwatch data carried by the walker, represented in the form $(c,s)$,
has to be stored in the internal registers for later use.
As walkers can enter a dark place via two different ports, four internal
registers are necessary to store this data.
The most simply rules that are adequate for the task at hand read:
\begin{subequations}
\label{UPDATErule1.N}
\begin{eqnarray}
y_0 &\leftarrow& c\quad,\quad y_1 \leftarrow s\quad\hbox{if } p=0 \;,
\label{UPDATErule1.Na}
\\
y_2 &\leftarrow& c\quad,\quad y_3 \leftarrow s\quad\hbox{if } p=1 \;.
\label{UPDATErule1.Nb}
\end{eqnarray}
\end{subequations}
Again, Eq.~(\ref{UPDATErule1.N})) is just a particular case of
Eq.~(\ref{UPDATErule0.N}), namely the one in which $\gamma=0$.
Obviously, the effect of these rules is to copy the stopwatch data into the
internal variables $y_0$ and $y_1$ if the walker arrives at port $p=0$
and into $y_2$ and $y_3$ if the walker arrives at port $p=1$.
Thus, this part of the unit learns and forgets instantaneously.
Note that if a walker enters at port $p=0,1$, only $y_{2p}$ and $y_{2p+1}$
are being updated. The other $y$'s, namely $y_{2-2p}$ and $y_{3-2p}$, do not change.

The final processing step in the input stage is to build, out of the six numbers
$x_0$, $x_1$, $y_0$, $y_1$, $y_2$, and $y_3$, the vector (array) of length four
defined as.
\begin{eqnarray}
\mathbf{a}=\left(
\begin{array}{c}
y_0\sqrt{x_0}\\
y_1\sqrt{x_0}\\
y_2\sqrt{x_1}\\
y_3\sqrt{x_1}\\
\end{array}
\right)\;.
\label{Zvector}
\end{eqnarray}
There is no reason to introduce $\mathbf{a}$ other than that in end,
it serves the purpose of winning the NEUTRON game.

The inspiration for introducing the vector $\mathbf{a}$ comes from recognizing that in the stationary regime,
the first two components of $\mathbf{a}$ ($a_0$ and $a_1$),
and the last two components of $\mathbf{a}$ ($a_2$ and $a_3$),
can be interpreted as the wave amplitude on port 0 and port 1 of the beam splitter, respectively~\cite{RAED05d,MICH11a}.

Readers familiar with quantum theory will no doubt have recognized the similarity between
a two-component, complex-valued wave function and the vector $(a_0 + i a_1, a_2 + i a_3)$.
Indeed, in the stationary regime, they are mathematically identical but have a very different
meaning. In quantum theory, the wave function is associated with an ensemble
of particles whereas $\mathbf{a}$ encodes the data, gathered by the splitter in S3.

{\bf Transformation stage:\ }
Whereas the input stage is generic, simply processing the data that walkers bring by
entering through the ports, the second stage implements the actual splitter operation
through an analog-to-analog transformation.
The simplest transformation of a vector to another similar vector is
by a linear transformation, most conveniently defined by a matrix.
Assuming that the matrix used in the wave mechanical description of a beam splitter
(\url{https://en.wikipedia.org/wiki/Beam_splitter})
may be of use to winning the NEUTRON game too,
the transformation matrix is defined by
\begin{eqnarray}
\mathbf{T}=\left(
\begin{array}{cccc}
\sqrt{T} &  0& 0& -\sqrt{R} \\
0 &  \sqrt{T}& \sqrt{R}&  0 \\
0 & -\sqrt{R}& \sqrt{T}&  0\\
\sqrt{R}&  0& 0&  \sqrt{T}\\
\end{array}
\right)\;.
\label{Tmatrix}
\end{eqnarray}
where $R=1-T$ and $T$ and $R$ are real numbers.
Using $T+R=1$, it can be shown that
the vector $\mathbf{b}=\mathbf{T}\mathbf{a}$
has the same length as $\mathbf{a}$, namely $(a_0^2+a_1^2+a_2^2+a_3^2)^{1/2}$.
In other words, the matrix $\mathbf{T}$ represents a rotation in the four-dimensional
space of real-valued vectors such as $\mathbf{a}$.

With the help of some elementary algebra, it can be shown
that the value of each element of the vector $\mathbf{b}=\mathbf{T}\,\mathbf{a}$
depends on the difference between the stopwatch times of the walkers
entering through port 0 and port 1 only.
In other words, only time differences matter.

{\bf Output stage:\ }
The purpose of the output stage is to use the
output $\mathbf{b}=\mathbf{T}\,\mathbf{a}$ of the transformation stage
to send the walker through port 0 or 1 and while also changing the hand of the stopwatch.

In order to mimic a random process, the splitter generates a (pseudo-) random number $r$
and acts as follows:

\begin{enumerate}
\item
If $r < b_0^2+b_1^2$, send the walker though port 0.
\item
Otherwise, if $r \ge b_0^2+b_1^2$, send the walker though port 1.
\end{enumerate}
In essence, this rule is the same as the one used in the YOUNG game, see Eq.~(\ref{UPDATErule2}).

\subsection{Rules by which bookkeepers operate}\label{bookkeeper}
In the NEUTRON game, bookkeepers placed along the O and H beam simply count the number of walkers that pass by.
Per point 19 in the description of the game, walkers that pass a bookkeeper leave the game.
Recall that point 5 stipulates that a walker cannot leave the room unless the previous walker has left the game.


\subsection{Conditions on the model parameters}\label{APP1} 

As in the YOUNG game, to observe simple, nice in interference patterns, it is necessary to
choose the model parameters  $f$, $w$, and $v_{\mathrm{mud}}$ well.
Consider a walker taking the path from S1 to S3
(the same arguments hold for a walker moving from S2 to S3, with  $-\chi$ replaced by $\chi$).

Using elementary geometry, it is easy to show that $l=w/\cos(\theta-\chi)$ is the shortest distance
through the mud, see Fig.~\ref{NeutronFig4} for the meaning of the symbols.
If $v_{\mathrm{mud}}$ denotes the constant walking speed in the mud
(which is assumed to be smaller than the normal walking velocity $v$),
$\tau=(w/v_{\mathrm{mud}})/\cos(\theta-\chi)$
is the corresponding time for a walker to pass the stretch of mud of width $w$.

Next, assume that $0\le \chi \le \chi_{\mathrm{max}}$ where $\chi_{\mathrm{max}}$ is (very) small.
For all $\chi$ in this range, using
$\cos(\theta\pm\chi)=\cos\theta\cos\chi\mp\sin\theta\sin\chi$,
$\sin\chi\approx\chi$, $\cos\chi\approx 1$ and $1/(1 - x) \approx 1 + x$ for $|x|\ll1$,
yields $\tau\approx (w/v_{\mathrm{mud}}) [1/\cos\theta + \chi\sin\theta/\cos^2\theta]$.
The $\chi$-independent contribution  $(w/v_{\mathrm{mud}})/\cos\theta$ to $\tau$
is the same for both the S1-S3 and S2-S3 path and can therefore not induce a modulation
as a function of $\chi$.

The importance of the value of the frequency $f$ can be made clear now.
Under the assumption that $\chi_{\mathrm{max}}$ is small,
the variation of $f\tau$ is linear in the angle $\chi$.
But, even though $\chi$ is small, the variation of $f\tau$ with $\chi$
can be made very large if the frequency $f$ is such that $wf/v_{\mathrm{mud}}$ is large.
Thus, in order to obtain the simple interference signals Eq.~(\ref{app3}) as a function of $\chi$,
it is necessary that  $f\gg v_{\mathrm{mud}}/w$.

In typical neutron interferometry experiments, $w \sim 1 \mathrm{cm}$,
$v_{\mathrm{mud}}\sim 2000\,\mathrm{m}/\mathrm{s}$, and $f\sim 10^{13} \mathrm{Hz}$~\cite{RAUC15}.
Accordingly, $wf/v_{\mathrm{mud}}$ is of the order of $10^{8}$, a large number indeed.
The expression of $\tau$ then shows that a
minute change of $\chi$ induce a changes of $\tau$ that is amplified by a factor of about $10^{8}$.
In the simulation software, the values of $f$, $w$, and $v_{\mathrm{mud}}$ are ``encapsulated''
in the choice of the step in the phase shift $\phi$.
This step also depends on whether the software makes a comparison with the experiment (see Fig.~\ref{neutronresults3})
or quantum theory (see Fig.~\ref{neutronresults1}).
For details, see the Python code.

\section{Quantum theory}\label{APP2}
\newcommand\T{t}
\newcommand\R{r}

A detailed quantum theoretical treatment of the interferometer depicted in Fig.~\ref{NeutronFig1}
is given in Ref.~\cite{RAUC15}.
Assuming that the incident plane wave satisfies the Bragg condition for scattering by the
first crystal plate (BS0), the device acts as a two-path interferometer~\cite{HORN86}.
The simplest description of the crystal plates BS0, BS1, BS2, and BS3
assumes that these plates act as beam splitters with transmission $T$ and refraction $R=1-T$.
Then, the probabilities to observe a particle leaving the interferometer
in the O- and H-beam are given by~\cite{RAUC15}.
\begin{subequations}
\label{app2}
\begin{eqnarray}
p_\mathrm{O}&=&2R^2T\left( 1+ \cos\varphi \right)
,
\label{app2o}
\\ 
p_\mathrm{H}&=&R(T^2+R^2)\left(1 -\frac{2 RT}{T^2+R^2}\cos\varphi \right)
,
\label{app2h}
\end{eqnarray}
\end{subequations}
where $\varphi$ is the relative phase shift acquired by the plane waves when
following the two different paths through the interferometer.

Note that the model Eq.~(\ref{app2}) has only one free parameter (e.g. $T>0$)
and that $p_{\mathrm{O}}+p_{\mathrm{H}}=R$ is less than one
because all neutrons transmitted by BS1 or BS2 (see Fig.~\ref{NeutronFig1})
leave the interferometer without being counted.

According to Eq.~(\ref{app2o}), the O-beam counts should be close to zero if $\varphi=\pi,3\pi,\ldots$,
in disagreement with the experimental data shown in Fig.~\ref{NeutronFig1}.
Also shown by Fig.~\ref{NeutronFig1} is that Eq.~(\ref{app3}) fits the experimental data rather well.
However, the numerical factors that appear in Eq.~(\ref{app3}) cannot be made compatible
with the expressions predicted by quantum theory~\cite{WILL20c}.

\end{subappendices}
\fi


\chapter{Mach-Zehnder interferometer}\label{MACHZEHNDER}
In the preceding chapters, we formulated games that can in principle be played by people
while still yielding the results of genuine quantum physics experiments.
By making physical events rather than mathematical symbols our primary focus,
we find that the ``strangeness'' typically associated with interpreting these results evaporates.
For brevity, we will no longer frame each experiment as a game.
However, readers should find it relatively simple to adapt the quantum experiments
in the coming chapters into game-based formulations themselves.
Throughout this chapter, we adopt dimensionless quantities,
which simplifies the notation and highlights the essential structure of the results.

\section{Mach-Zenhder interferometer (MZI) experiments}

\begin{figure}[ht]
\centering
\includegraphics[width=0.70\hsize]{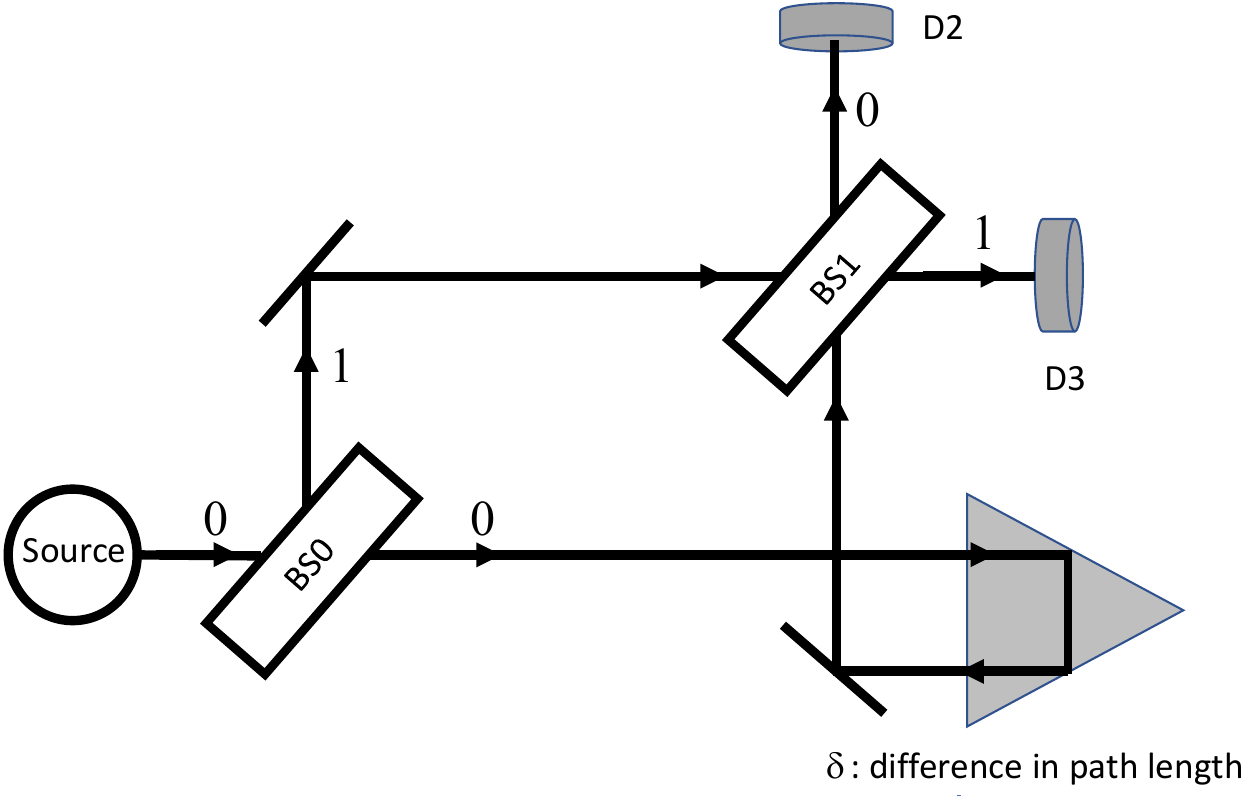} 
\caption{
Schematic of a MZI experiment~\cite{BORN64}.
}
\label{MZIfig1}
\end{figure}

Mach-Zehnder interferometry is often used to determine
the relative phase shift variations between two collimated beams of light
originating from a single source~\cite{BORN64}.
Unlike the Michelson interferometer~\cite{BORN64}, which reflects light back on itself,
the MZI uses a ``pass-through'' configuration, see Fig.~\ref{MZIfig1}.
In the standard experiment~\cite{BORN64}, a single beam of coherent light originating from the source
hits the beam splitter BS0. Half the light is reflected (path 1) and half is transmitted (path 0).
Light following path 1 is reflected by a mirror toward the second beam splitter (BS1).
Light following path 0 is passes through a movable prism
and is then reflected by a mirror toward BS1.
The purpose of the movable prism is to be able to change the length of path 0 for the light to reach BS1.
The partial light beams overlap within BS2 and interfere.
The intensity of the outgoing light beams depends on the difference in path lengths.
The sum if the intensities recorded by detectors D2 and D3 does not depend on the difference in path lengths.
In extreme cases, the interference is constructive or destructive.
In the extreme constructive (destructive) case, the intensity in path 0 will be zero maximal (minimal).

In general, the intensity exhibits a sinusoidal dependence as a function of the difference in path length $\delta$~\cite{BORN64}.
Note that qualitatively, the neutron interference experiment discussed in Chapter~\ref{NEUTRONINTRO} shows the same features.
For a Mach-Zehnder interferometer with input port 0 illuminated with intensity $I_0$,
and identical beam splitters with transmission $T$,
Maxwell's wave theory predicts that the output intensities at the two detectors are given by~\cite{BORN64}
\begin{align}
I_2=&I_0\left[T^2 + (1-T)^2 - 2T(1-T)\cos2\pi\delta\right]\;,
\nonumber \\
I_3=&2I_0T(1-T)\left(1+\cos2\pi\delta\right)
\;.
\label{MZI1a}
\end{align}
where $2\pi\delta$ is the phase difference of the partial waves in the two arms

\subsection{The experiment of Grangier et al.}

\begin{figure}[ht]
\centering
\includegraphics[width=0.50\hsize]{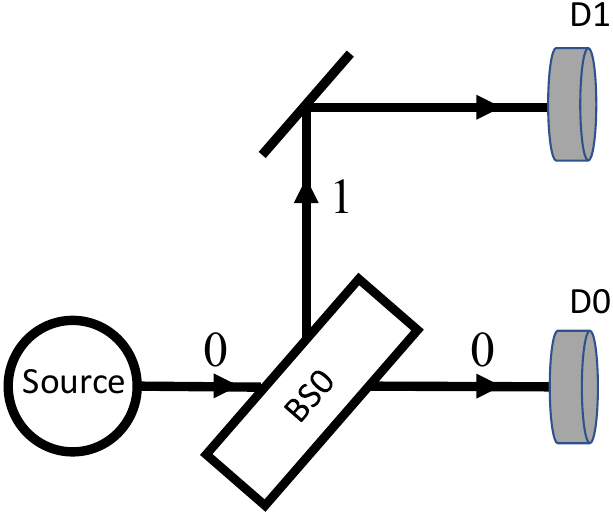} 
\caption{
Schematic of the first experiment by Grangier et al.~\cite{GRAN86}.
Using a calcium atomic cascade as a source,
Grangier et al. proved that light manifests itself as discrete photons:
when a single photon hits a beam splitter,
it is either reflected or transmitted, but never both, preventing coincident detections
(disregarding accidentals~\cite{GRAN86}).}
\label{MZIfig2}
\end{figure}

The MZI is one of the most elegant tools in quantum optics,
demonstrating the counterintuitive consequences of interpreting quantum theory.
Within quantum theory, it transforms the abstract debate of ``wave vs. particle'' into a measurable reality.
The experiments conducted by Philippe Grangier, G\'erard Roger, and Alain Aspect in 1986 were revolutionary~\cite{GRAN86}.
They showed that single photons seem to behave as indivisible particles
and interfering partial waves in the same experimental setup, but under different measurement conditions.
Phrased differently, while the MZI is fully described and understood within the wave theory of light,
as soon as the source emits single photons one at a time, within quantum theory, a mystery appears:
the photon appears to travel both paths simultaneously to interfere with itself.

Grangier et al. used a calcium atomic cascade to produce single photons~\cite{GRAN86}.
To prove that a photon is a indeed discrete ``chunk'',
they used the setup where the photons pass through BS0
and are recorded at detectors D0 or/and D1, see Fig.~\ref{MZIfig2}.
They observed that the detectors never fired at the same time
(disregarding accidentals~\cite{GRAN86}).
This proved the ``particle'' behavior of the photon; it does not split in half, it goes one way or the other
(like a neutron).

As the next step, they combined the single photon source with the full Mach-Zehnder setup, see Fig.~\ref{MZIfig1}.
Even though photons were sent one at a time, after many trials,
as a function of the difference in path length $\delta$, a sinusoidal interference pattern emerged.

Within the quantum theoretical framework, these observations seem to confirm that each individual
photon ``samples'' both paths, behaving as a wave.
The Grangier et al. experiment could be regarded as a manifestation of the
Principle of Complementarity~\cite{Howard2004}:
the nature of the photon is determined by the experiment one chooses to perform.

However, a key question remains:
if the photon is a particle that travels only one path
(as the first step of the experiment seems to prove),
how can it possibly interfere with itself?
We are left with a fundamental conundrum:
if the results confirm a particle restricted to a single path,
what is the underlying mechanism that builds the interference pattern event by event?

A solution to this conundrum is to shift focus from the particle's trajectory to the events themselves.
By treating the experiment as a series of possible detection events rather than the movement of a single entity,
the need for ``interference'' as a physical wave disappears.
In essence, we are repeating the reasoning that led to the construction of the EBES models
for the Young and neutron interference experiments, see Chapters~\ref{YOUNGINTRO} and~\ref{NEUTRONINTRO}.

\section{The EBES of the Grangier et al. experiments}

\begin{figure}[ht]
\centering
\includegraphics[width=0.70\hsize]{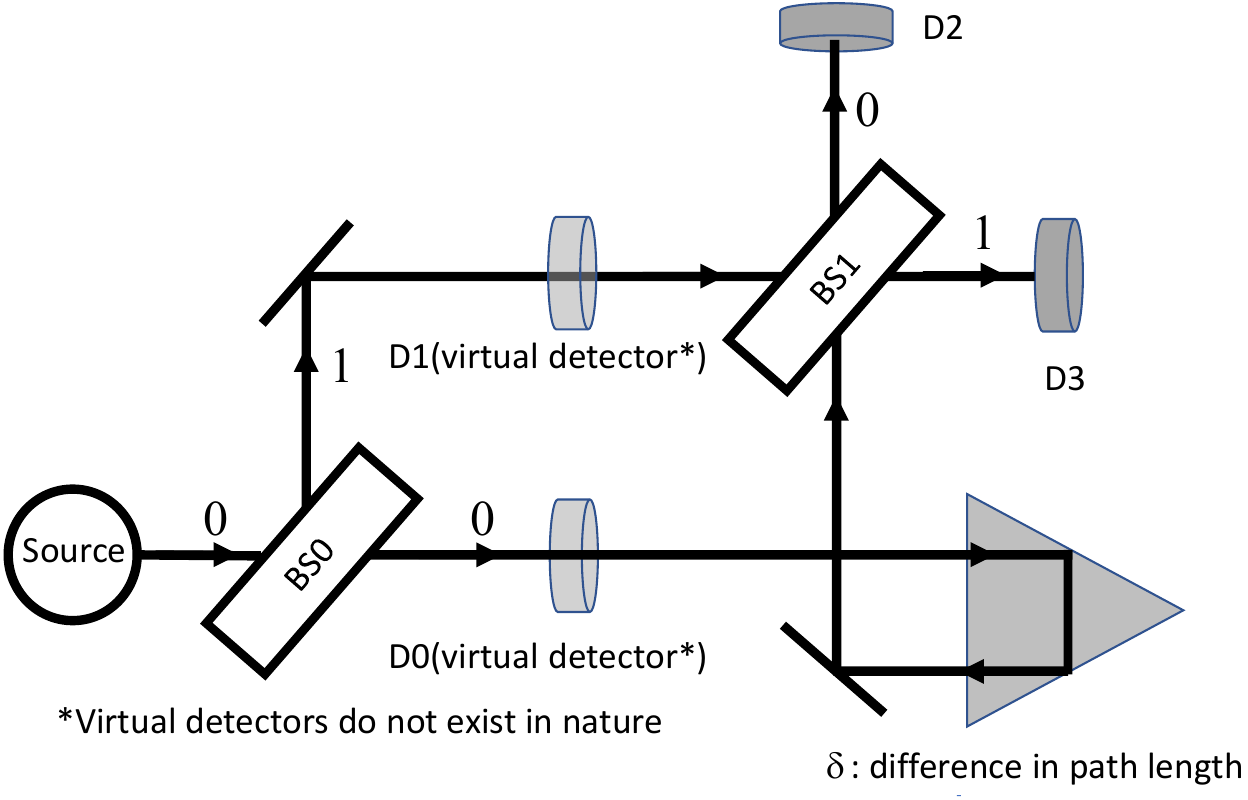} 
\caption{
Schematic of the EBES model for the second experiment by Grangier et al.~\cite{GRAN86}}
\label{MZIfig3}
\end{figure}

Since the EBES model for the MZI experiment is very similar
to the one used for the neutron interference experiment in Chapter~\ref{NEUTRONINTRO},
we will omit the specific details of the Python code here.
We trust that the interested reader will recognize that
rather than simply evaluating the standard quantum theoretical equations,
this code simulates the physical processes directly,

A conceptual advantage of the EBES model is the ability to track each individual particle as it wanders
through the MZI, providing particle trajectories that are absent in the standard wave-function description.
Moreover, the EBES framework allows us to simulate the two distinct stages of the Grangier et al. experiment simultaneously,
providing a unified view that the standard interpretation typically treats as mutually exclusive.
In this particular case, all that is required is to add ``virtual detectors'' that count the number of particles
on the two different paths connecting BS0 and BS1, see Fig.~\ref{MZIfig3}.

\begin{figure}[ht]
\centering
\includegraphics[width=0.70\hsize]{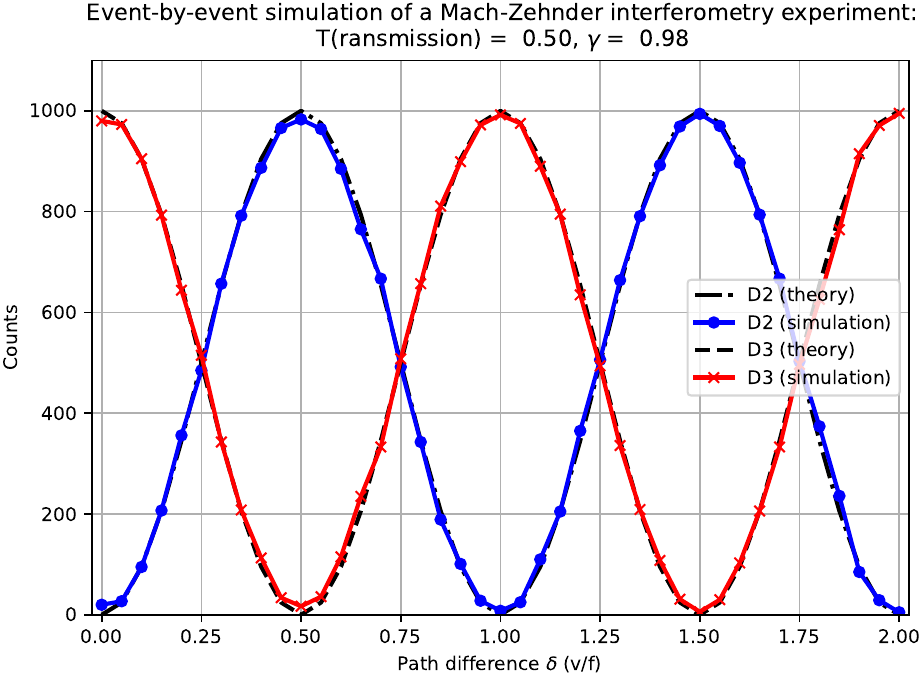} 
\caption{
The EBES and theoretical results for the second experiment by Grangier et al.~\cite{GRAN86}.
For 50/50 beam splitters ($T=1/2$), wave and quantum theory predicts that
the relative frequencies recorded by $D_2$ and $D_3$ are
given by $\cos^2(\pi \delta)$ and $\sin^2(\pi \delta)$,
respectively (see Eq.~(\ref{MZI1a}).
}
\label{MZIfig5}
\end{figure}

The procedure to run a simulation is as follows:
\begin{enumerate}
\item
Open a command window on your computer and change the directory to the directory that
contains the quantum optics demos.
\begin{figure}[!htp]
\centering
\includegraphics[width=0.95\hsize]{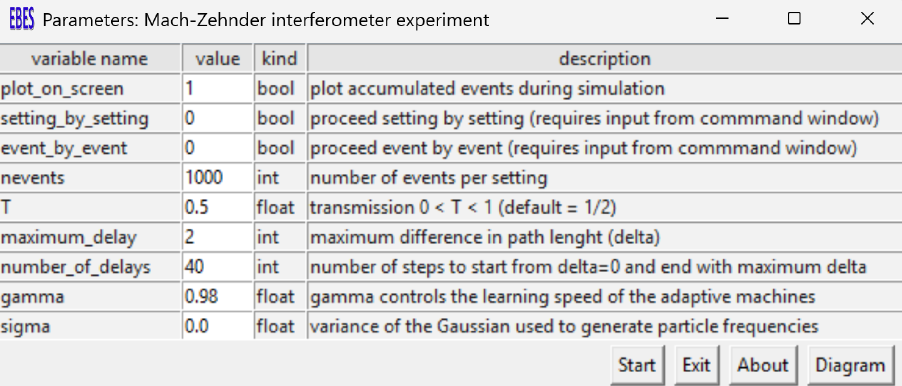} 
\caption{
Graphical interface to control the event-by-event simulation of the MZI experiment.
For simplicity, and because specific units are irrelevant to the underlying logic,
all quantities are treated as dimensionless.}
\label{MZIfig4}
\end{figure}
\item
Windows users, type 
``{\small \tt py run\_mach\_zehnder\_interferometer\_experiment.py}''
(adding the {\tt -w} option activates noninteractive mode).
Linux or Mac users type {\tt python3} instead of {\tt py}.
The window with default parameter settings, shown in Fig.~\ref{MZIfig4}, appears.
\item
Press RETURN or click on the START button.
\item
Another window with a plot appears, showing the counts on detector D2.
The counts of D0, D1, D3 appear in the command window.

The window is updated every {\tt nevents/10} times and if the setting $\delta$ changes.
With the default settings, $\delta$ varies from zero to two ({\tt = maximum\_delay})
in steps of 1/40 ({\tt = number\_of\_delays/maximum\_delay}).
After all events have been processed, the simulation
stops and draws a solid line on top of the histogram.
This solid line is the theoretically expected result
$\mathrm{nevents}\times\cos^2(\pi\delta)$.
\item
Click on the small window asking you to continue.
\item
Figure~\ref{MZIfig5} appears, showing the theoretically expected results (solid lines)
and the final counts (bullets).
The differences between the solid lines and the counts disappear if the total number
of events increases.
Simply changing the number of events from 1000 to 10000 and restarting the simulation
shows that this is indeed the case.
\end{enumerate}

In terms of events, in Eq.~(\ref{MZI1a}) $I_0$ stands for the total number of events
emitted by the source ({\tt nevents}) and $I_2$ ($I_3$) represent the counts
registered by detector $D_2$ ($D_3$).
The reader can easily check that the EBES yields the
predictions of quantum theory for the first and second experiment of Grangier et al.
by inspecting the output file {\tt mach\_zehnder\_interferometer\_experiment.out}.
Similarly, the reader can readily verify that the EBES reproduces the theoretical prediction of Eq.~(\ref{MZI1a})
by varying, for example, the transmission $T$ and the number of events {\tt nevents}.

The expression in Eq.~(\ref{MZI1a}) is derived under the assumption that the incident photons
exhibit no frequency fluctuations;
in other words, the source is assumed to emit perfectly monochromatic light.
By increasing the value of {\tt sigma} from zero to, for example, $0.2$, the EBES model
generates messengers with randomly varying frequencies
(see the accompanying Python code for implementation details).
This modification can have a pronounced impact on the detector counts as a function of $\delta$:
the ideal sinusoidal interference pattern is replaced by a sinusoid with a decaying envelope.

\section{One-beam-splitter thought experiment}\label{MZI.OB}
\begin{figure}[ht]
\centering
\includegraphics[width=0.70\hsize]{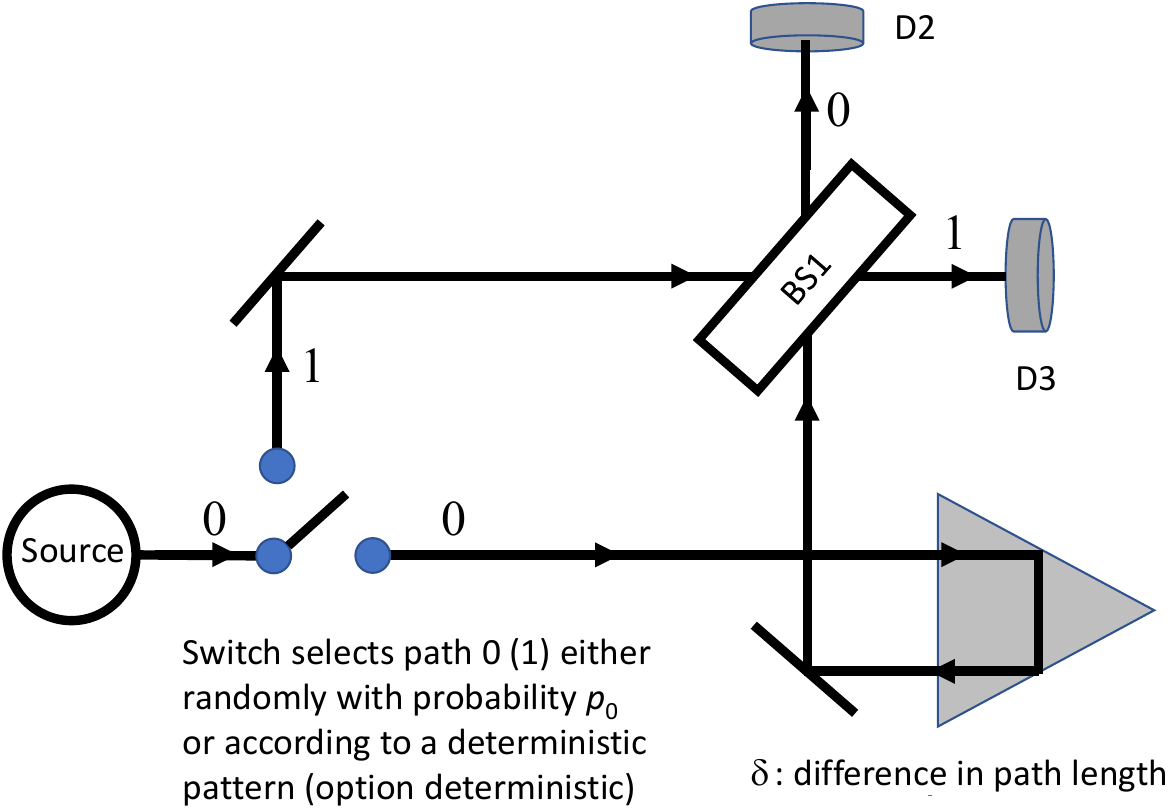} 
\caption{
Schematic of the EBES model for the combination of the first and second experiment by Grangier et al.~\cite{GRAN86}}
\label{MZIfig6}
\end{figure}

In the EBES model of the Mach-Zehnder interferometer, the first beam splitter acts as a randomizer,
directing each incoming particle into path 0 or path 1 with equal likelihood.
To examine how the randomizer influences the interference pattern, we consider the setup shown in Fig.~\ref{MZIfig6}.
The EBES implementation for this configuration permits adjusting the randomization frequency
for directing particles into path 0, or alternatively,
imposing a deterministic pattern (using the -d option or by setting {\tt deterministic} to a positive number)
that selects path 0 or path 1.
Although Eq.~(\ref{MZI1a}) can be adapted without difficulty to describe the first case,
the second case produces EBES outcomes that do not map in any simple way onto
the usual quantum-theoretical formulation.
For instance, set {\tt deterministic = 100} (generating
100 particles along path 0, followed by
100 particles along path 1, etc.)
and observe how the interference pattern
deviates from the quantum-theoretical prediction Eq.~(\ref{MZI1a}).
This deviation is not a real surprise: quantum theory does not account for the internal dynamics
of the EBES model.
We are not aware of any attempt to realize the one-beam-splitter thought experiment in the laboratory,
but it would certainly be interesting to see the data such an experiment would produce.

\section{Summary}
In essence, the EBES simulations of the MZI and the neutron interference experiments
rely on the same underlying principles.
In both models, the interference pattern emerges from the way the ``messengers''
interact with the internal state of the beam splitter.
The probability that a messenger follows a particular path is therefore not a ``choice''
made by the particle itself, but a consequence of the current internal state of the beam splitter.
Being outside the framework of quantum theory,
the EBES approach leaves no room for discussing wave-particle duality;
a particle-based description is entirely sufficient to account for the two experiments of Grangier et al.
For an extensive discussion of the various aspects involved, see Section~\ref{CONC.N}.

\begin{subappendices}
\section{Historical note}
The experiments of Grangier et al.~\cite{GRAN86} have been instrumental in the development of
the EBES approach~\cite{RAED05b,RAED05c,RAED05d}
because they offer the conceptually simplest demonstration of interference emerging
from individual particles being detected one at a time.
\end{subappendices}

\chapter{Wheeler's delayed-choice experiment}\label{DELAYEDCHOICE}
According to common lore, Wheeler's delayed-choice experiment shatters the idea that a quantum particle
``decides'' to behave as a wave or a particle at the start of its journey.
By delaying the choice of measurement until after the particle has passed the point of no return,
the experiment demonstrates that the notion of predetermined behavior is fundamentally flawed.
The results consistently confirm a mind-bending reality:
the measurement choice in the present appears to dictate the particle's prior history.
This ``quantum weirdness'' suggests that the past possesses no objective existence until it is recorded in the present.
Ultimately, it challenges our classical understanding of causality,
proving that the linear flow of time, where the past strictly precedes and causes the future,
does not fully apply to the quantum realm.

Of course, the preceding paragraph takes for granted that quantum theory provides
the sole framework for modeling these experiments.
As we demonstrate in this chapter by means of the EBES,
the apparent strangeness of delayed-choice experiments
is merely an interpretational artifact,
a pseudo-paradox arising from a specific, unnecessary, reading of the quantum formalism.

Indeed, by shifting the perspective such that the wavefunction or, in event-by-event language,
the internal state, is associated with the material rather
than being an intrinsic property of the particle,
the long-standing mysteries of wave-particle duality effectively vanish.
This conceptual shift demonstrates that the ``paradoxes'' of quantum theory
are not inherent to the events themselves,
but are instead consequences of an misplaced attribution of properties to the particle.
Notably, this realization is not a recent development;
it has been recognized for at least a century \cite{DUAN23, LAND65, LAND15}.

In this chapter, the EBES framework is used to implement two different realizations
of Wheeler's delayed choice experiment.
First, we integrate the two experimental protocols of Grangier et al. (see previous chapter)
by introducing a random number generator to determine the presence or absence of the second beam splitter.
Detector counts are subsequently recorded and categorized based on this stochastic choice.
This configuration realizes an idealized Wheeler's delayed-choice experiment,
provided that the decision to toggle the second beam splitter is made only after the particle
has traversed the first beam splitter and is en route along path 0 or path 1.
Second, we extend the EBES model to incorporate polarization,
enabling a rigorous analysis of a genuine laboratory implementation of Wheeler's proposal.

In both cases, the EBES framework causes the conceptual puzzles typically associated with Wheeler's delayed-choice
scenario to dissipate, demonstrating that the experimental results are fully consistent with a transparent,
intuitive, and strictly cause-and-effect description.

\section{A thought experiment}
\begin{figure}[ht]
\centering
\includegraphics[width=0.70\hsize]{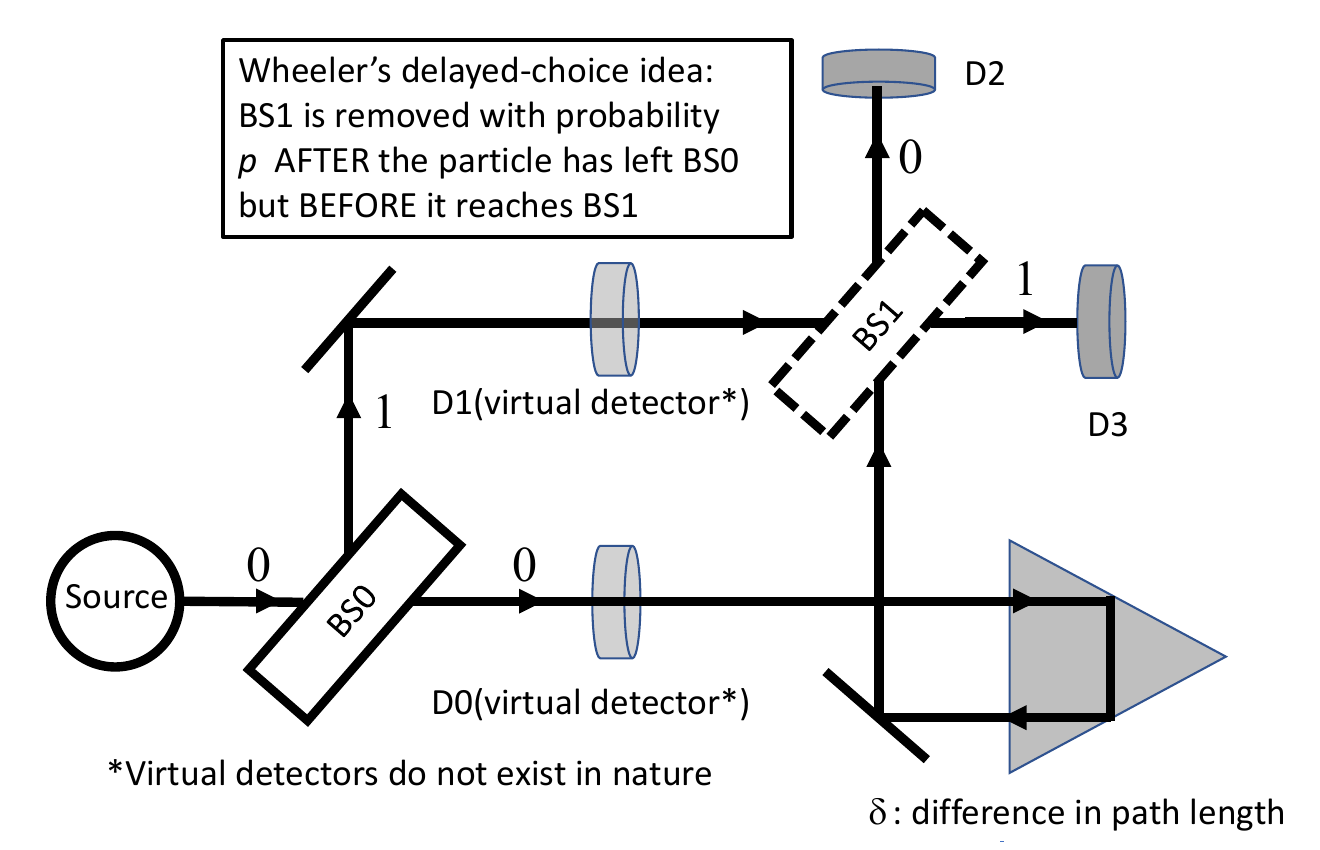} 
\caption{
Schematic of Wheeler's delayed-choice experiment with a modified Mach-Zehnder interferometer setup.
}
\label{DCfig1}
\end{figure}
Figure~\ref{DCfig1} illustrates the configuration of Wheeler's delayed-choice experiment,
realized here through the integration of the two experimental protocols by Grangier et al.
In the absence of the second beam splitter (BS2), the setup reduces to the first Grangier et al. experiment.
As established in the preceding chapter, the results of that configuration demonstrate
that individual photons traverse either path 0 or path 1 exclusively,
rather than following both paths simultaneously.
In this configuration, the counts registered by detectors D2 and D3 are expected to be independent
of the path-length difference $\delta$.
This lack of dependence is characteristic of the particle-like regime,
where the absence of a recombining element precludes the formation of an interference pattern,
regardless of the relative shift between the two trajectories.

When the second beam splitter is present, the setup functions as a Mach-Zehnder interferometer (MZI).
In this case, the counts registered by detectors D2 and D3 are expected to exhibit
a dependence on the path-length difference $\delta$,
behavior typically identified as the signature of the wave-like regime.

If the random number generator selects the configuration after the photon has traversed BS0,
but before it reaches the position of BS1, the conventional quantum-theoretical framework generates a pseudo-paradox.
Under this interpretation, the photon is seemingly forced to retroactively ``change its character'',
reconstituting itself as either a localized particle or an extended wave,
based on a decision made while it was already in flight.

Of course, within the EBES framework, the particle never ``changes its character''.
Its nature remains fundamentally constant throughout the experiment.
Nevertheless, the EBES accurately reproduces the statistical distributions predicted by quantum theory,
provided the detector counts are categorized according to the presence or absence of BS1.
This demonstrates that what is conventionally viewed as a temporal paradox is, in fact,
a straightforward result of local, event-based interactions.

The procedure to run the EBES is as follows:
\begin{enumerate}
\item
Open a command window on your computer and change the directory to the directory that
contains the quantum optics demos.
\begin{figure}[!htp]
\centering
\includegraphics[width=0.95\hsize]{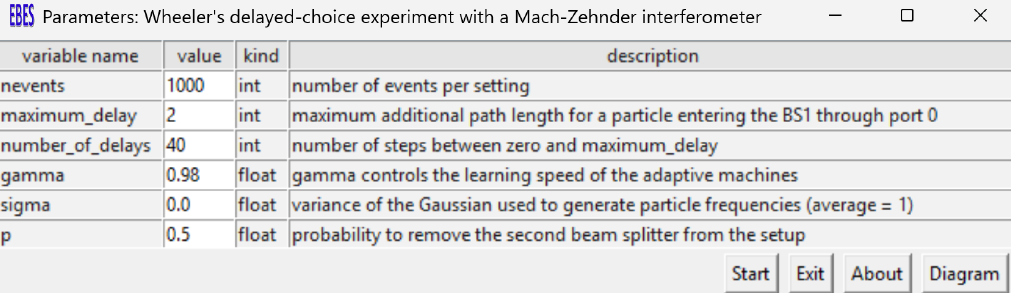} 
\caption{
Graphical interface to control the EBES of
Wheeler's delayed-choice experiment with the modified Mach-Zehnder interferometer setup
shown in Fig.~\ref{DCfig1}.}
\label{DCfig2}
\end{figure}
\item
Windows users, type 
``{\small \tt py run\_delayed\_choice\_experiment.py}''
(adding the {\tt -w} option activates noninteractive mode).
Linux or Mac users type {\tt python3} instead of {\tt py}.
The window with default parameter settings, shown in Fig.~\ref{DCfig2}, appears.
\item
Press RETURN or click on the START button.
\item
Figure~\ref{DCfig3} shows the theoretically expected results together with the final counts for both configurations.
The differences between the former and the latter disappear if the total number
of events increases.
Simply changing the number of events from 1000 to 10000 and restarting the simulation
shows that this is indeed the case.
\item
By adjusting the probability $p$ of removing the second beam splitter and restarting the simulation,
demonstrates that the EBES framework accounts for the full spectrum of experimental configurations.
This parameter sweep confirms that the model is robust across all intermediate cases,
from the strictly particle-like regime to the fully interferometric, wave-like setup.
\end{enumerate}

\begin{figure}[ht]
\centering
\includegraphics[width=0.90\hsize]{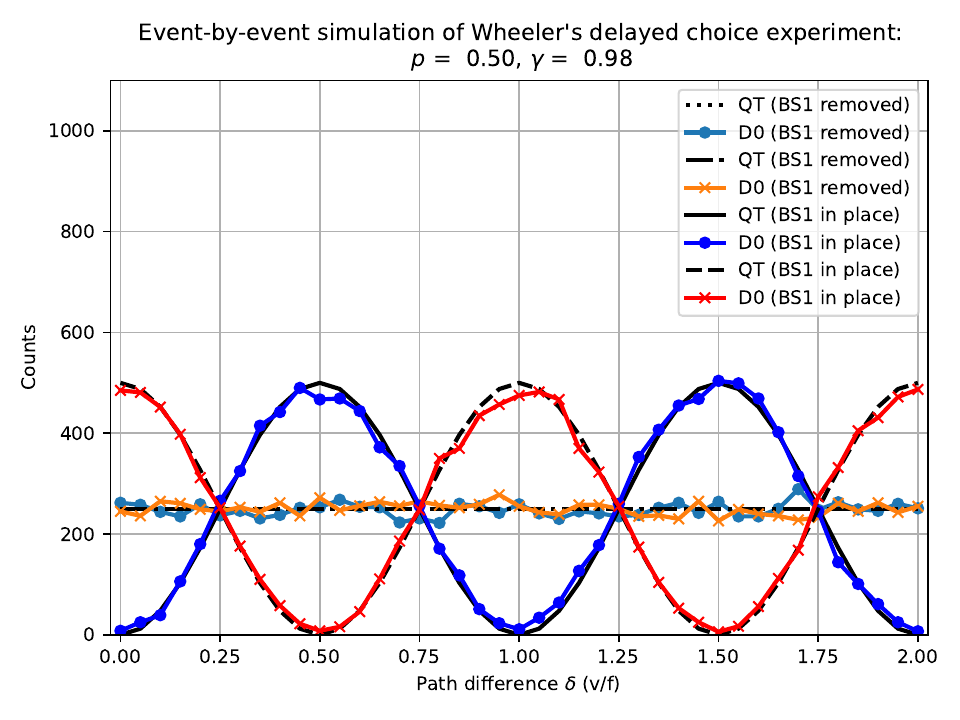} 
\caption{
The EBES results, generated by initiating the simulation via the graphical interface shown in Fig.~\ref{DCfig2}.
Even though the EBES simulates only 1000 events per setting of $\delta$,
the resulting data demonstrate excellent agreement with the quantum theoretical predictions
for both configurations.
This  confirms that the event-based approach successfully recovers the statistical expectations
of the standard formalism without implying the non-local or retrocausal assumptions
typically associated with the delayed-choice paradox.}
\label{DCfig3}
\end{figure}

\section{Delayed-choice experiment of Jacques et al.}

\begin{figure}[ht]
\centering
\includegraphics[width=0.90\hsize]{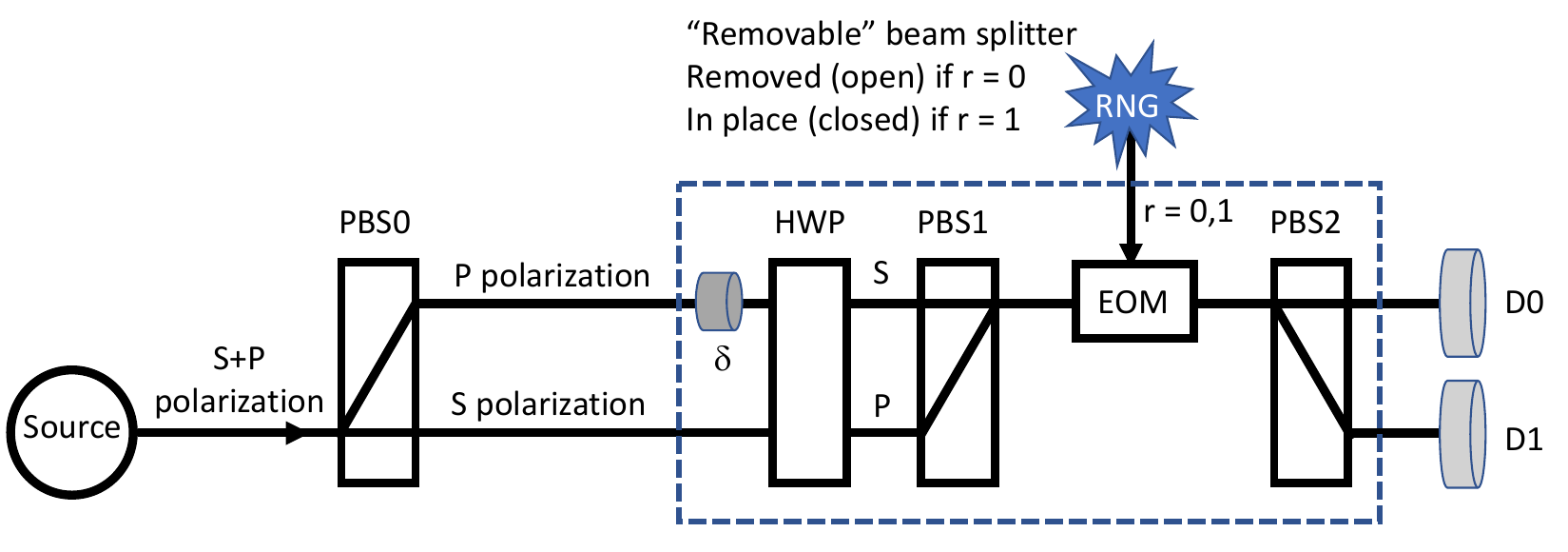} 
\caption{
Schematic of Wheeler's delayed-choice laboratory experiment
performed by Jacques et al.~\cite{JACQ07}.
PSB0, PSB1 and PSB2: polarizing beam splitters; HWP: half-wave plate; EOM:
electro-optic modulator; RNG: random number generator.
}
\label{DCfig4}
\end{figure}

While the configuration in Fig.~\ref{DCfig1} serves as a foundational thought experiment,
it has not, to our knowledge, been implemented in this precise form in a laboratory setting.
A genuine laboratory experiment that performs Wheeler's delayed-choice experiment
has been reported by Jacques et al.~\cite{JACQ07}, see Fig.~\ref{DCfig4}.
It is widely considered the first ``clean'' laboratory realization
of Wheeler's delayed-choice thought experiment.
It closed several loopholes that had plagued earlier attempts.

The experiment of Jacques et al., the ``delayed choice'' is not implemented by the mechanical movement of a beam splitter,
but rather through the high-speed control of the photon's polarization state.
The setup utilizes Wollaston prisms (polarizing beam splitters~\cite{BORN64})
and an Electro-Optic Modulator (EOM) to define the measurement basis:
\begin{itemize}
\item
Polarization Encoding: The two paths within the 48-meter interferometer are associated with orthogonal polarization states (e.g., S and P polarization~\cite{BORN64},
see Fig.~\ref{DCfig4}).
\item
The EOM, a voltage-controlled wave plate, acts as a ``switch''.
When the EOM is OFF (open configuration), the orthogonal polarizations remain unchanged,
allowing the detectors to provide definitive ``which-path'' information.
When the EOM is ON (closed configuration), it rotates the polarization of both paths by $45^\circ$.
This rotation effectively ``mixes'' the path information before it reaches the final polarizing beam splitter,
leading to the emergence of an interference pattern.
\end{itemize}

The experiment by Jacques et al. utilized a single-photon source, generated from a nitrogen-vacancy center in diamond,
to ensure that only one quantum of light occupied the apparatus at any given time,
thereby eliminating the possibility of multi-photon interference.

The decision to engage the second beam splitter was governed by a quantum random number generator.
Crucially, the spatial separation was sufficient to ensure that the command to ``switch''
reached the EOM only after the photon had traversed the first beam splitter.

\subsection{EBES extension to account for polarization}

To perform a faithful simulation of the experiment by Jacques et al.,
the EBES must be extended to incorporate polarization degrees of freedom.
From the EBES perspective, this extension is quite straightforward.
Just as a neutron carries a vector representing its magnetic moment,
the particle in this simulation carries an internal polarization vector.
Within this framework, optical elements such as wave plates and polarizing beam splitters (PBS) do not act on a global wave field;
instead, they locally alter the delay and trajectory of individual particles based on the specific internal states of the optical elements.
This allows the model to reproduce the complex correlations of the Jacques et al. setup
while maintaining a strictly local, particle-based ontology.
The technical details of this implementation can be found in the source file: {\tt polarizing\_beam\_splitter.py}.

\begin{figure}[ht]
\centering
\includegraphics[width=0.90\hsize]{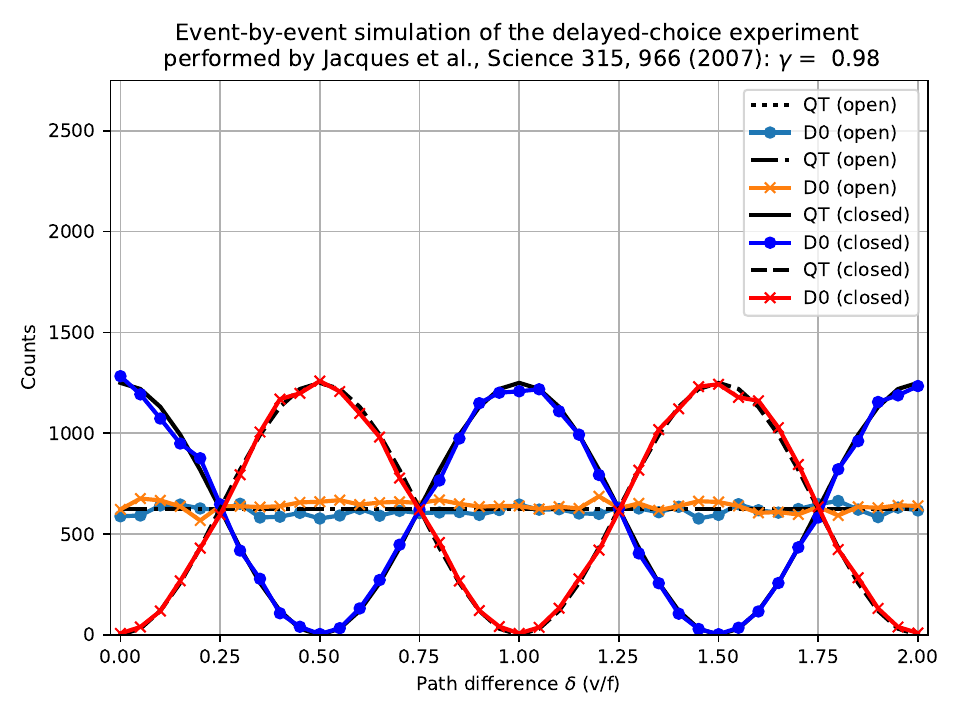} 
\caption{
The EBES results for the experiment of Jacques et al.~\cite{JACQ07},
generated by initiating the simulation via the graphical interface shown in Fig.~\ref{DCfig4}.
Even with a limited sample of 2600 events per $\delta$ setting (approximately the same as in Ref.~\cite{JACQ07}),
the EBES accurately reproduces the experimental observations found in Ref.~\cite{JACQ07}.
The data show a high degree of fidelity to quantum theoretical expectations in both configurations.
This confirms that the event-based approach successfully recovers the statistical expectations
of the standard formalism without implying the non-local or retrocausal assumptions
typically associated with the delayed-choice paradox.}
\label{DCfig6}
\end{figure}

The procedure to run the EBES is as follows:
\begin{enumerate}
\item
Open a command window on your computer and change the directory to the directory that
contains the quantum optics demos.
\begin{figure}[!htp]
\centering
\includegraphics[width=0.95\hsize]{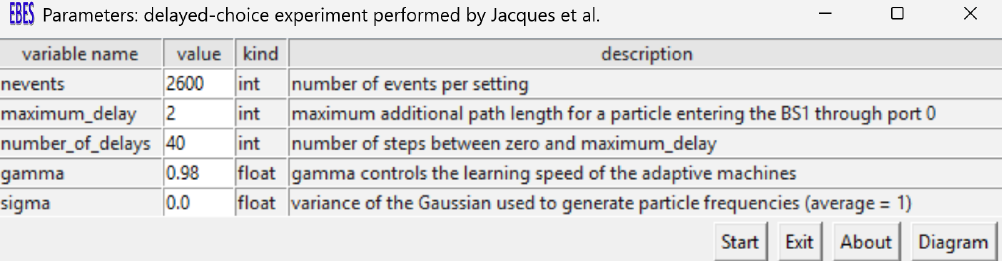} 
\caption{
Graphical interface to control the event-by-event simulation of
Wheeler's delayed-choice laboratory experiment
performed by Jacques et al.~\cite{JACQ07}, see Fig.~\ref{DCfig4}.}
\label{DCfig5}
\end{figure}
\item
Windows users, type \hfill\break
``{\small \tt py run\_delayed\_choice\_experiment\_polarization.py}''
(adding the {\tt -w} option activates noninteractive mode).
Linux or Mac users type {\tt python3} instead of {\tt py}.
The window with default parameter settings, shown in Fig.~\ref{DCfig5}, appears.
\item
Press RETURN or click on the START button.
\item
Figure~\ref{DCfig6} shows the theoretically expected results together with the final counts for both configurations.
The differences between the former and the latter disappear if the total number
of events increases.
Simply changing the number of events from 1000 to 10000 and restarting the simulation
shows that this is indeed the case.
\item
By adjusting the probability $p$ of removing the second beam splitter and restarting the simulation,
demonstrates that the EBES framework accounts for the full spectrum of experimental configurations.
This parameter sweep confirms that the model is robust across all intermediate cases,
from the strictly particle-like regime to the fully interferometric wave-like setup.
\end{enumerate}

In EBES the ``wave-like'' interference is not a change in the particle's character,
but a result of the local interaction between the polarization vector and the EOM's state at the moment of arrival.
By calculating the detection probabilities per setting of $\delta$
the EBES demonstrates that the experimental data is fully reproducible without
invoking any non-local or retrocausal ``decisions'' by the particle.


\section{Conclusion}

The standard quantum-theoretical interpretation of Wheeler's delayed-choice experiment
introduces a significant causal tension:
the photon is seemingly forced to ``adjust'' its prior history,
manifesting as either a single-path particle or a dual-path wave,
based on a measurement choice made while the particle is already in flight.
This retrocausal implication suggests that the nature of the past is dependent on the configuration of the present.

The EBES framework removes this tension entirely.
By attributing the ``wave-like'' behavior to the physical properties and internal states of the material
(the EOM and polarized beam splitters)
rather than to a shifting character of the particle itself,
the experiment is revealed as a sequence of local, deterministic events.
In this view, no ``decision'' by the photon is required;
the observed statistics emerge naturally from the interaction between the particle's internal degrees of freedom
and the experimental apparatus at the point of detection.
The EBES thus restores a transparent, intuitive causality to one of quantum theory's most enduring paradoxes.

\chapter{Diffraction and interference}\label{DIFFRACTION}
Contemporary physics teaching tells us that diffraction and interference are both consequences
of the wave nature (of light) and often occur together.
They represent two distinct ways in which waves interact with their environment.
While often categorized separately for simplicity,
diffraction as the bending of waves around obstacles
and interference as the superposition of multiple waves,
the distinction is largely a matter of convention.
As Richard Feynman noted~\cite{FEYN6} (The Feynman Lectures on Physics, Vol. I, 30-1),
both are fundamentally the result of the same principle:
the summation of amplitudes originating from various parts of a coherent source.
Conceptually:
\begin{itemize}
\item
    Diffraction is what allows a wave to ``see'' an obstacle.
\item
    Interference is what happens when two waves ``talk'' to each other through a material medium.
\end{itemize}

Young's double-slit experiment, see Fig.~\ref{YoungFig1},
is the most famous example showing the interplay of these two phenomena.
\begin{itemize}
\item
    Diffraction occurs first: As light hits the two narrow slits,
    it diffracts (spreads out) from each slit.
    Without diffraction, the light would just form two bright spots on the wall.
\item
    Interference occurs second:
    Because the light from both slits is now spreading into the same space, the partial waves overlap.
    Where a ``crest'' meets a ``crest'', a detector would show a bright fringe (constructive interference);
    where a ``crest'' meets a ``trough'', a detector would a dark fringe (destructive interference).
\end{itemize}
In general, the ``envelope'' or the overall shape of the light is determined by diffraction,
while the fine details (the individual stripes) are determined by interference.

In our EBES treatment of the Young game, we deliberately ignore diffraction for simplicity.
Following the standard wave mechanical description of Young's two-beam interference experiment~\cite{BORN64},
we replaced the physical double-slit with two virtual point sources.
The present chapter introduces a more comprehensive EBES model that incorporates diffraction,
moving beyond the idealized two-beam interference approximation.

\section{Fraunhofer diffraction by a grating}

\begin{figure}[!htp]
\centering
\includegraphics[width=0.90\hsize]{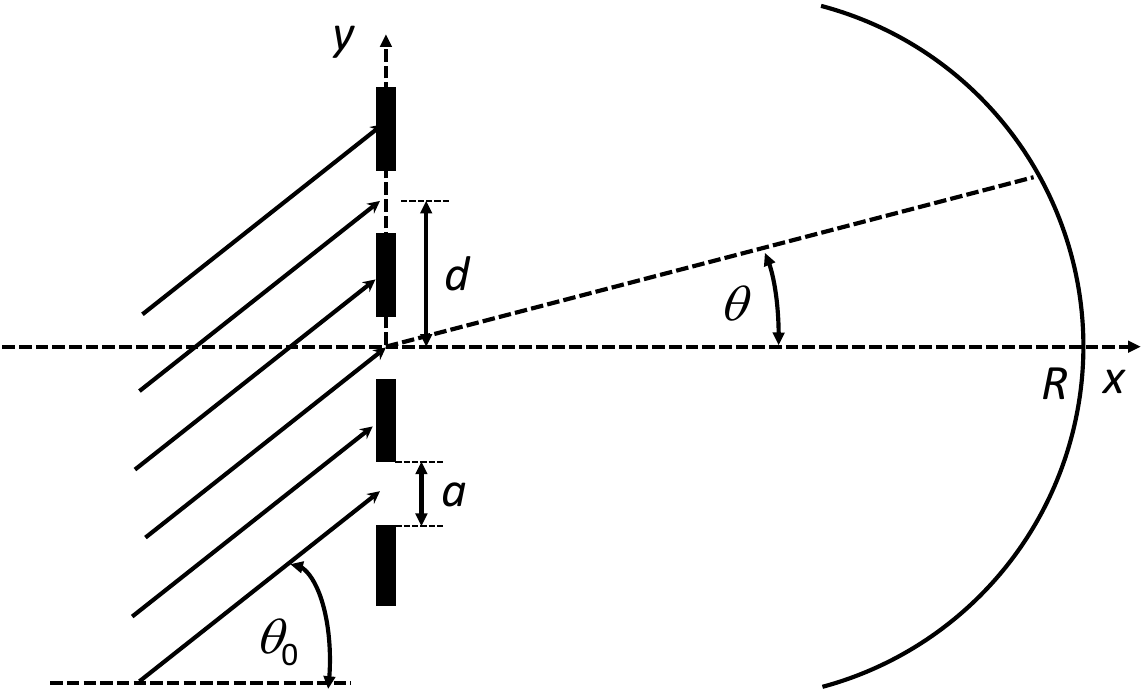} 
\caption{
Geometrical representation of an experiment showing Fraunhofer diffraction and interference by a grating.
Light impinges on the grating with the angle of incidence $\theta_0$.
Slits of aperture $a$ are separated by a distance $d$
The angle $\theta$ determines the point of observation at the screen
which is located a distance $R$ away from the plane of the sources.
}
\label{GratingFig1}
\end{figure}

Fraunhofer diffraction (or far-field diffraction) occurs when the light source and the observation screen
are effectively at an infinite distance from the aperture.
In this regime and in the wave picture, the wavefronts arriving at the slit are planar,
and the secondary wavelets reaching any point on the screen are also considered parallel.
Fraunhofer diffraction from a grating is the result of combining single-slit diffraction with multi-slit interference.
A diffraction grating consists of a number ($\ge2$) of parallel, equally spaced slits,
see Fig.~\ref{GratingFig1}.

\subsection{Wave theory}

In the Fraunhofer diffraction regime, the intensity profile of the detector signals is given by
the product of the single-slit diffraction envelope and the multi-slit interference pattern~\cite{BORN64}, namely
\begin{align}
I(\theta)&=N_s^2\left(\frac{\sin aq}{aq}\right)^2 \left(\frac{\sin dqN_s}{dqN_s}\right)^2
\;,
\label{DIFF1}
\end{align}
where $q=\pi (\sin\theta-\sin\theta_0)/\lambda$,
$N_s$ is the number of slits, and $\theta_0$ is the angle of incidence of the impinging plane wave
with wavelength $\lambda$.

{\color{red}
It is worth emphasizing that, within the framework of this book,
there is no point of constructing a Monte Carlo process that generates events based on Eq.~(\ref{DIFF1}).
While such a simulation creates the illusion that the diffraction pattern emerges event-by-event,
it relies on an (approximate) solution of the wave equation as input.
In contrast, as shown below, the EBES generates events with the correct distribution without requiring
a prior solution to the wave equation.
For completeness, the EBES code {\tt grating\_experiment.py}
includes a Metropolis Monte Carlo (MMC) sampling routine based on Eq.~(\ref{DIFF1}).
The numerical results of this simulation are printed on screen and to {\tt grating\_experiment.out},
while the graphical results are displayed on-screen and saved as {\tt grating\_experiment.3.pdf}.
Unlike rejection-based Monte Carlo methods~\cite{PRES03}, where random events are discarded,
MMC sampling ensures no events are lost.
This makes it easier to align the results with EBES data.
}

\subsection{EBES of diffraction by a grating}

Logically, if we want the two beams of particles in the EBES of the Young experiment to exhibit
interference pattern, there is only one way to accomplish this :
the detectors must be capable to process the time-of-flight information
carried along with the particles.
In this framework, the 'wave-like' behavior is an emergent property of the detector collective.
This interpretation aligns with Duane's perspective that wave-particle duality is
not an intrinsic attribute of the particle itself,
but rather a phenomenon arising from the interaction between the particle and the material structure,
which generates the observed interference~\cite{DUAN23,LAND65,LAND15}.
In contrast, to construct an EBES for grating diffraction, two primary methodologies exist:
\begin{enumerate}
\item
The virtual-source approach: treat each slit as an independent virtual source,
analogous to Young's double-slit experiment.
While computationally efficient, this model is only valid for scenarios where the detection efficiency
is not absolute, specifically, cases where not every incident particle contributes to the final count.
\item
The grating-material-interaction approach:
explicitly model the physical interaction between the particles and the grating structure
to account for the momentum transfer between the incident particles and the grating material.
Also this approach aligns with Duane's perspective~\cite{DUAN23,LAND65,LAND15}.
In this configuration, the detectors function as simple counters;
consequently, the time-of-flight information carried by the particles becomes redundant data,
as the interference pattern is already established in the spatial distribution.
\end{enumerate}

In this chapter, we perform EBES based on the grating-material-interaction approach.
The reader is encouraged to analyze lines 54--74 of {\tt grating\_experiment.py}
to gain a deeper understanding of how the EBES is implemented in the Python code.

The procedure to run the EBES is as follows:
\begin{enumerate}
\item
Open a command window on your computer and change the directory to the directory that
contains the quantum optics demos.
\begin{figure}[!htp]
\centering
\includegraphics[width=0.95\hsize]{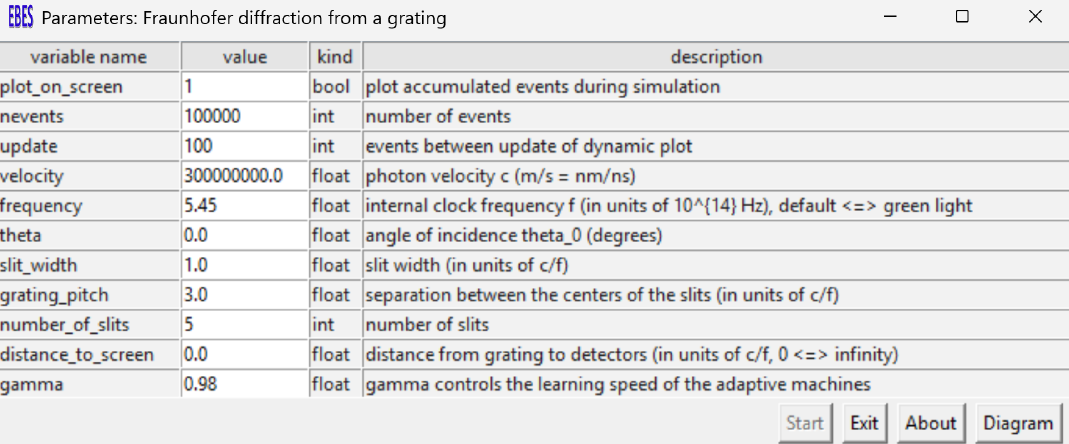} 
\caption{
Graphical interface to control the event-by-event simulation of
Fraunhofer diffraction by a grating, see Fig.~\ref{GratingFig1}.}
\label{GratingFig2}
\end{figure}
\item
Windows users, type \hfill\break
``{\small \tt py run\_grating\_experiment.py}''
(adding the {\tt -w} option activates noninteractive mode).
Linux or Mac users type {\tt python3} instead of {\tt py}.
The window with default parameter settings, shown in Fig.~\ref{GratingFig2}, appears.
\item
Press RETURN or click on the START button.
\item
Figure~\ref{GratingFig3} shows the theoretically expected results together with the final counts for both configurations.
Varying the total number of events reveals that the discrepancies between
the two models are subject to statistical fluctuations; as the sample size increases,
these differences evolve according to the underlying stochastic nature of the simulation.
\item
Restarting the EBES with different values of the incidence angle {\tt theta},
or grating parameters {\tt slit\_width}, {\tt grating\_pitch}, or {\tt number\_of\_slits}
demonstrates that the EBES framework accommodates the full range of physical configurations.
\item
The Fraunhofer diffraction formula, given in Eq.~(\ref{DIFF1}),
is derived under the far-field assumption.
When the {\tt distance\_to\_screen} parameter is set to a finite value (e.g., 50),
the EBES diffraction pattern diverges from the analytical predictions of Eq.~(\ref{DIFF1}).
Whether this emergent pattern remains consistent with the full solutions to Maxwell's equations,
specifically in the Fresnel regime,
remains an open question.
If a discrepancy exists, it only implies that the EBES methodology for modeling particle-material interactions
requires further refinement to accurately capture near-field effects.
\end{enumerate}

\begin{figure}[ht]
\centering
\includegraphics[width=0.90\hsize]{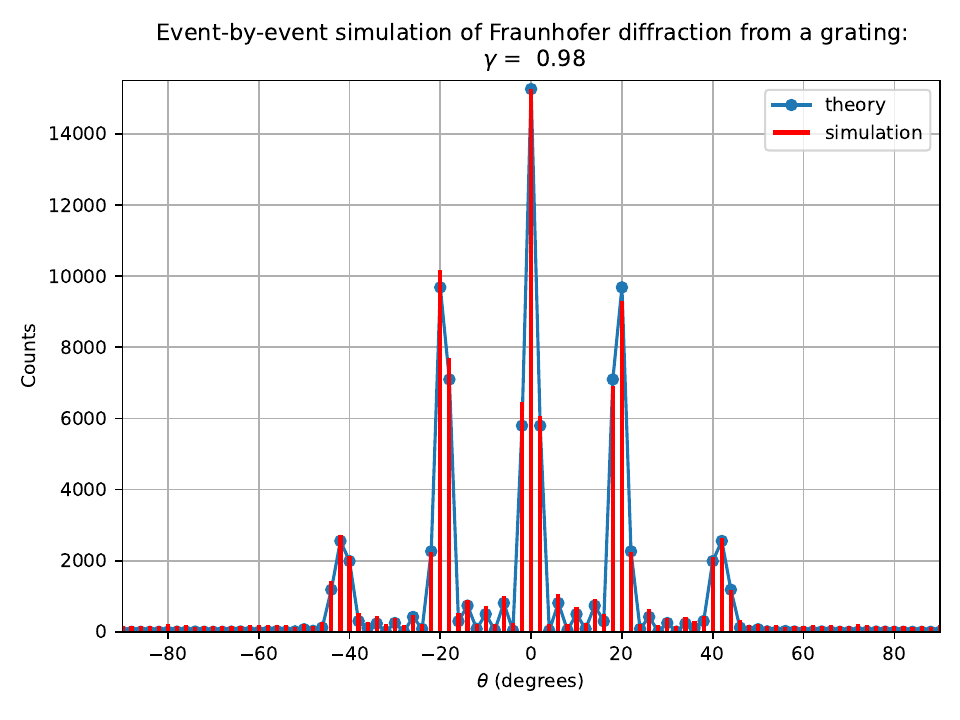} 
\caption{
The EBES results, generated by initiating the simulation via the graphical interface
shown in Fig.~\ref{GratingFig2}, demonstrate excellent agreement with quantum theoretical predictions.
This correspondence confirms that the event-based, material-interaction approach successfully
recovers the statistical expectations of the standard formalism without necessitating
an intrinsic 'wave-like' character for the particles.}
\label{GratingFig3}
\end{figure}

\section{Conclusion}

Consistent with previous EBES implementations, the 'wave-like' interference observed in
grating diffraction does not necessitate a fundamental change in the particle's nature.
Instead, it emerges from local interactions between the particles and the material,
a medium inherently capable of supporting diverse wave phenomena.
In other words,
the EBES approach suggests that the particle remains a discrete entity throughout the process;
it is the material substrate that facilitates wave-like distributions.
This perspective effectively demystifies particle-wave duality by reinterpreting
it as an emergent property of particle-material interactions rather than an intrinsic paradox of the particle itself.


\chapter{Einstein-Podolsky-Rosen-Bohm experiments}\label{EPRBINTRO}
\begin{figure}[H]
\centering
\includegraphics[width=0.90\hsize]{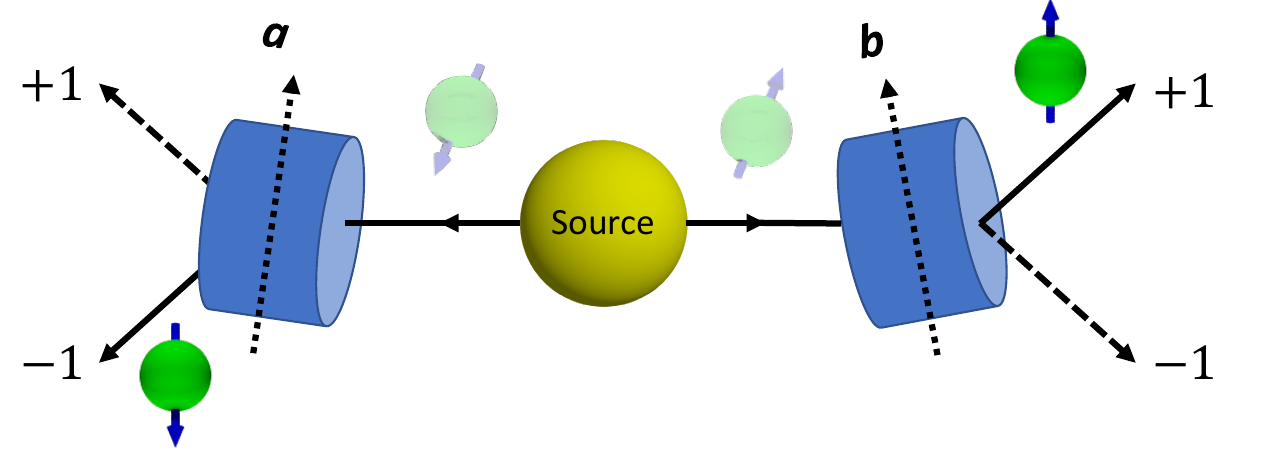}
\caption{%
Conceptual representation of the Einstein-Podolsky-Rosen thought experiment~\cite{EPR35},
in the modified form proposed by Bohm~\cite{BOHM51}.
A source emits pairs of particles whose magnetic moments are opposite.
Consequently, each pair leaves the source with perfectly correlated magnetic moments.
Each particle then enters a Stern-Gerlach apparatus (represented by the cylindrical objects),
whose magnetic field gradient, together with a uniform field component
oriented along the unit vectors $\ba$ and $\bb$,
deflects the particle into one of two spatially separated output channels labeled $+1$ and $-1$.
The chosen pair of settings $(\ba,\bb)$ specifies the measurement conditions
under which the data pair $(A_{n},B_{n})$ is recorded.
The values $A_{n}$ and $B_{n}$ correspond to the labels of the respective output
channels into which the two particles are deflected.
Repeating this procedure for all emitted particle pairs yields the dataset
${\cal D}=\{ (A_{1},B_{1}),\ldots, (A_{N},B_{N})\}$
where $N$ denotes the total number of pairs emitted by the source.
}%
\label{eprbidea}
\end{figure}

The Einstein-Podolsky-Rosen (EPR) thought experiment was originally proposed to challenge
the completeness of quantum mechanics~\cite{EPR35},
with ``completeness'' specifically defined within that foundational work.
Bohm later introduced a simplified configuration using spin-1/2 particles
rather than the continuous coordinates and momenta of the original two-particle system~\cite{BOHM51}.
This modified setup, commonly known as the Einstein-Podolsky-Rosen-Bohm (EPRB) experiment,
has since become the primary framework for numerous experimental tests of
quantum foundations~\cite{BELL64,PEAR70,PENA72,FINE74,FINE82,FINE82a,FINE82b,MUYN86,KUPC86,BRAN87,JAYN89,BROD89,BROD93,PITO94,FINE96,KHRE09z,
SICA99,BAER99,HESS01a,HESS01b,HESS05,ACCA05,KRAC05,SANT05,
KUPC05,MORG06,KHRE07,ADEN07,Khrennikov2008,NIEU09,MATZ09,KARL09,KHRE09,GRAF09,KHRE11,NIEU11,Brunner2014,HESS15,KUPC16z,KUPC17,HESS17a,NIEU17,
Adenier2017,Khrennikov2018,Drummond2019,Lad2020,Blasiak2021,Cetto2021,Lad2022}.

The essence of the EPRB thought experiment is shown and described in Fig.~\ref{eprbidea}.
Performing the EPRB thought experiment under the condition
defined by the directions $(\mathbf{a},\mathbf{b})$ yields the data set of pairs
\begin{eqnarray}
{\cal D}=\{(A_{1},B_{1}),\ldots,(A_{N},B_{N})\}
\;,
\label{DATA0}
\end{eqnarray}
where $N$ is the number of pairs emitted by the source
and all $A$'s and $B$'s are either $+1$ or $-1$.

Inspired by the quantum-theoretical description of the EPRB thought experiment, 
the empirical data set ${\cal D}$ is {\sl imagined} to exhibit the following features:
\begin{itemize}
\item[1.]
For each $n$, the numerical values of $A_n$ and $B_n$ are unpredictable but
their values may be correlated, meaning that if, say the value of $A_{n}$ is known, the probability
to correctly predict the observed value of $B_{n}$ is not zero.
\item[2.]
For all $n'\not=n$, there is no relation between the numerical values of $(A_{n},B_{n})$
and $(A_{n'},B_{n'})$ (but according to feature 1, there can be a relation between the values of $A_n$ and $B_n$).
This implies that, based on the knowledge of $A_{m}$ and $B_{m}$
it is impossible to predict any of the $A_{n\not=m}$ or $B_{n\not=m}$ with certainty.
In other words, pairs $(A_{n},B_{n})$ are unpredictable.
\item[3.]
The averages $\langle A \rangle$, $\langle B \rangle$ and the correlation $\langle AB \rangle$ defined by
\begin{eqnarray}
\langle A \rangle=\frac{A_1+\ldots+A_n}{N}
\;,\;
\langle B \rangle=\frac{B_1+\ldots+B_n}{N}
\;,\;
\langle AB \rangle=\frac{A_1B_1+\ldots+A_nB_n}{N}
\;,
\label{CORR}
\end{eqnarray}
are invariant under simultaneous rotation of the Stern-Gerlach magnets,
that is they can only depend on the directions of Stern-Gerlach magnets through the inner product $\ba\cdot\bb$
of their respective unit vectors.
\item[4.]
The averages $\langle A \rangle\approx0$, $\langle B \rangle\approx0$, and
the correlation $\langle AB \rangle\approx-\ba\cdot\bb$.
The latter implies that
if the directions of the Stern-Gerlach magnets are the same ($\ba=\bb$) or opposite ($\ba=-\bb$),
the data shows perfect anticorrelation or correlation,
that is for each $n=1,\ldots,N$, $A_{n}=-B_{n}$ or $A_{n}=+B_{n}$, respectively.
Then, the $A$'s and $B$'s are said to be perfectly anticorrelated or correlated, respectively.
\end{itemize}

In Feature 4, the two zeros and the expression $-\ba\cdot\bb$
represent the single-particle averages and the two-particle correlation
of two spin-1/2 objects in the singlet state.
These values are derived from the quantum-theoretical description of the EPRB thought experiment.
Notably, understanding the core of the EPRB experiment does not require expertise in quantum theory;
one simply needs to accept that the quantum-theoretical framework
requires the data to satisfy the conditions in Feature 4.
The primary conceptual challenge regarding the experiment is addressed in Appendix~\ref{ISSUE},
while a concise summary of the corresponding laboratory implementation is provided in Appendix~\ref{LABEXP}.

In Feature 4, and in the sections that follow, the symbol $\approx$ indicates that the quantities
are expected to be equal up to ordinary statistical fluctuations.
The size of these fluctuations decreases with increasing sample size,
and in the limit of a large number of pairs $N$, they are expected to vanish.

\section{A game with people}\label{EPRBgame}

The primary objective of the EPRB game is to demonstrate that the data,
collected after finishing this game manifests features typically and exclusively attributed to quantum systems.
Specifically, the statistical averages and the correlation defined by Eq.~(\ref{CORR})
align precisely with the predictions of the quantum-theoretical treatment of the EPRB experiment.
For those well-versed in Bell's theorem, this result might appear paradoxical or even disruptive to established intuition.
However, as detailed in Appendix \ref{ISSUE}, while Bell's theorem remains mathematically unassailable,
it does not apply to the data generated by the EPRB game or to empirical data in general~\cite{RAED23,RAED24}.

In this game, we use human actors to represent the system, rather than photons or massive particles.
One can easily identify the parallels between the setup in Figure~\ref{eprbgame}
and the physical EPRB experiment described in Appendix~\ref{LABEXP} (Figure~\ref{eprbexp}).

Figure~\ref{eprbgame} does not display the second, equally essential \PHASE\ of the EPRB game,
and of laboratory experiments more generally,
namely the processing of the data produced and collected during the first \PHASE.
As in the YOUNG and NEUTRON games, the EPRB game must be repeated under different experimental conditions.
It is this third \PHASE, also absent from Fig.~\ref{eprbgame},
that ultimately yields the final results, which can then be compared with the predictions of quantum theory
or with alternative models such as the one proposed by Bell.

\begin{figure}[!bp]
\centering
\includegraphics[width=0.90\hsize]{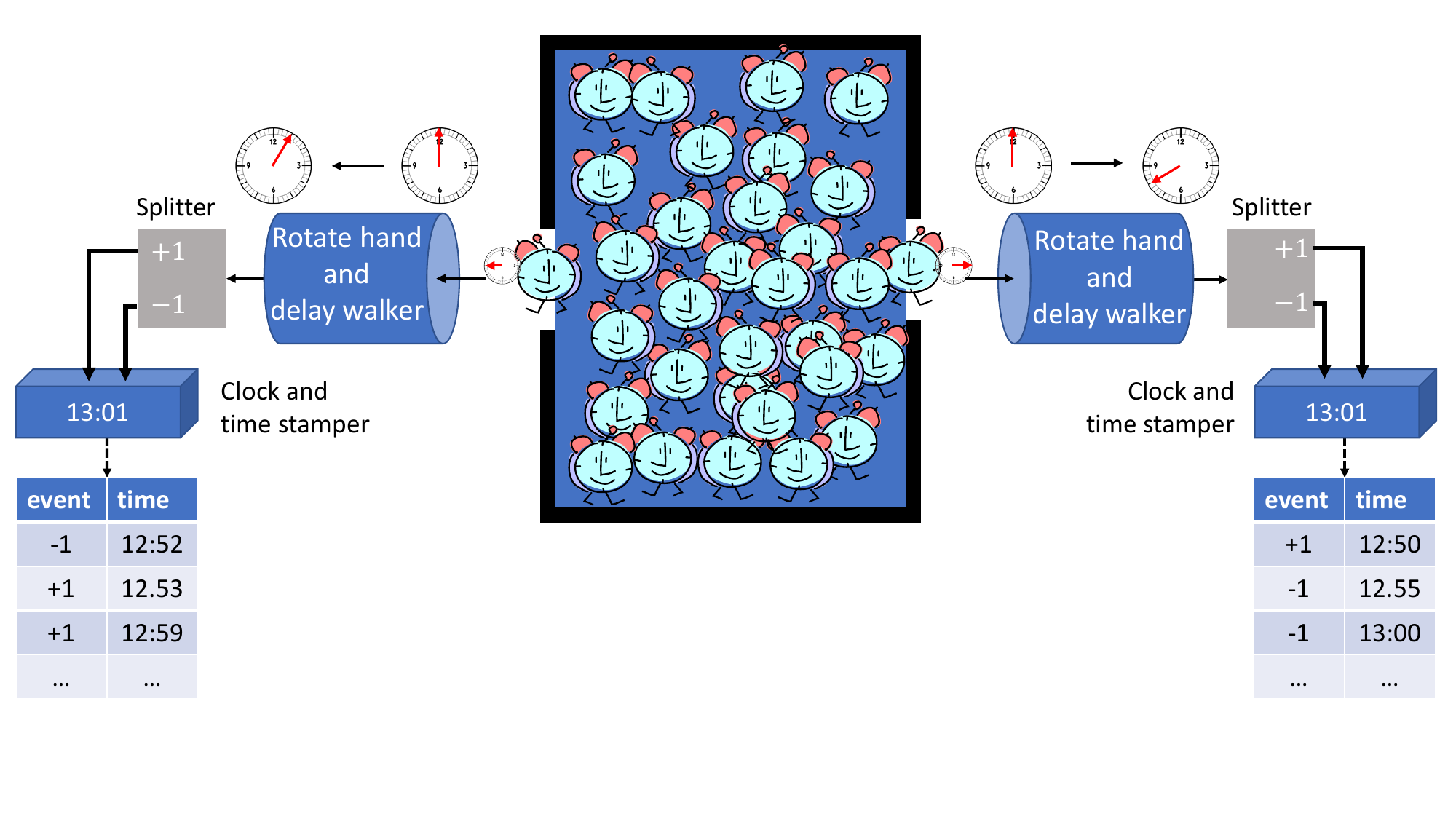}
\caption{
Picture of a game played with people, inspired by laboratory realizations of EPRB experiments, see Fig.~\ref{eprbexp}.
Following the rules of the game as described demonstrates that the generated data display correlations
commonly considered a hallmark of quantum physics.}
\label{eprbgame}
\end{figure}

\subsection{First \PHASE}\label{PART1}

The elements and rules of the first \PHASE\ of the EPRB game are as follows:

\renewcommand{\labelenumi}{A\arabic{enumi}.}

\begin{enumerate}
\item
There is a room filled with walkers.
\item
Walkers can leave through one of the two doors but they always have to leave in pairs.
Pairs of walkers leave the room one-by-one 
but not before the previous pair of walkers has left the game.
The number of pairs of walkers in the room is denoted by $N$.
\item
A walker never communicates with another one.
\item
When a pair of walkers leaves the room, each walker is given an artefact, similar to the Austrian parking disc of 1994.
The artifact contains a disc that can rotate.
The symbol $\varphi_{1}$ ($\varphi_{2}$)
denotes the angle of hand of the disc carried by the walker moving
to the left (right), relative to the upward direction of the artefact.
\item
The angle $\varphi_{1}$ is chosen randomly between zero and $360^\circ$ (clockwise)
and $\varphi_{2}=\varphi_{1}+90^\circ$.
Clearly, there is a fixed relation between the values of the angles $\varphi_{1}$
and $\varphi_{2}$.
Note that rotating a hand by $360^\circ$ leaves the position of the hand unchanged.
Therefore, should $\varphi_{2}=\varphi_{1}+90^\circ$
exceed $360^\circ$, a counterclockwise rotation by $360^\circ$,
that is setting $\varphi_{2}\leftarrow\varphi_{2}-360^\circ$,
brings $\varphi_{2}$ back between zero and $360^\circ$.
For example, if $\varphi_{1}=300^\circ$ then $\varphi_{2}=30^\circ$.
Thus, all angles may always be considered to lie in the interval of zero to $360^\circ$.
\item
The walkers are directed to proceed to the adjacent unit labeled ``splitter'' in a straight line.
The walking pace of all walkers is the same.
The distances from each room exit to the adjacent splitter are also the same.
\item
On their way to a splitter, walkers pass through a kind of tunnel (the blue cylinders in Fig.~\ref{eprbgame})
hosting a guard.
The guard in the left-most tunnel rotates, counterclockwise, the hand of the disc by an angle $a_{1}$.
For simplicity, it is assumed that the angle $a_{1}$ is fixed during the course of a game.
This changes $\varphi_{1}$ to $\varphi_{1}-a_{1}$.
Furthermore, the guard keeps the walker from moving for a time $t_{1}=T(\varphi_{1},a_{1})$.
The precise form of the time delay function $T(\varphi,a)$ is discussed later.
The rightmost guard does exactly the same as the leftmost guard, with the subscripts ${1}$
replaced by and ${2}$ of course.
Note that $a_{1}$ and $a_{2}$ can be different.

\item
After leaving the tunnel, the walkers proceed to the splitter units.
The leftmost splitter instructs the walker to proceed to the exit labeled $+1$ (see Fig.~\ref{eprbgame})
if two times the angle of the hand, which has changed from $2\varphi_{1}$
to $2(\varphi_{1}-a_{1})$ (modulo $360^\circ$) by now, lies
in the first or fourth quadrant of the disc.
Otherwise, the walker proceeds to the exit labeled $-1$ (see Fig.~\ref{eprbgame}).

The superficially looking factor of two only serves
to make contact to description of laboratory experiments that employ the polarization of photons (to be discussed later).
It is not essential for the EPRB game itself
but has the effect that it is sufficient to consider angles in the range of zero to $180^\circ$ only.

Expressed in terms of mathematical symbols, the leftmost splitter performs the operation
\begin{eqnarray}
x_{1} = \sign\left[\cos2(\varphi_{1}-a_{1})\right]
=
\left\{
\begin{array}{ccc}
+1 & \hbox{if}& \cos2(\varphi_{\mathit{l,n}}-a_{1}) \ge 0\\
\\
-1 & \hbox{if}& \cos2(\varphi_{\mathit{l,n}}-a_{1}) < 0\\
\end{array}
\right.
\;.
\label{x}
\end{eqnarray}
The rightmost splitter operates in exactly the same manner (with the subscripts ${1}$
replaced by ${2}$).

Note that the cosine in Eq.~(\ref{x}) automatically takes care of the fact that the rotation
of the disc is defined up to multiples of $360^\circ$ only (see A.5).
Functions $\cos2(\varphi_{1}-a_{1})$ also appear
quite naturally in description of the photon polarization~\cite{BORN64}.

The model defined by Eq.~(\ref{x}) is a two-dimensional version of the model used by J.S. Bell in
his famous paper on the Einstein-Podolsky-Rosen paradox~\cite{BELL93}.
Therefore, it will be referred to as the Bell-model in this chapter and in the accompanying software.

\item
The walkers then move to the next adjacent units labeled ``clock and time stamper''.
Again, both units operate in exactly the same manner.
The purpose of such a unit is to fill (as indicated by the dashed line connecting to the table)
a database with records containing the label of the splitter exit ($+1$ or $-1$)
and the time of arrival at the ``clock and time stamper'' unit.
Once the record has been added to the database, the walker leaves the game.
\item
If both walkers have left the game, a new pair of walkers may leave the room,
as long as there are pairs left in the room.
Note that this requirement is only necessary to eliminate the possibility that
walker of one pair can communicate with walkers of another pair.

\end{enumerate}
\renewcommand{\labelenumi}{\arabic{enumi}.}

\subsection{Second \PHASE}\label{PART2}

The second \PHASE\ of the EPRB game consists of processing the data stored in the two databases.
Recall that each record contains two items, namely the event label and the timestamp.
It is expedient to denote the event labels by $x_{1,n}$ and  $x_{2,n}$ where
the subscript refers to the number of the $n$th pair of walkers or, equivalently, the $n$th
record in both databases.
Likewise, the timestamp data is denoted by $t_{1,n}$ and $t_{2,n}$.
The $x$'s take values $+1$ or $-1$ only while the $t$'s are equal to or larger than zero.
Because of rules A.7 and A.8, both $x_{1,n}$ and $t_{1,n}$
depend on $\varphi_{\mathit{l,n}}-a_{1}$ and the same for the subscript ${1}$ replaced by ${2}$.

\renewcommand{\labelenumi}{\Roman{enumi}.}
\begin{center}
\framebox{
\parbox[t]{0.9\hsize}{%
The analysis of the data produced in the first \PHASE\ of the EPRB game is greatly simplified by imposing
the following additional conditions:
\begin{enumerate}
\item
Each time the EPRB game starts, the stopwatches are reset, both databases are cleared,
and the room is filled with $N$ pairs of walkers.
\item
The time interval, denoted by $T$, between successive pairs leaving the room is constant.
Thus, the $n$th pair leaves at the wall clock time $(n-1)T$.
In the absence of a delay, the time for walkers to reach a timestamp unit and leave the game is denoted by $T'$,
independent from which door the walker leaves the room.
\item
The time for walking from a door to the adjacent timestamp unit
is always shorter than $T$, irrespective of the delay imposed by a guard.
This assumption guarantees that both walkers of the $n$th pair have
left the game before the $(n+1)$th pair leaves the room.
\end{enumerate}
}}%
\end{center}
\renewcommand{\labelenumi}{\arabic{enumi}.}

The first set of quantities to compute are the averages and correlation of the $x$'s according
to the standard definitions
\begin{subequations}
\label{eprbave}
\begin{eqnarray}
\langle x_{1}\rangle &=& \frac{x_{1,1}+\ldots+x_{1,N}}{N}
\;,
\label{eprbave1}
\\ 
\langle x_{2}\rangle &=& \frac{x_{2,1}+\ldots+x_{2,N}}{N}
\;,
\label{eprbave2}
\\ 
\langle x_{1}\,x_{2}\rangle &=&
\frac{x_{1,1}\,x_{2,1}+\ldots+x_{1,N}\,x_{2,N}}{N}
\label{eprbave3}
\;.
\end{eqnarray}
\end{subequations}
Obviously, the three quantities in Eq.~(\ref{eprbave}) ignore all information that might be
contained in the timestamps.

In laboratory EPRB experiments, the timestamps serve to discard database records.
There are essentially two different ways to accomplish this.
The Python program that simulates the EPRB game implements both.

\subsubsection{Selection by a coincidence window}

First note that because of rule A.7 and condition III, the arrival time of the $n$th left going walker
is given by $T_{1,n}=(n-1)T + T' + t_{1,n}\le nT$,
where $t_{1,n}\ge 0$ is the time delay caused by passing through the leftmost tunnel.
Similarly, $T_{2,n}=(n-1)T + T' + t_{2,n}\le nT$.
Therefore, for all $n$, the time difference
$T_{1,n}-T_{2,n}=t_{1,n}-t_{2,n}$,
independent of $T$ and $T'$.

The rule to discard records is simple and reads as follows
\renewcommand{\labelenumi}{B.\arabic{enumi}.}
\begin{enumerate}
\item
For the calculation of the averages and correlations,
discard all records $n$ for which $|t_{1,n}-t_{2,n}|> W$
where $W$ is a so-called time window.
\end{enumerate}
\renewcommand{\labelenumi}{\arabic{enumi}.}

The arithmetic operations involved are simple.
It is expedient to introduce a new variable by
\begin{eqnarray}
w_{n}&=&
\left\{
\begin{array}{ccc}
0 & \hbox{if}& |t_{1,n}-t_{2,n}| > W \\
\\
1 & \hbox{if}& |t_{1,n}-t_{2,n}|\le W \\
\end{array}
\right.
\label{w}
\;.
\end{eqnarray}
Taking into account that the number of records is no longer necessarily
equal to $N$, the averages and correlation are calculated as
\begin{subequations}
\label{eprbwin}
\begin{eqnarray}
\langle x_{1}\rangle_{\mathrm{C}} &=& \frac{w_{1}\,x_{1,1}+\ldots+w_{N}\,x_{1,M}}{
w_{1}+\ldots+w_{N}}
\;,
\label{eprbwin1}
\\ 
\langle x_{2}\rangle_{\mathrm{C}} &=& \frac{w_{1}\,x_{2,1}+\ldots+w_{N}\,x_{2,M}}{
w_{1}+\ldots+w_{N}}
\;,
\label{eprbwin2}
\\ 
\langle x_{1}\,x_{2}\rangle_{\mathrm{C}} &=&
\frac{w_{1}\,x_{1,1}\,x_{2,1}+\ldots+w_{N}\,x_{1,N}\,x_{2,N}}{
w_{1}+\ldots+w_{N}}
\;,
\label{eprbwin3}
\end{eqnarray}
\end{subequations}
where the subscript $\mathrm{C}$ indicates
that averages and correlation have been computed by looking for coincidences in the timestamp data.

By condition III, the maximum delay time, denoted by $t_{\mathrm{max}}$, is smaller than
the time interval $T$ between pairs leaving the room.
In symbols $|t_{1,n}-t_{2,n}|\le t_{\mathrm{max}}<T$.
Therefore, if the time window $W$ is larger than $t_{\mathrm{max}}$, it follows from Eq.~(\ref{w}) that $w_n=1$ for all $n$,
implying that Eqs.~(\ref{eprbwin}) yield the same numerical values as the corresponding Eqs.~(\ref{eprbave}).
In other words, if the time window is chosen such that $W<t_{\mathrm{max}}$ the effect of the time delay
will show itself through a difference between the
numerical values of Eqs.~(\ref{eprbwin}) and the corresponding ones of Eqs.~(\ref{eprbave}).

\subsubsection{Selection by a local window}

Some EPRB experiments discard data not through a coincidence window but through a local window,
which in practice may be implemented as a voltage threshold~\cite{GIUS15,SHAL15}.
A nearly trivial modification of the EPRB game, also included in the EBES software,
handles this case as well.
Introducing two new variables by
\begin{eqnarray}
w_{1,n}&=&
\left\{
\begin{array}{ccc}
0 & \hbox{if}& t_{1,n} > W \\
\\
1 & \hbox{if}& t_{1,n} \le W \\
\end{array}
\right.
\label{lw}
\;,
\end{eqnarray}
where $w_{2,n}$ is given by Eq.~(\ref{lw}) with all subscripts ${1}$ replaced by ${2}$
and taking into account that the number of records is no longer necessarily
equal to $N$, the averages and correlation are calculated as
\begin{subequations}
\label{eprbloc}
\begin{eqnarray}
\langle x_{1}\rangle_{\mathrm{L}} &=& \frac{w_{1,1}x_{1,1}+\ldots
+w_{1,N}x_{1,N}}{
w_{1,1}\,w_{2,1}+\ldots+w_{1,N}\,w_{2,N}}
\;,
\label{eprbloc1}
\\ 
\langle x_{2}\rangle_{\mathrm{L}} &=& \frac{w_{2,1}x_{2,1}
+\ldots+w_{2,N}x_{2,N}}{
w_{1,1}\,w_{2,1}+\ldots+w_{1,N}\,w_{2,N}}
\;,
\label{eprbloc2}
\\ 
\langle x_{1}\,x_{2}\rangle_{\mathrm{L}} &=&
\frac{w_{1,1}\,w_{2,1}x_{1,1}\,x_{2,1}+\ldots+
w_{1,N}\,w_{2,N}y_{1,N}\,y_{2,N}}{
w_{1,1}\,w_{2,1}+\ldots+w_{1,N}\,w_{2,N}}
\label{eprbloc3}
\;.
\end{eqnarray}
\end{subequations}
where the subscript $\mathrm{L}$ is a reminder
that averages and correlation have been computed by using local windows to select data.

\subsubsection{Common features}
A direct consequence of the rule A.5 and the choice Eq.~(\ref{x}) is that the averages in Eqs.~(\ref{eprbave}),~(\ref{eprbwin})
and~(\ref{eprbwin2})
are, within the usual statistical fluctuations, equal to zero.
This fact follows from the observation that
$x_{1,n}$ ($x_{2,n}$) only depends on
$\varphi_{\mathit{l,n}}-a_{1}$
($\varphi_{\mathit{r,n}}-a_{2}$).
As the latter covers the full circle in a random manner, the frequencies with which the $x$'s take the values
$+1$ or $-1$ are almost the same and therefore the average of these $x$'s is approximately zero.

Another immediate consequence of the rules A.5 and A.8 and the choice Eq.~(\ref{x})
is that the choice $a_{1}=a_{2}$ implies that
$x_{1,n}=+1\,(-1)$ and $x_{2,n}=-1\;(+1)$.
Thus, if $a_{1}=a_{2}$,
the correlations Eq.~(\ref{eprbave3}), Eq.~(\ref{eprbwin3}), and Eq.~(\ref{eprbloc3})
are exactly minus one, for any $n$.
In other words, $x_{1,n}$ and $x_{2,n}$ are perfectly anticorrelated
if the angles $a_{1}$ and $a_{2}$ are the same,
in accordance with the requirements set out in beginning of this chapter.

\subsubsection{Malus law polarizer}
From the perspective of real-world EPRB experiments, the Bell model is limited by more than
just its failure to account for a critical experimental parameter:
the coincidence (or local) time window.
It also fails fundamentally to describe the physical operation of a single polarizer.

This basic mechanism is governed by Malus's Law, discovered by \'Etienne-Louis Malus in the early 19th century.
The law states that the intensity of a polarized beam of light passing
through a perfect polarizer is proportional to the square of the cosine of the angle
between the light's initial polarization and the polarizer's transmission axis: $I=I_0\cos^2\alpha$
where $I_0$ is the incident intensity and $\alpha$ is the angle between
the polarization direction and the polarizer's axis.

Simulating a polarizer that obeys Malus law on an event-by-event based is easy.
The rule is 
\begin{eqnarray}
x_{\mathit{1,n}} = \sign\left[\cos2(\varphi_{\mathit{l,n}}-a_{1})-r_n\right]
=
\left\{
\begin{array}{ccc}
+1 & \hbox{if}& \cos2(\varphi_{\mathit{l,n}}-a_{1}) \ge r_n\\
\\
-1 & \hbox{if}& \cos2(\varphi_{\mathit{l,n}}-a_{1}) < r_n\\
\end{array}
\right.
\;,
\label{ML}
\end{eqnarray}
where $r_n$ is a random number between minus and plus one.
Note that if $r_n=0$ for all $n$, the rule Eq.~(\ref{x}) of the Bell model is recovered.
As always, $x_{\mathit{2,n}}$ is defined by replacing all first subscripts ${1}$ by ${2}$.

\subsection{Third \PHASE}\label{PART3}

The third \PHASE\ consists of repeating the EPRB game for a fixed value of $W$
and different choices of the rotation angles $a_{1}$ and $a_{2}$.

At first sight, presenting the numerical values of Eqs.~(\ref{eprbave}) and Eqs.~(\ref{eprbwin})
as a function of these two angles would require a three-dimensional plot.
Fortunately, the data can equally well be represented by a simple two-dimensional plot
as a function of the difference $\theta=a_{1}-a_{2}$.

This simplification is made possible by the fact that $\varphi_{1}$ is chosen randomly, see rule A.5.
The argument goes as follows.
Recall that because of rules A.7 and A.8, both $x_{1,n}$ and $T_{1,n}$
depend on $\varphi_{\mathit{l,n}}-a_{1}$ and the same for the subscript ${1}$ replaced by ${2}$.
Also recall that $\varphi_{\mathit{r,n}}=\varphi_{\mathit{l,n}}+90^\circ$
and that all angles are to be taken modulo $360^\circ$.
As $\varphi_{1}$ is random in the range of zero to $360^\circ$,
$\varphi'_{\mathit{l,n}}=\varphi_{\mathit{l,n}}-a_{1}$ is random in the range of zero to $360^\circ$ also.
If $N$ is large there is, within the usual statistical fluctuations (vanishing as $1/\sqrt{N}$),
no difference between using $\varphi_{\mathit{l,n}}-a_{1}$ or $\varphi'_{\mathit{l,n}}$.
Both angles randomly sample the interval of zero to $360^\circ$.

Introducing $\varphi'_{\mathit{r,n}}=\varphi'_{\mathit{l,n}}+90^\circ$,
it follows that $\varphi_{2}-a_{\mathit{r,n}}=\varphi'_{\mathit{r,n}}+a_{1}-a_{2}$.
Therefore, always within the usual statistical fluctuations,
the averages and correlations Eqs.~(\ref{eprbave}) and~(\ref{eprbwin})
only depend on $a_{1}$ and $a_{2}$ through their difference $\theta=a_{1}-a_{2}$.
This significantly simplifies the presentation and analysis
of the numerical values of Eqs.~(\ref{eprbave}) and Eqs.~(\ref{eprbwin}).

\begin{figure}[!htp]
\centering
\includegraphics[width=0.90\hsize]{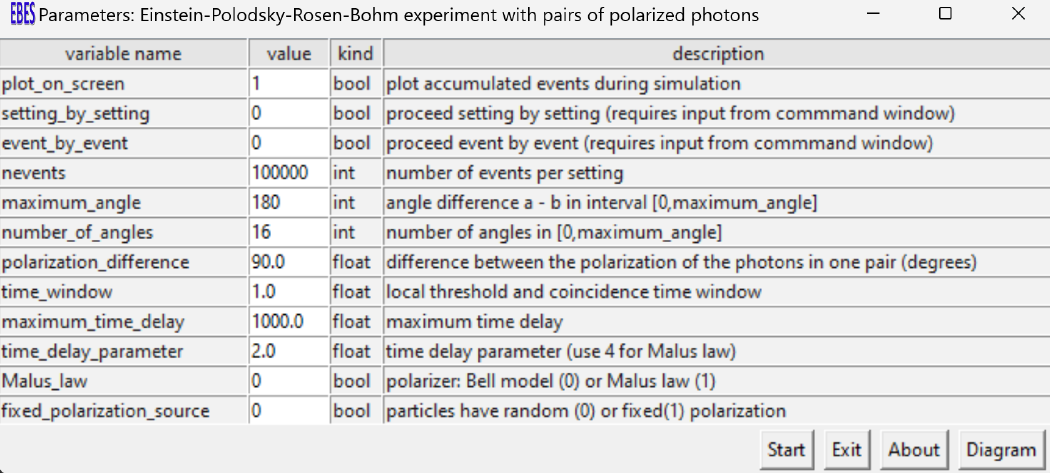}
\caption{
Graphical interface to control the event-by-event simulation of the EPRB game shown in Fig.~\ref{eprbgame}.
}
\label{eprbFig5}
\end{figure}

\section{Event-by-event simulation of the EPRB game}\label{EPRBresults}

\begin{figure}[!htp]
\centering
\includegraphics[width=0.80\hsize]{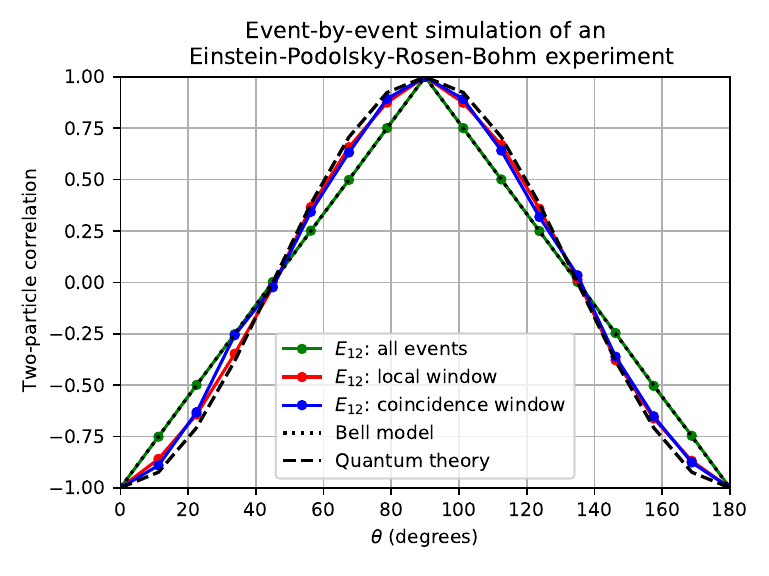}
\caption{
The correlations obtained by simulating the EPRB game.
All events: Eq.~(\ref{eprbave3});
local window: Eq.~(\ref{eprbloc3});
coincidence window: Eq.~(\ref{eprbwin3});
dotted line: Bell model Eq.~(\ref{eprbave4});
dashed line: quantum theoretical result $-\cos\theta$ for the correlation of two polarizations in the singlet state.
}
\label{eprbresults1}
\end{figure}

Implementing the EPRB game described in Section~\ref{EPRBgame} is fairly straightforward.
In fact, several of its rules are more difficult to express in words than to encode in a program.
Moreover, the EPRB game is simple enough to permit an analysis using traditional methods of theoretical physics,
at least in several important limiting cases~\cite{RAED06c,RAED07b,RAED23}.
This analysis shows that the EPRB game reproduces the quantum-theoretical results of the original
EPRB thought experiment in the limit of a vanishing time window,
and that it agrees with the predictions of Bell's model
when the time-window data are omitted from the analysis or, equivalently,
when the time window exceeds the maximum time delay~\cite{RAED06c,RAED07b}.
Historically, the simulation code was found to reproduce the expected results
before the subsequent theoretical treatment of the model confirmed them.

The EPRB game simulator accompanying this book was designed not only with the EPRB game
described in Section~\ref{EPRBgame} in mind, but also to simulate,
simultaneously, several variants of the game, such as Bell's model, coincidence windows,
local windows, and Malus polarizers,
some of which emulate laboratory experiments with polarized photons.
As the code snippets shown later illustrate, writing the core program to
simulate the EPRB game itself is hardly a major programming effort.
However, as Section~\ref{PART3} already suggests,
in contrast to the YOUNG and NEUTRON games,
the post-processing of the data constitutes a substantial part of playing the EPRB game.
Unfortunately, processing and plotting these data requires a considerable amount of code.

A brief discussion, primarily intended to help readers start with carrying out their own simulations follows.
The simulation is started by typing \hfill\break{\tt py(thon3) run\_eprb\_experiment.py} on the command line.
The parameter window Fig.~\ref{eprbFig5} appears.

Pressing the {\tt Start} button on the parameter window produces Fig.~\ref{eprbresults1},
in steps of $(180/16)^\circ$.
For each value of $\theta$, the number of pairs $N=100000$ generated was chosen such that the simulation finishes in
about a minute. Increasing $N$ will reduce the statistical fluctuations on the simulation data.
Disregarding these statistical fluctuations,
there can be little doubt that the EPRB game simulation generates data that are in excellent agreement
with predictions of the corresponding theoretical (Bell and quantum) models.

\begin{figure}[!htp]
\centering
\includegraphics[width=0.90\hsize]{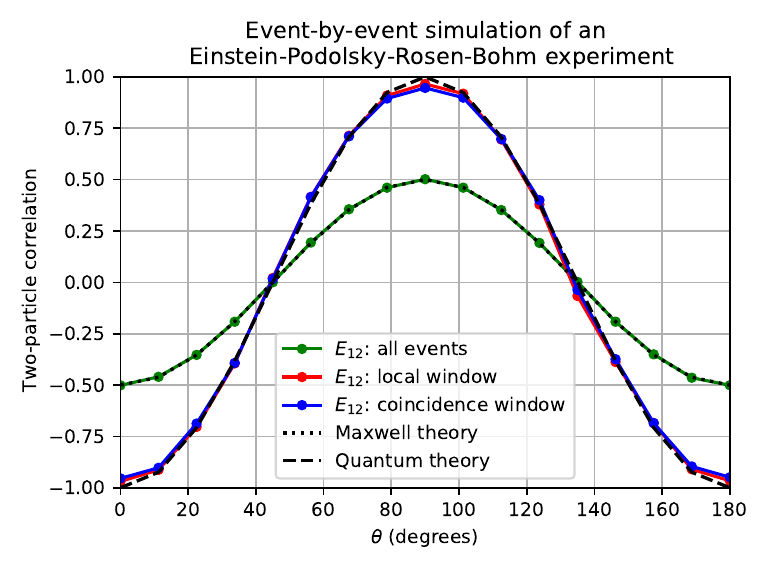}
\caption{
The correlations obtained by simulating the EPRB game
using Malus law instead of Bell's model for the polarizers.
All events: Eq.~(\ref{eprbave3});
local window: Eq.~(\ref{eprbloc3});
coincidence window: Eq.~(\ref{eprbwin3});
dotted line: Maxwell's theory;
dashed line: quantum theoretical result $-\cos\theta$ for the correlation of two polarizations in the singlet state.
}
\label{eprbresults3}
\end{figure}

To select the Malus law polarizer instead Bell model, modify the entry {\tt Malus\_law} in the parameter window
and also change the value of the time\_delay\_parameter from 2 to 4 (see later) and press {\tt Start}.
Alternatively, start the simulator using the command \hfill\break{\tt py[thon3] run\_eprb\_experiment.py -M -d 4}.
The reason for using $d=4$ instead of $d=2$ (the role to the parameter $d$ is discussed
in Section~\ref{EPRBwin}) for the time-delay function $T(\varphi,a)=T_0 |\sin2(\varphi-a)|^d$
is that it has been shown analytically (after evidence produced by simulation)
that in the case of Malus law, using $d=4$ (and in the limit of large $N$ and vanishing $W$),
exactly yield the correlation $-\cos\theta$ predicted by quantum theory~\cite{ZHAO08,RAED07c,RAED23}.
The results are shown in Figs.~\ref{eprbresults3}.
They demonstrate that also for this, physically more relevant model, the EBES
reproduces the results of Maxwell's and quantum theory, without using any of the concepts of these
wave theories, of course.

\section{Python code for simulating the EPRB game}\label{EPRB.COMPUTERCODE}

As with the implementations of the other games in this book,
the GUI and the plotting routines constitute the bulk of the Python code for the EPRB game.
In the following, we discuss only the core simulation code,
assuming that the remaining implementation details are readily accessible to readers familiar with Python.

Table~\ref{eprbtab1} show the one-to-one mapping between
the actors and their action in the game with people (see section~\ref{EPRBgame}),
the objects and their functions in the simulation code
and objects in the laboratory EPRB expedient (see Fig.~\ref{eprbexp}).
This mapping may help to understand how the simulation code is build and functions.
As always, when referring to the code, a word such as ``particle''
is merely a convenient label for an object/class/..., and does in no way refer to the concept
of a particle as it is used in physics.

\begin{table}[H]
\caption{%
Correspondence between the actors in the EPRB game (Fig.~\ref{eprbgame}), the objects used in the Python code,
and the elements in a real EPRB experiment with polarized photons (Fig.~\ref{eprbexp})~\cite{WEIH98}.
}
\begin{tabular}{ccc}
\noalign{\medskip}
\hline\hline\noalign{\smallskip}
game with people       &  simulation code &  experiment           \\
\hline\noalign{\smallskip}
walker                 & particle        &  photon              \\
room                   & two-particle source       &  two-photon source    \\
hand of the disc       & message                   &  polarization         \\
tunnel+guard           & plane rotation            &  electro-optic modulator        \\
splitter               & polarizer                 &  polarizing beam splitter        \\
clock and time stamper & time delay                &  clock and time stamper       \\
arrival time           & time tag                  &  time tag       \\
data base              & arrays to hold data       &  files on a computer     \\
averages & averages &  averages     \\
correlations& correlations & correlations     \\
\hline\noalign{\smallskip}
\end{tabular}
\label{eprbtab1}
\end{table}

\renewcommand\code{{\bf experiment/eprb\_experiment.py}}
Listing~\ref{EPRB.Lmain} presents the code snippet (slightly modified from \break\hfill\code\ for presentation purposes)
that performs the simulation (see section~\ref{PART1}) of the EPRB game.

The Python language interpreter destroys the {\tt particle1}
and {\tt particle2} objects when it reaches the end of the loop, as required by rule A.10.
From the code itself, it is clear that the processing of the particles representing the leftgoing
walkers ({\tt particle1}) cannot have any effect whatsoever on the
messages carried by the particles representing the rightgoing walkers ({\tt particle2}) and vice versa.
Denying this elementary fact is merely a proof that one has not come to grips with how a computer language
is translated into the instructions carried out by a CPU.
In any case, the code complies with A.3.

Lines 10 and 11 show how the static elements, in this case the tunnel + guard and the clock + time stamper
in Fig.~\ref{eprbgame}, are represented by objects
created by reference to the class of {\tt MalusLawPolarizer} which is discussed in more detail later.
The {\tt pol1} and {\tt pol2} objects combine the tasks of the tunnel + guard and the clock + time stamper.

The variables {\tt crotation0(1)} and {\tt srotation0(1)}
correspond to $\cos a_{\mathrm{l}}$ ($\cos a_{\mathrm{r}}$) and
$\sin a_{\mathrm{l}}$ ($\sin a_{\mathrm{r}}$), respectively,
and are passed to the {\tt do\_the\_simulation(...)} function by the main code looping over
the difference $\theta=a_{\mathrm{l}}-a_{\mathrm{r}}$.

The values of ${\tt maximum\_time\_delay}$ (= $t_{\mathrm{max}}$)
and {\tt time\_delay\_parameter} (a parameter of the time delay function $T(\varphi,a)$, see later)
are taken from the GUI parameter window (see Fig.~\ref{eprbFig5}).

The statement in line number 13 initiates a loop over {\tt nevents} events.
The value of {\tt nevents} is taken from the GUI parameter window (see Fig.~\ref{eprbFig5}).

Lines 15 and 16 create particle objects with random polarizations, as indicated by the comments.
The arrays {\tt cos0(1)} and {\tt sin0(1)}
contain the {\tt nevents} (random) values of $\cos\varphi$ ($\sin(\varphi+\pi/2)$)
and $\sin\varphi$ ($\cos(\varphi+\pi/2)$).
For reasons of computational efficiency and also to illustrate the use of Occam's razor principle,
the {\tt Particle} class, discussed later,
is implemented in a slightly different manner that in the case of the YOUNG and NEUTRON game.

{\scriptsize
\lstset{style=mystyle,linewidth=0.95\textwidth,caption={Python code for simulating the EPRB game, described in section~\ref{EPRBgame}.},label={EPRB.Lmain}}
\hbox{\hbox to 0.3cm{}
\begin{lstlisting}{ht}
def do_the_simulation(self, nevents,
                      crotation0, srotation0,  # setting of observation station 1 (fixed in this demo)
                      crotation1, srotation1,  # setting of observation station 2 (variable in this demo)
                      cos0, sin0,              # cos and sin of random angle
                      cos1, sin1,              # cos and sin of random angle + pi/2 (unless specified differently)
                      event1, timestamp1,      # arrays to store type of events and time stamps of station 1
                      event2, timestamp2,      # arrays to store type of events and time stamps of station 2
                      ran1, ran2, ran3, ran4): # arrays of random numbers used by MalusLawPolarizer
# create instances of the polarizers, pol1 <=> left side, pol2 <=> right side
    pol1 = MalusLawPolarizer(crotation0, srotation0, Tmax=self.settings.options.maximum_time_delay, d=self.settings.options.time_delay_parameter)
    pol2 = MalusLawPolarizer(crotation1, srotation1, Tmax=self.settings.options.maximum_time_delay, d=self.settings.options.time_delay_parameter)
# loop over all events
    for j in range(nevents):
# create a pair particles with random, correlated polarization, see dlm/eprb_particle.py
        particle1 = Particle(cos0[j], sin0[j])  # left going particle, (cos0,sin0) <=> (cos(phi),sin(phi))
        particle2 = Particle(cos1[j], sin1[j])  # right going particle, (cos1,sin1) <=> (cos(phi+pi/2),sin(phi+pi/2))
# perform the measurements, see dlm/eprb_polarizer.py
        event1[j] = pol1.process(particle1, ran1[j], ran2[j])  # store the j-th event
        timestamp1[j,:] = particle1.time_of_flight  # j-th time delays for the local and coincidence counting
        event2[j] = pol2.process(particle2, ran3[j], ran4[j])
        timestamp2[j,:] = particle2.time_of_flight

\end{lstlisting}
}}
\medskip

\renewcommand\code{{\bf dlm/eprb\_particle.py}}
{\scriptsize
\lstset{style=mystyle,linewidth=0.95\textwidth,caption={Python class definition of a particle object.},label={Lparticle}}
\hbox{\hbox to 0.3cm{}
\begin{lstlisting}{ht}
class Particle:  # defines class of particle objects
    def __init__(self, cpolarization, spolarization):
        self.time_of_flight = np.zeros(2)  # to be used in conjuction with the local/coincidence time window
        self.message = np.array([cpolarization, spolarization])  # message
\end{lstlisting}\label{LISTING\thelstlisting}
}}
\medskip

Line 18 instructs the polarizer object {\tt pol1} to process the message carried by the {\tt particle1} object.
The arrays {\tt ran1} and {\tt ran2} contain {\tt nevents} random numbers
in the range of $-1$ to $+1$ and of $0$ to $+1$, respectively.
The $j$th pair of these random numbers is used to decide through with exit ($+1$ or $-1$)
the $j$th instance of {\tt particle1} leaves the splitter and to compute the $j$th time delay.
The latter is stored in {\tt time\_of\_flight} property of the $j$th instance of {\tt particle1} object.
Line 19 copies this number to the $j$th element of the array {\tt timestamp1}.
Lines 20 and 21 process {\tt particle2} in the same manner as lines 18 and 19 did for {\tt particle1}.

Listing~\ref{Lparticle} shows the code (taken from \code) to create a particle object
with the time-of-flight and the cosine and sine of the hand's angle as its properties.
In essence, this is the same code as the ones used in the YOUNG and NEUTRON game.

{\scriptsize
\lstset{style=mystyle,linewidth=0.95\textwidth,caption={Python class definition of a polarizer object.},label={Lpolarizer}}
\hbox{\hbox to 0.3cm{}
\begin{lstlisting}
class MalusLawPolarizer():
    """
    MalusLawPolarizer: see Front. Phys. 8:160 (2020), doi: 10.3389/fphy.2020.00160
    """
    def __init__(self, crotation=1, srotation=0, Tmax=5000, d=4):  # creates an instance of a Malus law, beam-splitting polarizer
        self.Tmax = Tmax  # maximum time delay
        self.d = float(d)  #  time-delay exponent
        self.cDLM = 1  #  internal state of the adaptive machine
        self.sDLM = 0  #  see Front. Phys. 8:160 (2020), doi: 10.3389/fphy.2020.00160
        self.dDLM = 1 if d <= 0  else 1 / float(d)  #  adaptive machine exponent, trial and error
        self.crotation = crotation
        self.srotation = srotation

    def process(self, particle, r0, r1):  # process incoming event, r0 and r1 are input, random numbers

        cinp = particle.message[0]
        sinp = particle.message[1]

        z = abs( (1 - ( self.cDLM * cinp + self.sDLM * sinp ) ) / 2 ) ** self.dDLM
        self.cDLM = cinp # use the current message to set the internal state of the adaptive machine
        self.sDLM = sinp #

        c2 = self.crotation * cinp + self.srotation * sinp  # plane rotation  , phi --> phi - angle
        s2 = self.crotation * sinp - self.srotation * cinp

        time_delay = r1 * self.Tmax * abs(2 * c2 * s2)**self.d  # 2*c2*s2 = sin(2(x-a))
        particle.time_of_flight[0] = z * z * time_delay  # used for local time windows
        particle.time_of_flight[1] = z * time_delay  # coincidence time window

        if c2 * c2 - s2 * s2 >= r0:  # -1<=r0<=1; c2*c2-s2*s2 = cos(2(x-a)); if r0=0 ==>  sign(..) model
            port = 0  # corresponds to S=+1        ; according to Eq.(11), Frontiers of Physics (2020).
            particle.message[0] = self.crotation
            particle.message[1] = self.srotation
        else:
            port = 1  # corresponds to S=-1
            particle.message[0] = -self.srotation
            particle.message[1] = self.crotation

        return port
\end{lstlisting}\label{LISTING\thelstlisting}
}}
\medskip

Upon creation of an object of the class {\tt MalusLawPolarizer}, the {\_\_init\_\_} function
initializes the variables that belong to this particular object.

The {\tt process} function performs the follows operations
(to simplify the writing, we temporarily drop the subscripts ${1}$ and ${2}$).
\begin{itemize}
\item
Lines 19 to 21 implement a Deterministic Learning Machine (DLM)
that distinguishes between random and fixed sequences of $\varphi$~\cite{RAED20a}.
This is accomplished as follows.
First note that lines 20 and 21 copy the content {\tt cinp} and {\tt sinp} into the
and the internal state variables {\tt self.cDLM} and {\tt self.sDLM} of the DLM.
Thus, except for the very first invocation of {\tt process},
the internal state of the DLM is just the pair {\tt cinp} and {\tt sinp} passed
to the {\tt process} during the previous invocation.
In order to detect if the input angles are fixed, it is sufficient
to compute the difference with the angle of the previous invocation.
If this difference is zero for all (except possibly the first) invocation,
the input angles considered to be fixed.
Line 19 employs $\cos(x-y)=\cos x\cos y + \sin x \sin y$ to use data that is already available.
If the input angles are always the same, $x-y=0$ and {\tt z1} will be zero.
If the input angles are random, {\tt z1} is a random number between zero and one.
The exponent {\tt self.dDLM} is, in principle, a free parameter of the model.
From many simulations, it follows that the choice made in the {\_\_init\_\_} section yields very satisfactory results.
It should be mentioned that lines 19 to 21 are not essential for the EPRB game itself
but are essential to reproduce the quantum theoretical results
of an extended EPRB experiment~\cite{RAED20a}.
\item
Lines 23 and 24 implement the rotation of the disc by the guard.
The values of $\cos\varphi$ and $\sin\varphi$ are taken from the message of the particle object,
and a plane rotation by the angle $a$ specified through the {\tt process} arguments
{\tt crotation} and {\tt srotation} is carried out.
\end{itemize}

\begin{itemize}
\item
Line 26 defines one part of the expression of the time delay.
The value of $r_1$ is passed to the {\tt process} code as an argument and
changes with each instance of the particle object.
The values of other two variables are set upon creation of the polarizer object (see the {\tt \_\_init\_\_} function)
and are fixed during the simulation.

Line 28 stores the value of the time delay, modified by the value of {\tt z},
in the array element {\tt particle.time\_of\_flight[1]}
that is used to compute the value of $w_n$ according to Eq.~(\ref{w}).
The array element {\tt particle.time\_of\_flight[0]} contains the time delay that
is used for the local time window procedure, to be discussed later.
\item
Line 30 computes $\cos2(\varphi-a)=\cos^2(\varphi-a) - \sin^2(\varphi-a)$ and compares it to the input variable {\tt r0}.
In the case of the Bell-model, used in the EPRB game, {\tt r0=0}, implying that
the {\tt if} statement in line 30 implements Eq.~(\ref{x}).
In the case of the Malus law model (explained later),
the input variable {\tt r0} is a random number in the range of $-1$ to $+1$,
which changes with each instance of the particle object.
\item
Depending on the result of the {\tt if} statement in line 30, the {\tt port} variable is set of zero or one and,
for completeness (but not used in this application),
the message is updated to mimic the operation of a real polarizer.
\end{itemize}

This completes the description of the essential elements of the EPRB game simulator.
Actually, as mentioned earlier, the code simultaneously
simulates several variants of the EPRB game that are close to
laboratory experiments with polarized photons, adding to the complexity of the code.

\renewcommand\code{{\bf dlm/eprb\_polarizer.py}}
Listing~\ref{Lpolarizer} shows the code (taken from \code) to create an object
that performs the combined operation of the tunnel + guard, splitter, and time stamper, see Fig.~\ref{eprbgame}.
Combining these three steps is for computational efficiency only.

The class is called {\tt MalusLawPolarizer} because the {\tt process} function can simulate
the Bell-model Eq.~(\ref{x}) and also, for laboratory experiments, the more relevant model
that complies with Malus law (to be discussed later).

\section{Conclusion}\label{EPRB.CONC}

The EPRB game, and by extension the EBES simulation,
is uniquely adaptable, allowing for the adjustment or even the removal of features
such as the time window or voltage thresholds for pair selection.
This flexibility transforms the simulation into an ``ideal experiment'',
bypassing the physical limitations of a laboratory while providing a robust platform
to compare theoretical assumptions with empirical data.

Acting as a rigorous test bed for theoretical models,
the EBES simulation reveals a critical oversight in the standard EPRB thought experiment:
the assumption that separated particles are automatically identified as a pair.
While Bell's model takes this identification for granted,
laboratory experiments must actively perform this task to generate relevant results.
Consequently, Bell's oversimplified model lacks the necessary complexity to accurately represent experimental data.
Ignoring the concrete mechanics of pair identification creates a fundamental ``blind spot''~\cite{BLINDSPOT}
that renders any direct comparison between such theories and actual EPRB experimental data impossible.

Most significantly, games played with human participants,
designed to replicate laboratory Einstein-Podolsky-Rosen-Bohm experiments,
can exhibit correlations typically attributed to entangled quantum objects.
These results are achieved through strictly local interactions, without necessitating
``spooky action at a distance''.
The core takeaway is that simple, non-quantum systems can produce a series of causal events that,
when viewed collectively, are indistinguishable from the traditional signatures of quantum entanglement.


\ifFULLBOOK
\begin{subappendices}

\section{Winning the EPRB game}\label{EPRBwin}

From the rules of the EPRB game laid out in sections~\ref{PART1}~--~\ref{PART3},
it follows that the number pairs $N$, the size of the window $W$, and
the time delay function $T(\varphi,a)$ determine the outcome of the game.
The other two parameters, $T$ and $T'$ merely serve to guarantee that in a real-life
scenario, walkers can never communicate with each other. They are redundant
for the formulation in terms of a computer program.
Moreover, $T$ only sets the time scale and can be eliminated by redefining
$W\rightarrow W/T$, etc.
Therefore, it is legitimate to consider $W$ as a free dimensionless parameter.

How to choose the two relevant parameters and function such
that the simulation reproduces the results of the theoretical models is considered next.
In view of the random character of the processes involved, it is necessary to collect statistics
about many pairs $N$. Thus, $N$ should be large, say $N=100000$ or larger.

In the penultimate paragraph of section~\ref{PART2}, it was argued that
for large $N$ and all possible differences $\theta$ of the two angles,
$\langle x_{1}\rangle\approx 0$, $\langle x_{2}\rangle\approx0$,
$\langle x_{1}\rangle_{\mathrm{W}}\approx0$, and $\langle x_{2}\rangle_{\mathrm{W}}\approx 0$,
independent of the choice of $t_{\mathrm{max}}$ and $W$.
Clearly, this is a rather boring result.
However, for appropriate choices of $T(\varphi,a)$ and $W$,
the $\theta$ dependence of the correlations can be more interesting (as show above).
Therefore:

{\color{blue}
\begin{center}
\framebox{
\parbox[t]{0.9\hsize}{%
Winning the EPRB game is to establish the conditions, meaning
the values of the window $W$ and the function $T(\varphi,a)$
such that the correlation Eq.~(\ref{eprbwin3}) or
Eq.~(\ref{eprbloc3}) plotted as a function
of $\theta$ closely follows the curve  $-\cos2\theta$.
}}%
\end{center}
}
Recall that the only reason for requiring the correlation Eq.~(\ref{eprbwin3}) or Eq.~(\ref{eprbloc3})
to follows the curve  $-\cos2\theta$
is that the latter is the correlation obtained from the quantum theoretical description
of an ensemble of two photons with their polarizations in the singlet state~\cite{BALL03}
and that some laboratory EPRB experiments have produced data that fits well to $-\cos2\theta$, see Fig.~\ref{fig.A5}.
The reason for {\bf not} imposing a similar requirement to the correlation Eq.~(\ref{eprbave3})
is that by using Eq.~(\ref{x}) and straightforward calculus, it can be shown that as $N$ tends to infinity~\cite{BELL64}
\begin{eqnarray}
\langle x_{1}\,x_{2}\rangle
=\left\{
\begin{array}{ccc}
-1 + 4\theta/\pi & \hbox{if}& 0\le \theta \le {\pi}/{2}\\
\\
-1 + 4(\pi-\theta)/\pi &\hbox{if}& \pi/2\le \theta \le \pi\\
\end{array}
\right.
\label{eprbave4}
\;,
\end{eqnarray}
which, when plotted, is a triangular function (the dotted line called Bell model in Fig.~\ref{eprbresults1})
rather than the desired function $-\cos2\theta$.
Therefore, if the timestamp data is excluded from the analysis,
the data generated by playing the EPRB game with a large number of pairs {\bf cannot} yield
a good approximation to the function $-\cos2\theta$, in concert with Bell's theorem~\cite{BELL93}.

On the other hand, with a little more calculus, it can be shown that for particular values of the parameter $d$,
choosing $T(\varphi,a)=T_0 |\sin(\varphi-a)|^d$ for the time-delay function
exactly yields the correlation $-\cos2\theta$ if
$N$ tends to infinity and $W$ tends to zero afterwards~\cite{RAED06c,RAED07b}.
With all this in mind, the simulation program of the EPRB game implements
$T(\varphi,a)=T_0 |\sin2(\varphi-a)|^d$ for the time-delay function.
The user can set
the parameters $T_0$ (variable name {\tt maximum\_time\_delay}) and $d$ ({\tt time\_delay\_parameter})
via the GUI or command line options.

\section{What is the issue really?}\label{ISSUE}

Referring to Fig.~\ref{eprbidea}, imagine that the particle traveling to the left (right)
is already very close to the leftmost (rightmost) magnet
but has not yet interacted with it,
and that any faster-than-light communication between the particles is explicitly excluded.
Also assume that the distance between the two Stern-Gerlach magnets is so large that a signal
sent by one particle cannot reach the other before both particles complete
their trajectories and arrive at their respective detectors.
Under these conditions, changing the orientation of the left (right) magnet
cannot influence the particle passing through the right (left) magnet.
Consequently, knowledge of $A_{n}$'s ($B_{n}$'s) cannot affect the value of $B_{n}$'s ($A_{n}$'s).
The behavior of the particle going left is therefore completely independent of the behavior of the particle going right.

Next, consider the case in which
the directions of the two magnets are either parallel or antiparallel, that is $\ba=\pm\bb$.
Then, according to the features of the data listed above,
\begin{enumerate}
\item[a.]
The values of the $A_{n}$ and $B_{n}$ for the $n$th pair,
are unpredictable, randomly taking values $\pm1$.
\item[b.]
The value of the product $A_{n}B_{n}=\mp1$ for all $n=1,\ldots,N$ pairs,
depending on the direction $\ba=\pm\bb$ of the magnets.
\end{enumerate}
Thus, even though the value of, say $A_{n}$ is unpredictable (see feature a),
once it is known, the assumed perfect (anti)correlation (see feature b)
allows us to predict the value $B_{n}$, even before it is actually recorded.

This feature is often highlighted in popular and scientific discussions of the EPRB experiment,
typically invoking the supposed magic of ``spooky action at a distance'',
where two particles appear to influence each other instantaneously,
an idea frequently presented as a hallmark of the ``quantum world'' and one that Einstein famously rejected.
It is somewhat ironic that, in his now-classic paper,
Bell~\cite{BELL64} introduced a remarkably simple model that (i) describes the two-particle system
in terms of two independent one-particle systems and
(ii) provides a scientifically sound explanation of the very feature just mentioned
(see the two lines above Eq.~(8) in Ref.~\cite{BELL64}).
Bell's model thus offers a perfectly rational account of the alleged magic
by introducing a variable that predetermines the outcomes of the supposedly random measurements.
What, then, is the issue?

Recall that the dataset in Eq.~(\ref{DATA0}) corresponds to a specific choice of $(\ba,\bb)$.
Naturally, the averages and correlations are expected to vary with these settings.
In this scenario, the two particles are spatially separated;
the outcome for a single particle depends strictly on its local Stern-Gerlach orientation
and the shared variables established at the source.

To implement the notion of separability, J.S. Bell proposed to model the correlations
of an EPRB thought experiment by~\cite{BELL71,BELL93}
\begin{align}
{\cal C}(\ba,\bb)=\int {A}(\ba,\lambda){B}(\bb,\lambda)\,\mu(\lambda)\,d\lambda
\;,\;&
|{A}(\ba,\lambda)|\le1
\;,\;
|{B}(\bb,\lambda)|\le1\;,
\nonumber \\
&0\le\mu(\lambda)
\;,\;\int \mu(\lambda)\,d\lambda=1
\;,
\label{IN0}
\end{align}
where ${A}(\ba,\lambda)$ and ${B}(\bb,\lambda)$ are
functions of the conditions $\ba$ and $\bb$, respectively,
and the common variable $\lambda$ denoting an arbitrary set of ``hidden'' variables.
Bell gave a proof that ${\cal C}(\ba,\bb)$
cannot arbitrarily closely approximate the correlation $-\ba\cdot\bb$ {\it for all}
unit vectors $\ba$ and $\bb$~\cite{BELL64}.
According to Bell himself (see Ref.~\cite{BELL93}(p.65)), {\bf this is the theorem.}
In discretized form,  Bell's theorem says
that there do not exist functions $A_n(\ba)=\pm1$ and $B_n(\bb)=\pm1$
such that for any choice of $N$, the correlation $\left[A_1(\ba)B_1(\bb)+\ldots+A_n(\ba)B_n(\bb)\right]/N$ can
arbitrarily closely approximate the function $-\ba\cdot\bb$ for all possible choices of $\ba$ and $\bb$.
Consequently, although Bell's simple, so-called ``local realistic'' model can reproduce the main features (a) and (b),
it contradicts the quantum-theoretical prediction of the EPRB thought experiment,
namely $-\ba\cdot\bb=-\cos\theta$ where $\theta$ is the angle between the directions
$\ba$ and $\bb$. That is the issue.

\section{Laboratory experiments}\label{LABEXP}

\begin{figure}[!bp]
\centering
\includegraphics[width=0.90\hsize]{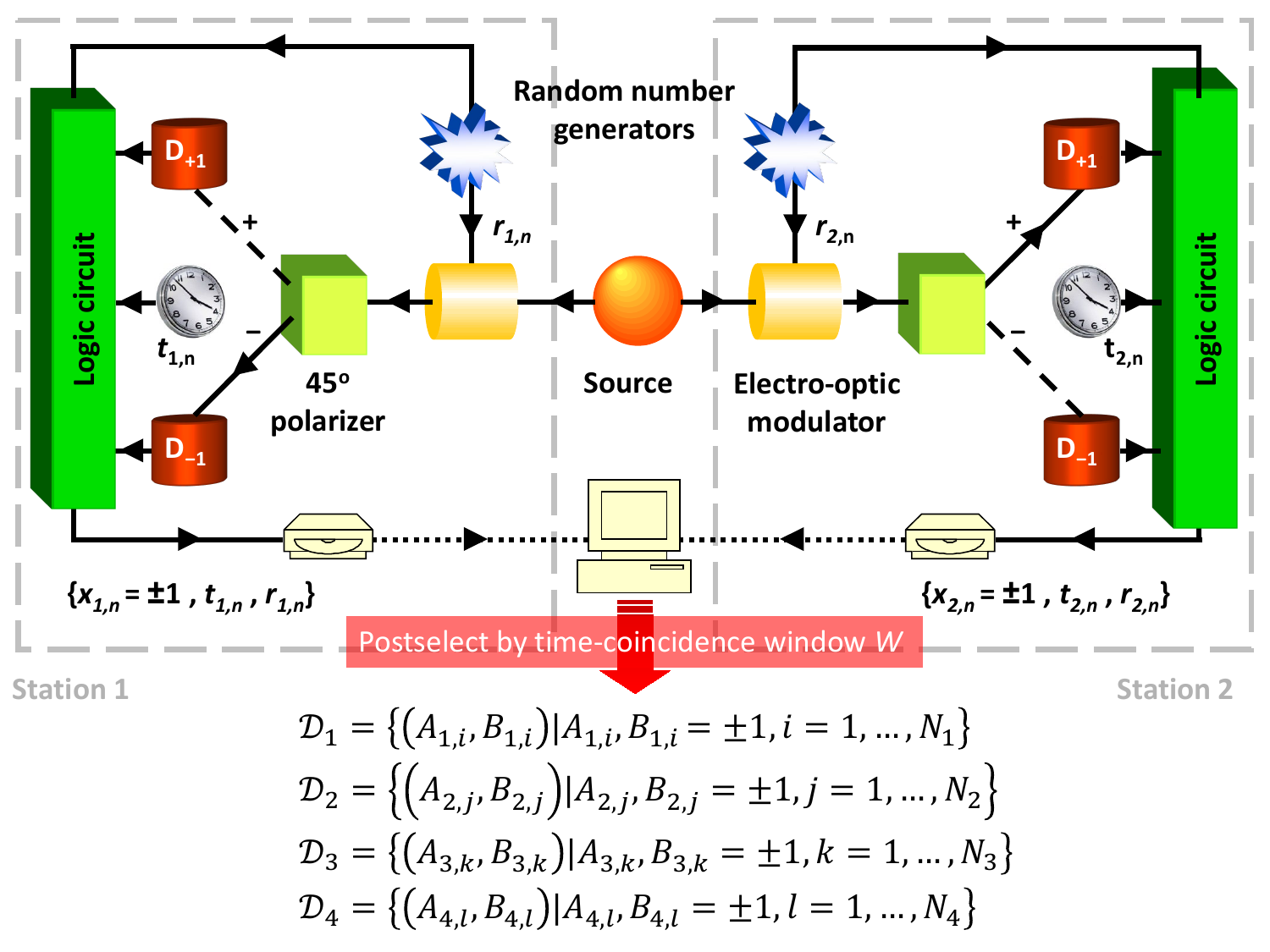}
\caption{Schematic diagram of an EPRB experiment with photons performed by Weihs {\sl et al.}~\cite{WEIH98,WEIH00}.
}
\label{eprbexp}
\end{figure}

\begin{figure}[!htp]
\centering
\includegraphics[width=0.90\hsize]{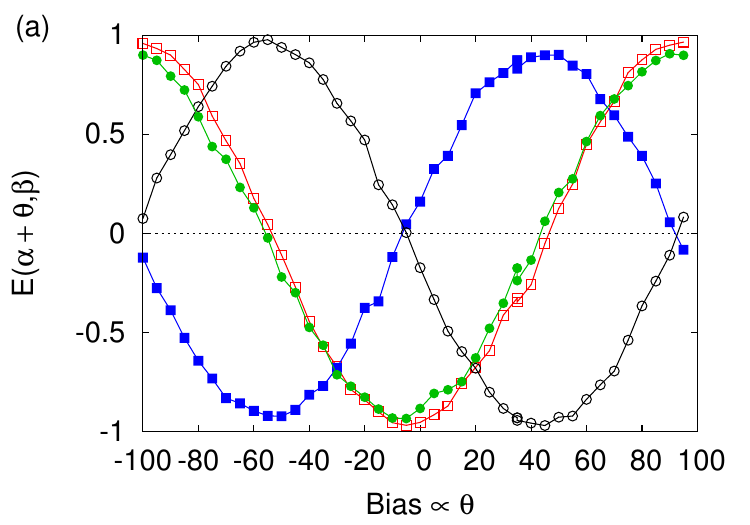} 
\caption{%
The correlation $E(\alpha+\theta,\beta)$ as a function of the bias applied to Alice's EOM is shown for a coincidence window $W=2\,\mathrm{ns}$, based on an analysis of the experimental data set {\bf scanblue1} recorded in the EPRB experiments of Weihs et al.~\cite{WEIH00}.
In this experiment, the angle $\theta$, proportional to the bias applied to the EOMs, is updated every 5 seconds~\cite{WEIH00}.
Each 5-second interval yields at least 7500 coincidence events.
The minimum average time between individual detection events is approximately $24000\,\mathrm{ns}$, which is four orders of magnitude larger than the coincidence $W=2\,\mathrm{ns}$ used to compute the correlations.
For each emitted photon pair, the measurement settings $(\alpha,\beta)$ are chosen randomly from four possible combinations~\cite{WEIH98,WEIH00}.
Open squares: $(\alpha,\beta)=(a,c)$;                
solid squares: $(\alpha,\beta)=(a,d)$;               
open circles: $(\alpha,\beta)=(b,c)$;                
solid circles: $(\alpha,\beta)=(b,d)$,               
where $a=0$, $b=\pi/4$, $c=\pi/8$ and $d=3\pi/8$.
These correlations cannot be obtained from Bell's model~\cite{RAED23} but, for instance the blue markers,
follow the curve obtained by the EBES, see Fig.~\ref{eprbresults3}.
}
\label{fig.A5}
\end{figure}

The central question is, of course, {\it what outcome does an EPRB laboratory experiment actually produce}.
Figure~\ref{eprbexp} presents a schematic of an EPRB setup, adapted from Ref.~\cite{WEIH98},
which provides a concrete basis for comparing experimental results with the predictions of quantum theory and
the EBES data.

The source emits pairs of photons in spatially separated directions.
Photons arriving at station $1$ pass through an electro-optic
modulator (EOM) which rotates the polarization of the photon that passes through it
by an angle corresponding to the voltage applied to that EOM.
The latter is controlled by a binary variable $r_{1,n}$, which is chosen at random~\cite{WEIH98,WEIH00}.
The two different angles to choose from in station 1 (2) are
denoted by $a$ ($c$) and $b$ ($d$).

As the photon leaves the EOM, a polarizing beam splitter
directs the photon to either detector $D_{+1}$ or detector $D_{-1}$.
Depending of the detection efficiency (about 5\%~\cite{WEIH98}),
one of these detectors fires, producing either a signal $x_{1,n}=+1$
or a signal $x_{1,n}=-1$ and a time stamp $t_{1,n}$.
Each triple $(x_{1,n},t_{1,n},r_{1,n})$ is written to a file.
The same holds for photons arriving at station $2$.
After all data has been written to the two files, that is when the experiment has finished,
a time window $W$ is used to remove all data that does not satisfy a time-coincidence criterion~\cite{WEIH98,WEIH00}.
The remaining data is organized in four data sets,
corresponding to the four different values of the pairs of random numbers $(r_{1,n},r_{2,n})$.
Note that the values of the $A$'s ($B$'s) that appear in say ${\cal D}_{\Cac}$ (${\cal D}_{\Cbc}$),
and those that appears in say ${\cal D}_{\Cad}$ may be different,
even though they may have been recorded for the same angle $a$ ($c$).
The four sets of discrete data ${\cal D}_{\Cac}$, ${\cal D}_{\Cad}$, ${\cal D}_{\Cbc}$, and ${\cal D}_{\Cbd}$
are the result of the experiment for the particular value of the time-coincidence window $W$.
As Fig.~\ref{fig.A5} shows, the experimental data for the correlations resemble a cosine quite well.
However, the data for the averages is in blatant contradiction
with the quantum-theoretical description of the EPRB thought experiment~\cite{ADEN09,RAED23}.

An in-depth discussion of the results of this and similar laboratory experiments,
as well as discussion of the (ir)relevance of Bell inequalities and Bell's theorem
from the perspective of data is out of the scope of the book, see Ref.~\cite{RAED23,RAED24}.

\end{subappendices}
\fi


\chapter*{Epiloge}\label{EPILOGE}
\addcontentsline{toc}{chapter}{Epilog}
\pagestyle{plain}
The journey through this book has traced a path that runs counter to much of the received wisdom in quantum physics.
From the earliest chapters, the central message has been clear:
when we shift our perspective from abstract amplitudes and operators to {\bf event-level processes},
the fog surrounding many ``mysterious'' quantum phenomena begins to lift.
What remains is not a diminished physics, but a more transparent one;
one in which individual clicks, rotations, and decisions form the raw material of understanding.

The event-by-event simulation (EBES) framework shows that the heart of quantum physics experiments can be reconstructed from simple rules,
executed one event at a time. No wavefunctions are invoked, no superpositions assumed, no entanglement postulated.
Yet the simulations reproduce the very patterns and correlations that have long been taken as evidence of quantum theory's indispensability.
This outcome does not invalidate quantum theory; rather, it reveals that its probabilistic structure
 may not be the only lens through which nature can be understood.

Throughout the book, experiments, some of them traditionally framed as paradoxes, have been recast as {\bf games}
played by agents following clear instructions.
This shift is more than pedagogical. It demonstrates that the explanatory power of quantum theory can, in many cases,
be matched by models grounded in classical logic and local information processing.
The EBES software, discussed in detail and made available for exploration, stands as a practical testament to this claim.
Anyone with curiosity and a computer can reproduce the simulations and verify the results for themselves.

Of course, EBES is not a replacement for quantum theory.
It does not address quantization, nor does it attempt to describe the microscopic structure of matter.
Instead, it highlights a conceptual gap: quantum theory predicts probabilities, but laboratory experiments produce events.
Bridging this gap requires a framework capable of describing individual occurrences without invoking statistical ensembles from the outset.
EBES provides such a framework, and in doing so, it invites us to reconsider what is truly ``classical'' and what is genuinely ``quantum''.

The implications of this shift are still unfolding.
If interference, diffraction, and entanglement can be modeled through event-level rules,
what other domains might benefit from this perspective?
Could computational electrodynamics, long constrained by wave-based formulations,
be reimagined through particle-level algorithms?
Might new numerical methods emerge from this granular viewpoint?
These questions remain open, and their answers will depend on the creativity and persistence of future researchers.

As you close this book, the hope is not that you adopt EBES as a new doctrine,
but that you carry forward a renewed sense of possibility.
Physics advances when its foundations are questioned with clarity and rigor.
The event-by-event approach is one such questioning, an invitation to explore, simulate, and rethink.
Whether you are a student, a researcher, or simply a curious mind, the tools and ideas presented here are yours to use, test, and challenge.

The final message is simple: the world of individual events is rich enough to tell its own story.
We need only listen, one click at a time.

\chapter*{Acknowledgments}\label{ACKN}
\addcontentsline{toc}{chapter}{Acknowledgments}
\pagestyle{plain}

Thanks to Koen De Raedt for teaching his father how to write Javascript, setting up the Python framework on which the
event-by-event simulation codes are based, for his help during the development of the
Python programs, and for his original contributions to the development of the event-by-event simulation approach.
Also thanks to brother Bart De Raedt for critical reading of the manuscript and
for making pertinent comments.

The bundling of the material collected in the book and software was inspired by years of extensive discussions
with Martin Warnke, Arianna Borrelli, and Wolfgang Hagen whose insights continually challenged and enriched my thinking.
Their perspectives helped me refine the questions that guided this work and illuminated paths I would
not have discovered on my own.

\medskip
\noindent
{\bf Declaration of AI Assistance}

\medskip
\noindent
The author acknowledge the use of Gemini and Microsoft Copilot
to assist in the linguistic refinement, structural organization,
and stylistic polishing of the technical descriptions presented in the various chapters and web pages.
While the AI provided support in optimizing the prose for clarity and academic tone,
the conceptual development of the event-by-event simulation, the underlying physical models,
the simulation algorithms, and the final presentation and interpretation of the simulation data
remain the original work and sole responsibility of the author.

\appendix
\chapter{Installing the software}\label{SOFTWARE}
Click on or go to
\href{\HOMEURL/DES/software.html}{\HOMEURL/DES/software.html}
for the latest information on how to download and install the software.

\chapter{EBES bibliography}\label{EBESBIB}
This appendix provides an annotated bibliography of selected publications
detailing EBES implementations for various experiments.
The majority of these expand upon the methodology described in this book,
illustrating the broad applicability of EBES to topics that lie beyond the scope of this book.
Links to arXiv papers are included for all publications that are not otherwise publicly accessible.

\medskip\noindent
{\bf The basic ideas of the EBES appear in:}
\begin{itemize}
\item
Deterministic event-based simulation of quantum phenomena,
K. De Raedt, H. De Raedt, and K. Michielsen, Comp. Phys. Comm. 171, 19 -- 39 (2005);
\hfill\break
\href{https://doi.org/10.48550/arXiv.quant-ph/0409213}{https://doi.org/10.48550/arXiv.quant-ph/0409213}
\item
Event-based simulation of single-photon beam splitters and Mach-Zehnder interferometers,
K. De Raedt, H. De Raedt, and K. Michielsen, Europhys. Lett. 69, 861 -- 867 (2005);
\hfill\break
\href{https://doi.org/10.48550/arXiv.quant-ph/0501141}{https://doi.org/10.48550/arXiv.quant-ph/0501141}
\item
Efficient data processing and quantum phenomena: Single-particle systems,
H. De Raedt, K. De Raedt, K. Michielsen, and S. Miyashita, Comp. Phys. Comm. 174, 803 -- 817 (2006);
\hfill\break
\href{https://doi.org/10.48550/arXiv.quant-ph/0512233}{https://doi.org/10.48550/arXiv.quant-ph/0512233}
\end{itemize}
{\bf with an early application to quantum computation:}
\begin{itemize}
\item
Simulation of quantum computation: a deterministic event-based approach,
K. Michielsen, K. De Raedt, and H. De Raedt, J. Comput. Theor. Nanosci. 2, 227 -- 239 (2005);
\hfill\break
\href{https://doi.org/10.48550/arXiv.quant-ph/0501140}{https://doi.org/10.48550/arXiv.quant-ph/0501140}.
\end{itemize}

\medskip\noindent
{\bf Two-beam interference:}
\begin{itemize}
\item
Corpuscular model of two-beam interference and double-slit experiments with single photons,
F. Jin, S. Yuan, H. De Raedt, K. Michielsen, and S. Miyashita, J. Phys. Soc. Jpn. 79, 074401 (2010);
\hfill\break
\href{https://doi.org/10.48550/arXiv.1005.0906}{https://doi.org/10.48550/arXiv.1005.0906}
\end{itemize}

\medskip\noindent
{\bf Delayed-choice experiments}
\begin{itemize}
\item
EBES of the delayed-choice experiment reported in Ref.~\cite{JACQ07}:

\noindent
Computer simulation of Wheeler's delayed choice experiment with photons,
S. Zhao, S. Yuan, H. De Raedt and K. Michielsen, Europhys. Lett. 82, 40004 (2008);
\hfill\break
\href{https://doi.org/10.48550/arXiv.0712.1606}{https://doi.org/10.48550/arXiv.0712.1606}
\item
EBES of the delayed-choice experiment reported in Ref.~\cite{JACQ08}:

\noindent
Coexistence of full which-path information and interference in Wheeler's delayed choice experiment with photons,
K. Michielsen, S. Yuan, S. Zhao, F. Jin, H. De Raedt, Physica E 42, 348 -- 353 (2010);
\hfill\break
\href{https://doi.org/10.48550/arXiv.0908.1032}{https://doi.org/10.48550/arXiv.0908.1032}

\item
EBES of the delayed-choice thought experiment with the movable beam splitter in a MZI configuration:
\noindent
Corpuscular event-by-event simulation of quantum optics experiments:
application to a quantum-controlled delayed-choice experiment,
H. De Raedt, M. Delina, F. Jin, and K. Michielsen,
Phys. Scr. 2012, T151, 014004 (2012); 
\hfill\break
\href{https://doi.org/10.48550/arXiv.1208.2369}{https://doi.org/10.48550/arXiv.1208.2369}

\item
Quantum delayed-choice laboratory experiment reported in Ref.~\cite{Tang2012}:

\noindent
Event-by-event simulation of a quantum delayed-choice experiment (includes Python code),
H. C. Donker, H. De Raedt, and K. Michielsen, Comp. Phys. Comm. 185, 3109 -- 3118 (2014);
\hfill\break
\href{https://doi.org/10.48550/arXiv.1408.5593}{https://doi.org/10.48550/arXiv.1408.5593}
\end{itemize}

{\bf Quantum eraser laboratory experiment, as reported in Ref.~\cite{SCHW99}:}
\begin{itemize}
\item
Event-by-event simulation of a quantum eraser experiment,
F. Jin, S. Zhao, S. Yuan, H. De Raedt, and K. Michielsen, J. Comp. Theor. Nanosci. 7, 1771 -- 1782 (2010);
\hfill\break
\href{https://doi.org/10.48550/arXiv.0908.1036}{https://doi.org/10.48550/arXiv.0908.1036}
\end{itemize}

{\bf Two-particle correlation experiments (excluding EPRB):}
\begin{itemize}
\item
Hanbury Brown-Twiss experiment:

\noindent
Event-by-event simulation of the Hanbury Brown-Twiss experiment with coherent light,
F. Jin, H. De Raedt, and K. Michielsen, Commun. Comput. Phys. 7, 813 -- 830 (2010);
\hfill\break
\href{https://doi.org/10.48550/arXiv.0908.1040}{https://doi.org/10.48550/arXiv.0908.1040}
\item
Hanbury Brown-Twiss and Ghosh-Mandel experiments:

\noindent
Nonclassical effects in two-photon interference experiments:
\hfil\break
event-by-event simulations,
K. Michielsen, F. Jin, and H. De Raedt, Proc. of SPIE Vol. 8832 88321L-1(2013);
\hfill\break
\href{https://doi.org/10.48550/arXiv.1312.6357}{https://doi.org/10.48550/arXiv.1312.6357}
%
\end{itemize}

{\bf Several experiments with individual neutrons, including Ref.~\cite{HAME75,ERHA12,SULY13,DENK14}:}
\begin{itemize}
\item
Event-Based Simulation of Neutron Interferometry Experiments,
H. De Raedt, F. Jin, and K. Michielsen, Quantum Matter 1, 20 -- 40 (2012);
\hfill\break
\href{https://doi.org/10.48550/arXiv.1208.2367}{https://doi.org/10.48550/arXiv.1208.2367}
\item
Discrete-event simulation of uncertainty in single-neutron experiments,
H. De Raedt and K. Michielsen, Front. Physics 2:14,
\hfill\break
\href{https://doi.org/10.3389/fphy.2014.00014}{https://doi.org/10.3389/fphy.2014.00014}
\item
Discrete-Event Simulation Unmasks the Quantum Cheshire Cat,
K. Michielsen, Th. Lippert, and H. De Raedt, J. Comp. Theor. Nanosci. 14, 2268 -- 2283 (2017);
\hfill\break
\href{https://doi.org/10.48550/arXiv.1707.04230}{https://doi.org/10.48550/arXiv.1707.04230}
\item
Classical, Quantum and Event-by-Event Simulation of a Stern-Gerlach Experiment with Neutrons,
H. De Raedt, F. Jin, and K. Michielsen, Entropy 24, 1143 (2022);
\hfill\break
\href{https://doi.org/10.3390/e24081143}{https://doi.org/10.3390/e24081143}
\end{itemize}

{\bf Many (quantum) optics experiments:}
\begin{itemize}
\item
Event-based Corpuscular Model for Quantum Optics Experiments,
\hfil\break
K. Michielsen, F. Jin, and H. De Raedt, J. Comp. Theor. Nanosci. 8, 1052 -- 1080 (2011);
\hfill\break
\href{https://doi.org/10.48550/arXiv.1006.1728}{https://doi.org/10.48550/arXiv.1006.1728}
\end{itemize}

{\bf Optics:}
\begin{itemize}
\item
Event-based simulation of light propagation in lossless dielectric media,
\hfil\break
B. Trieu, K. Michielsen, and H. De Raedt, Comp. Phys. Commun. 182, 726 -- 734 (2011);
\hfill\break
\href{https://doi.org/10.48550/arXiv.1012.2437}{https://doi.org/10.48550/arXiv.1012.2437}
\end{itemize}

{\bf Quantum cryptography:}
\begin{itemize}
\item
Event-by-event simulation of quantum cryptography protocols,
S. Zhao and H. De Raedt, J. Comp. Theor. Nanosci. 5, 490 -- 504 (2008);
\hfill\break
\href{https://doi.org/10.48550/arXiv.0708.1734}{https://doi.org/10.48550/arXiv.0708.1734}
\end{itemize}

{\bf EBES of the quantum walk experiment reported in Ref.~\cite{ROBE15}:}
\begin{itemize}
\item
Discrete-Event Simulation of Quantum Walks,
M. Willsch, D. Willsch, \hfil\break
K. Michielsen, and H. De Raedt, Front. Phys. 8:145 (2020);
\hfill\break
\href{https://doi.org/10.3389/fphy.2020.00145}{https://doi.org/10.3389/fphy.2020.00145}
\end{itemize}

{\bf Einstein-Podolsky-Rosen-Bohm experiments:}
\begin{itemize}
\item
A local realist model for correlations of the singlet state,
K. De Raedt, K. Keimpema, H. De Raedt, K. Michielsen, and S. Miyashita, Euro. Phys. J. B 53, 139 -- 142 (2006);
\hfill\break
\href{https://doi.org/10.1140/epjb/e2006-00364-9}{https://doi.org/10.1140/epjb/e2006-00364-9}

\item
A computer program to simulate Einstein-Podolsky-Rosen-Bohm experiments with photons
K. De Raedt, H. De Raedt, K. Michielsen, Comp. Phys. Comm. 176, 642 -- 651, (2007);
\hfill\break
\href{https://doi.org/10.1016/j.cpc.2007.01.007}{https://doi.org/10.1016/j.cpc.2007.01.007}
\item
Event-based computer simulation model of Aspect-type experiments strictly satisfying Einstein's locality conditions,
H. De Raedt, K. De Raedt, K. Michielsen, K. Keimpema, and S. Miyashita, J. Phys. Soc. Jpn. 76, 104005 (2007)
\hfill\break
\href{https://doi.org/10.48550/arXiv.0712.2565}{https://doi.org/10.48550/arXiv.0712.2565}
\item
Event-by-event simulation of Einstein-Podolsky-Rosen-Bohm experiments,
S. Zhao, H. De Raedt, and K. Michielsen, Found. Phys. 38, 322 -- 347 (2008);
\hfill\break
\href{https://doi.org/10.1007/s10701-008-9205-5}{https://doi.org/10.1007/s10701-008-9205-5}
\item
The digital computer as a metaphor for the perfect laboratory experiment:
\hfil\break
Loophole-free Bell experiments,
\hfil\break
H. De Raedt, K. Michielsen and K. Hess, Comp. Phys. Comm. 209, 42 -- 47 (2016);
\hfill\break
\href{https://doi.org/10.1016/j.cpc.2016.08.010}{https://doi.org/10.1016/j.cpc.2016.08.010}
\item
Discrete-event simulation of an extended Einstein-Podolsky-Rosen-Bohm experiment,
H. De Raedt, M.S. Jattana, D. Willsch, M. Willsch, F. Jin, and K. Michielsen, Front. Phys. 8:160 (2020);
\hfill\break
\href{https://doi.org/10.3389/fphy.2020.00160}{https://doi.org/10.3389/fphy.2020.00160}
\end{itemize}

{\bf Reviews:}
\begin{itemize}
\item
Event-by-event simulation of quantum phenomena,
\hfil\break
H. De Raedt and K. Michielsen, Ann. Phys. 524, 393 -- 410 (2012);
\hfill\break
\href{https://doi.org/10.1002/andp.201100299}{https://doi.org/10.1002/andp.201100299}
\item
Event-based simulation of quantum physics experiments,
K. Michielsen and H. De Raedt, Int. J. Mod. Phys. C 25, 1430003 (2014)
\hfill\break
\href{https://doi.org/10.48550/arXiv.1312.6942}{https://doi.org/10.48550/arXiv.1312.6942}
\end{itemize}

{\bf From discrete events to quantum theory:}
\begin{itemize}
\item
Quantum theory as the most robust description of reproducible experiments,
H. De Raedt, M.I. Katsnelson, and K. Michielsen,
Ann. Phys. 347, 45 -- 73 (2014);
\hfil\break
\href{https://doi.org/10.1016/j.aop.2014.04.021}{https://doi.org/10.1016/j.aop.2014.04.021}
\item
Quantum theory as a description of robust experiments: Derivation of the Pauli equation,
H. De Raedt, M.I. Katsnelson, H.C. Donker, and K. Michielsen,
Ann. Phys. 359, 166 -- 186 (2015);
\hfil\break
\href{https://doi.org/10.1016/j.aop.2015.04.017}{https://doi.org/10.1016/j.aop.2015.04.017}
\item
Logical inference approach to relativistic quantum mechanics: Derivation of the Klein-Gordon equation,
H.C. Donker, M.I. Katsnelson, H. De Raedt, and K. Michielsen,
Ann. Phys. 372 74-82 (2016);
\hfil\break
\href{https://doi.org/10.1016/j.aop.2016.04.018}{https://doi.org/10.1016/j.aop.2016.04.018}
\item
Quantum theory as plausible reasoning applied to data obtained by robust experiments,
H. De Raedt, M.I. Katsnelson, and K. Michielsen,
Phil. Trans. R. Soc. A 374: 20150233 (2016);
\hfil\break
\href{https://doi.org/10.1098/rsta.2015.0233}{https://doi.org/10.1098/rsta.2015.0233}
\item
Logical inference derivation of the quantum theoretical description of Stern-Gerlach and
Einstein-Podolsky-Rosen-Bohm experiments,
H. De Raedt, M.I. Katsnelson, and K. Michielsen,
Ann. Phys. 396, 96 -- 118 (2018);
\hfil\break
\href{https://doi.org/10.1016/j.aop.2018.07.014}{https://doi.org/10.1016/j.aop.2018.07.014}
\item
Separation of conditions as a prerequisite for quantum theory,
H. De Raedt, M.I. Katsnelson, D. Willsch, and K. Michielsen,
Ann. Phys. 403, 112-135 (2019);
\hfil\break
\href{https://doi.org/10.1016/j.aop.2019.01.012}{https://doi.org/10.1016/j.aop.2019.01.012}
\item
Einstein-Podolsky-Rosen-Bohm experiments: A discrete data driven approach,
H. De Raedt, M.I. Katsnelson, M.S. Jattana, V. Mehta, M. Willsch, D. Willsch, K. Michielsen, and F. Jin,
Ann. Phys. 453, 169314 (2023);
\hfill\break
\href{https://doi.org/10.1016/j.aop.2023.169314}{https://doi.org/10.1016/j.aop.2023.169314}
\hfill\break
\noindent
with a major correction in section 11.4.2:
\hfill\break
\href{https://arxiv.org/abs/2304.03962}{https://arxiv.org/abs/2304.03962}
\end{itemize}

%

\bibliography{/D/papers/all26}
\end{document}